\documentclass[]{aastex631}

\pdfoutput=1 
\usepackage{amsmath,amstext}
\usepackage[T1]{fontenc}
\usepackage{makecell}
\usepackage{gensymb}
\usepackage{subfigure}
\usepackage{longtable}
\usepackage{savesym}
\savesymbol{tablenum}
\usepackage{siunitx}
\restoresymbol{SIX}{tablenum}
\begin{document}

\title{Paving the Road to the Habitable Worlds Observatory with High-Resolution Imaging \\ I: New and Archival Speckle Observations of Potential HWO Target Stars}

\correspondingauthor{Zachary Hartman}
\email{zachary.hartman366@gmail.com}
\author[0000-0003-4236-6927]{Zachary D. Hartman}
\affiliation{NASA Ames Research Center, Moffett field, CA 94035, USA}

\author[0000-0002-2361-5812]{Catherine A. Clark}
\affiliation{NASA Exoplanet Science Institute, IPAC, California Institute of Technology, Pasadena, CA 91125, USA}

\author[0000-0003-2527-1598]{Michael B. Lund}
\affiliation{NASA Exoplanet Science Institute, IPAC, California Institute of Technology, Pasadena, CA 91125, USA}

\author[0000-0002-9903-9911]{Kathryn V. Lester}
\affiliation{Mount Holyoke College, South Hadley MA 01075, USA}

\author[0000-0002-7349-1387]{Jos\'{e} A. Caballero}
\affiliation{Centro de Astrobiología (CSIC-INTA), ESAC, Camino bajo del castillo s/n, 28692 Villanueva de la Cañada, Madrid, Spain}

\author[0000-0002-2532-2853]{Steve B. Howell}
\affiliation{NASA Ames Research Center, Moffett field, CA 94035, USA}

\author[0000-0002-5741-3047]{David Ciardi}
\affiliation{NASA Exoplanet Science Institute, IPAC, California Institute of Technology, Pasadena, CA 91125, USA}

\author[0009-0002-9833-0667]{Sarah Deveny}
\affiliation{Bay Area Environmental Research Institute, Moffett Field, CA 94035, USA}

\author[0000-0002-0885-7215]{Mark E. Everett}
\affiliation{NSF NOIRLab, 950 N. Cherry Ave., Tucson, AZ 85719, USA}

\author[0000-0001-9800-6248]{Elise Furlan}
\affiliation{NASA Exoplanet Science Institute, IPAC, California Institute of Technology, Pasadena, CA 91125, USA}

\author[0000-0002-4641-2532]{Venu Kalari}
\affiliation{International Gemini Observatory/ NSF NOIRLab, 670 A'ohoku Place, Hilo, HI 96720, USA}

\author[0000-0001-7746-5795]{Colin Littlefield}
\affiliation{Bay Area Environmental Research Institute, Moffett Field, CA 94035, USA}

\author[0000-0002-4434-2307]{Andrew W. Stephens}
\affiliation{International Gemini Observatory/ NSF NOIRLab, 670 A'ohoku Place, Hilo, HI 96720, USA}

\author[0000-0002-0040-6815]{Jennifer A. Burt}
\affiliation{Jet Propulsion Laboratory, California Institute of Technology, 4800 Oak Grove Drive, Pasadena, CA 91109, USA}


\author[0009-0001-3007-3855]{Guillaume Huber}
\affiliation{Institute for Astronomy, University of Hawai’i, 640 N. Aohoku Place, Hilo, HI 96720, USA}



\author[0000-0001-7233-7508]{Rachel Matson}
\affiliation{U.S. Naval Observatory, 3450 Massachusetts Ave. NW, Washington, D.C. 20392, USA}

\author[0000-0003-2008-1488]{Eric E. Mamajek}
\affiliation{Jet Propulsion Laboratory, California Institute of Technology, 4800 Oak Grove Drive, Pasadena, CA 91109, USA}
\author[0000-0003-3989-5545]{Noah Tuchow}
\affiliation{NASA Goddard Space Flight Center, Greenbelt, Maryland, USA}



\begin{abstract}

One of the key goals of the Habitable Worlds Observatory (HWO) is to directly image about 25 potentially habitable exoplanets and determine their properties.
This challenge will require a large survey of nearby, bright stars -- $\sim$100 according to the Astro2020 Decadel Survey. 
To ensure the success of the mission and to help guide design decisions, the stellar multiplicity of the target stars must be well-understood. 
To this end, we present optical speckle imaging of stars in the NASA Exoplanet Exploration Program (ExEP) provisional HWO star list, which is currently the Tier 1 target list for the HWO Target Stars and Systems Sub-Working Group. 
We obtained new observations using `Alopeke and Zorro at Gemini Observatory and queried the Exoplanet Follow-up Observing Program Archive for archival observations, resulting in speckle imaging data for 80 of the 164 stars. 
We confirmed one candidate companion detected previously by Gaia (HD 90089) and obtained an ambiguous detection of a known companion (HD 212330).
To examine our sensitivity to companions, we simulated stellar companions down to $\sim0.1 M_{\odot}$ for each target and found that 75\%-85\% would be detected in our speckle images; the remaining simulated companions are either too faint or too close-in, and will require follow-up using other methods such as long-term spectroscopic measurements and space-based techniques.
This work represents a first step towards surveying potential HWO targets for close-in stellar companions and helping to inform the target selection process for the HWO direct-imaging survey, bringing us closer towards the discovery of potential habitable worlds.

\end{abstract}



\section{Introduction} \label{sec:intro}
The Astro2020 Decadal Survey laid out a grand vision for the field of astronomy over the coming decades \citep{2020decadal}. 
One of its key recommendations called for NASA to begin planning a new Infrared/Optical/Ultraviolet (IR/O/UV) space mission for launch in the mid-2040s. 
Currently called the Habitable Worlds Observatory (HWO), one of the mission's primary goals is to directly image 25 potentially habitable exoplanets. 
Now, NASA and the astronomical community are beginning to prepare for HWO in a variety of ways, including maturing technology \citep{Scowen2025AAS...24530504S} and developing scientific working groups\footnote{\url{https://science.nasa.gov/astrophysics/programs/habitable-worlds-observatory/wgs/}} to inform the final design of the telescope and the target sample. 
For the direct-imaging survey, this preparation means beginning the process of assembling a large sample of nearby, bright stars with well-defined properties -- such as mass, age, and radius -- from which future scientists will determine the targets most amenable to a search for habitable-zone planets.

As a part of this process, the HWO target star candidates will need to be searched for close stellar companions.
Although the definition of ``close'' in multiple stellar systems is fluid and context-dependent, as described in \citet{2025cifuentes}, for this study, close means companions with angular separations $<3\arcsec{}$, or projected physical separations $\leq$ 30\,au for typical target distances of $\sim10$ pc.
This separation limit matches that of \citet{2024mamajek} and was selected as a reasonable limit for what HWO's starlight suppression techniques might be able to take into account.
Roughly half of solar-type stars (F-, G-, and early K-dwarfs) are expected to form in multi-star systems \citep{1991duquennoy,2010raghavan,2013duchene,2014tokovinin,2023offner}. 
Unbound chance alignments, where one or more stars appear in the background and are unassociated, are also possible.
From a technical perspective, these additional stars -- bound or unbound -- will add more light into the field-of-view, increasing the noise in the observations and potentially over-powering the light from any planets in the system. 
However, advances in multi-star wavefront control may allow HWO to reach the necessary noise floor to detect potentially habitable exoplanets, even in the presence of a close-in stellar companion \citep{2015belikov,2015thomas,2017sirbu,2023sirbu,2021bendek}. 
From a scientific perspective, bound companions with separations less than 100 au may truncate or inhibit the formation of disks around the stars in the system, or gravitationally excite and eject planetesimals, preventing the formation of the planets HWO is hoping to observe \citep{2006quintana,HaghighipourRaymond2007ApJ...666..436H, Jang-Condell2015ApJ...799..147J, RafikovSilsbee2015ApJ...798...69R, RafikovSilsbee2015ApJ...798...70R, 2015thebault, 2021hamers}.
As such, it has been shown that the stellar companions in planet-hosting systems are shifted to larger separations as compared to field multiples \citep{2016kraus, 2021lester, 2022clark, 2024gonzalez}.
Nonetheless, while the likelihood of finding planets in close binaries is lower than around single stars, it has been shown that multi-star systems with separations less than 100 au can host planets \citep{2004zucker,2020bohn,2019matson,2022feng,2024gonzalez,2024sullivan}.
On the other hand, wide stellar systems with separations of 10s to 100s of arcseconds and larger may be especially valuable for HWO as the wide companion would not interfere with the direct imaging observation, but could provide a wealth of information on the system ranging from the metallicity, age, or formation scenarios depending on the nature of the companion \citep{2017tokovinin,2020hawkins,2024heintz}.

To ensure the success of the mission, potential HWO targets must be searched for close-in stellar companions with periods of days to decades, either to remove the binaries from the target list, or to identify and characterize them for proper observation with HWO.
Additionally, observations examining the area where these stars will be in 2040 to identify potential background stars and determine what effect, if any, they would have on HWO observations are needed.
Multiplicity information will need to be gathered using a number of different methods, including long-term spectroscopic measurements and space-based techniques \citep{2018baroch,2019hutter,2021hirsch,2021moe}. 
However, while long-term spectroscopic observations, and subsequent radial velocity measurements, are sensitive to companions with separations less than 20 - 50 milliarcseconds, the detection sensitivity decreases significantly for systems with larger separations \citep{2015teske}. 
Even space-based observatories can miss close-in companions.
Gaia, for example, typically cannot resolve relatively faint companions closer than 0.6\arcsec - 0.8\arcsec as the resolution power strongly depends on the magnitude difference between the two stars \citep{Clark2024AJ....167...56C,2025cifuentes, 2025matson}.

Ground-based, high-resolution imaging, however, is sensitive to subarcsecond companions at moderate contrasts after only a single visit.
In particular, optical speckle imaging on the 8.1-meter Gemini telescopes offers some of the highest angular resolutions (0.025\arcsec{} at 832 nm) and largest magnitude contrasts at small separations (4-5 magnitudes at 0.1\arcsec) of any telescope and instrument combination currently available. 
This allows astronomers to search for faint companions across a wide range of possible separations quickly and efficiently.
As such, high-resolution imaging will play a key role in surveying the environments around potential HWO target stars, as well as characterizing the orbital parameters of any close-in binaries.


Over the past decade, the astronomical community has obtained, reduced, and published a large amount of data on nearby stars that may become HWO targets. 
In this paper, we publish new observations that were taken using the `Alopeke and Zorro speckle imagers.
We also examine a set of high-resolution imaging data that is publicly available on the NASA Exoplanet Follow-up Observing Program (ExoFOP) archive. 
In Section \ref{sec:method}, we show our new observations were obtained, how we retrieved the archival data, and how the archival observations were originally collected.
We also highlight our Gaia and Washington Double Star catalog cross-matches and our literature searches for more companions.
In Section \ref{sec:results}, we present our overall results and notes on individual systems, as well as an analysis of potential companions that may be missing from our observations.
In Section \ref{sec:Discussion}, we compare our 20- and 60-ms observations and discuss the implications of our results on target selection for HWO. 
We present our conclusions in Section \ref{sec:Conclusions}.

\section{Methods} \label{sec:method}

\subsection{Sample Selection}

Our sample is based on the NASA Exoplanet Exploration Program (ExEP) provisional star list for HWO \citep{2024mamajek}, which consists of 164 stars. 
The selection for the ExEP list was based on a number of criteria, including the potential star-planet brightness ratios, the possible albedoes of the planets, the magnitude of the potential planets, the angular size of the habitable zone, and the stellar multiplicity of the systems. 
In particular, \cite{2024mamajek} removed all systems with known companions within 3\arcsec{} that are listed in the Washington Double Star Catalog \citep[WDS;][]{2001mason}, as well as any spectroscopic binaries that are listed in the Ninth Catalogue of Spectroscopic Binary Orbits \citep[SB9;][]{2004A&A...424..727P}.
However, while extensive, neither catalog is complete to all separations or magnitude differences, and the observations are not uniform across all targets.
In particular, many of the observations in the WDS were taken using small telescopes or antiquated techniques and therefore, some stellar companions may have been missed.
A uniform multiplicity study, using the best available facilities, is needed for a full understanding of the stellar multiplicity of the targets on the ExEP list.

\begin{figure}
\centering
\includegraphics[width=\textwidth]{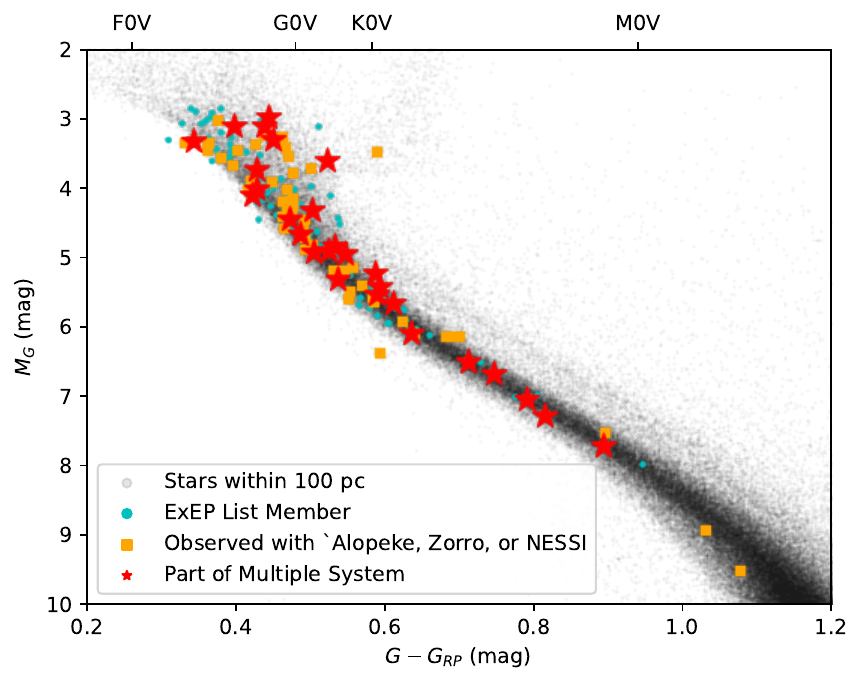}
\caption{\label{fig:hwohr} Color-magnitude diagram showing the targets in the ExEP target list for HWO. The grey points represent a sample of stars within 100 pc taken from Gaia DR3. The cyan points indicate the locations of the \citet{2024mamajek} targets that are in Gaia and have not been observed with speckle imaging. Orange squares highlight 80 ExEP list targets that have `Alopeke, Zorro, or NESSI observations. Red stars show the 27 targets that have been identified as members of stellar multiples by this work. The top axis provides approximate $G-G_{RP}$ values for F0V, G0V, K0V, and M0V stars.}
\end{figure}

As such, we obtained new speckle observations of stars and sought out archival speckle observations of stars on the ExEP provisional star list for HWO. 
Table \ref{tab:sample} includes the TESS Input Catalog (TIC, \citealt{2019stassun,2021paegert}) ID, Gaia DR3 ID, Simbad Name, R.A., decl., $\mu_{R.A.}$, $\mu_{decl.}$, Parallax, Gaia $G$ Magnitude, Gaia $G-G_{RP}$ color, and Gaia radial velocity ($V_r$) for each star targeted in this study. 
Figure \ref{fig:hwohr} shows a color-magnitude diagram of the targets surveyed throughout this work as compared to other stars within 100 pc.

\subsection{Archival and New Speckle Imaging}

We observed 34 stars using `Alopeke and Zorro on the 8.1-meter Gemini North and South telescopes, respectively \citep{2021scott}.
We note that one target, HD 95735, has two TIC IDs associated with it, TIC 166646191 and TIC 353969903.
We take any observations associated with either of the entries to be of HD 95735 and use TIC 166646191 as the TIC ID used in this study.
Four of these 34 were listed in Appendix E of \citet{2024mamajek}. 
These Appendix E targets are listed in the WDS as potential binaries, but subsequent studies have cast doubt on that determination. 
We therefore sought to clarify the multiplicity of these systems.

We also cross-matched the 164 targets on the ExEP provisional star list for HWO with the ExoFOP archive to search for publicly available high-resolution imaging data\footnote{\url{https://exofop.ipac.caltech.edu/tess/}}.
Of them, 46 systems had been previously observed with either `Alopeke or Zorro or the NN-Explore Exoplanet Stellar Speckle Imager (NESSI) on the 3.5-meter Wisconsin-Indiana-Yale-NOIRLab (WIYN) telescope \citep{2018scott}. 
In this work, we focused on observations from these instruments as they are nearly identical, and they offer high angular resolution and large magnitude contrasts at small separations.

The observing procedure was roughly the same for each of the 'Alopeke, Zorro, and NESSI archival observations. 
Speckle imaging observations consist of data cubes containing one thousand 40-ms (WIYN) or 60-ms (Gemini) exposures.
These exposure times were selected in order to obtain the best signal-to-noise ratio (S/N) for the observations \citep{2011Howell}.
These short exposure times allow one to "freeze" the atmospheric interference.
The number of data cubes is determined either by the PI or by the on-site observer following the tables on the `Alopeke and Zorro\footnote{\url{https://www.gemini.edu/instrumentation/alopeke-zorro}} or NESSI\footnote{\url{https://www.wiyn.org/Instruments/nessi/}} webpages, and depends on the magnitude of the target and weather conditions during the observation.
During observations, a dichroic splits the collimated beam of light at $\sim700$ nm and directs it through two filters wheels and onto two EMCCDs.
For the archival observations, the 562/54 nm and 832/40 nm narrowband filters were used ($\overline{\lambda} / \Delta \lambda$).
A point source standard is observed close on-sky and in-time to the target to provide an estimate of what the point spread function of a single star is in roughly the same conditions as the science was taken in.
Additionally, calibration binaries are observed periodically to determine the pixel scale and image orientation of the instrument.
These binaries are drawn from the Sixth Catalog of Orbits of Visual Binary Stars and are required to have grades 1 or 2 \citet{2001hartkopforb6}. 

For our new observations, we modified this observing procedure in two ways.
We first reduced the `Alopeke and Zorro exposure times to 20 ms rather than 60 ms to avoid potential saturation from these bright targets ($G\sim3-7$ mag).
We also used double the recommended amount of time on-target (six minutes) to improve the S/N of the observations and thus the contrasts achieved.
In two instances, HD 5015 and the November 2024 observation of HD 88230, we observed for the normal amount of time (three minutes).
We discuss the implications of using a shorter exposure time in Section \ref{sec:exposures}.

The speckle data sets were reduced using the data reduction pipeline detailed in \cite{2009Horch,2011Horcha,2011Horchb} and \citet{2011Howell}. 
This pipeline uses bi-spectral analysis to compute a reconstructed image for each target \citep{1983Lohmann}. 
Briefly, the pipeline produces a co-added power spectrum from the speckle frames.
After dividing by the power spectrum of the point source standard, fringes in the power spectrum indicate the presence of a stellar companion. 
The fringes are fit using a cosine-squared function to determine the angular separation ($\rho$), position angle ($\theta$), and an estimate of the magnitude difference ($\Delta m$) for the companion. 
The reconstructed image is then generated from the modulus of the object given in the power spectrum. 

Uncertainties on these properties are not provided by the pipeline for each individual target. 
Instead, \citet{2021scott} analyzed a sample of calibration binaries and found a pixel scale uncertainty of 0.21 mas/pixel, position angle uncertainty of 0.7\degree, and magnitude difference uncertainties of 0.02 and 0.04 mag in the 562\,nm and 832\,nm filters
We use these as our uncertainties on $\rho$, $\Delta m$, and $\theta$.

\begin{figure}
\centering
\subfigure{
\includegraphics[width=0.49\textwidth]{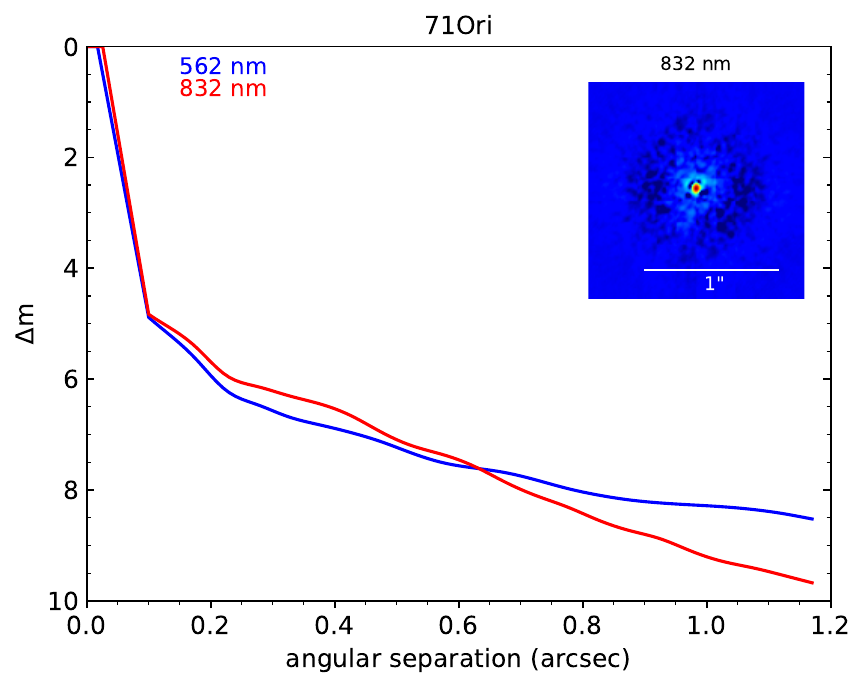}}
\subfigure{
\includegraphics[width=0.49\textwidth]{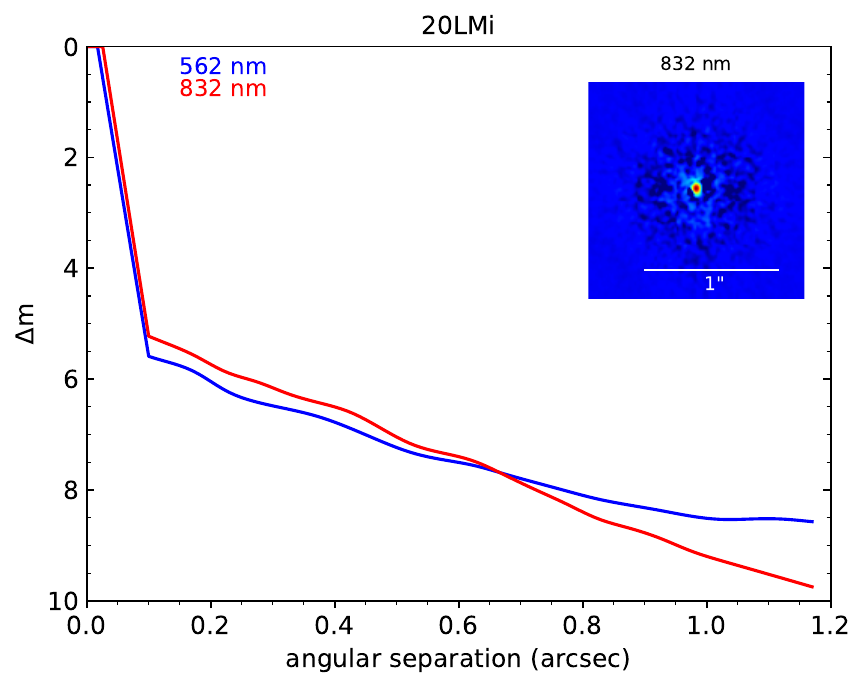}}
\caption{\label{fig:71ori} The contrast curves for the two narrowband filters centered at 562\,nm (blue line) and 832\,nm (red line) for our observation of 71 Ori and 20 LMi using `Alopeke at Gemini North. The insets are the reconstructed images at 832\,nm.} 
\end{figure}

Magnitude contrast curves for both filters are then constructed by examining the maximum and minimum background values in annuli centered on the primary star.
The detection limit within each annulus is estimated as the mean value of the maxima plus five times the average standard deviation of the maxima and minima.
These curves assess the sensitivity of the observation to stellar companions as measured in magnitudes fainter than the primary star and as a function of angular separation.
Two examples of the reconstructed images and contrast curves produced by this pipeline are shown in Figure \ref{fig:71ori}. 
These contrast curves demonstrate that we are sensitive to stellar companions that are much fainter and redder than their host star, down to the diffraction limit of the telescope.
All reconstructed images and contrast curves from the new observations are publicly available on the ExoFOP archive.

Table \ref{tab:speckleresults} provides results of the speckle observations and gives the TIC ID, instrument used, and PI of each observation, as well as the contrasts reached at 0.1" and 0.5" in the 562\,nm and 832\,nm filters. 
The $\rho$, $\Delta m$, and $\theta$ of any detected companions in the 832\,nm filter are provided in Table \ref{tab:specklebinresults}. 
Only the 832\,nm filter results are shown because no companion was detected in the 562\,nm filter.
We note that three multiple systems have speckle observations of both components: 61 Cyg A and B, $\zeta^{01}$ Ret and  $\zeta^{02}$ Ret, and $\xi$ Boo and $\xi$ Boo B.

\begin{figure}
\centering
\subfigure{
\includegraphics[width=0.49\textwidth]{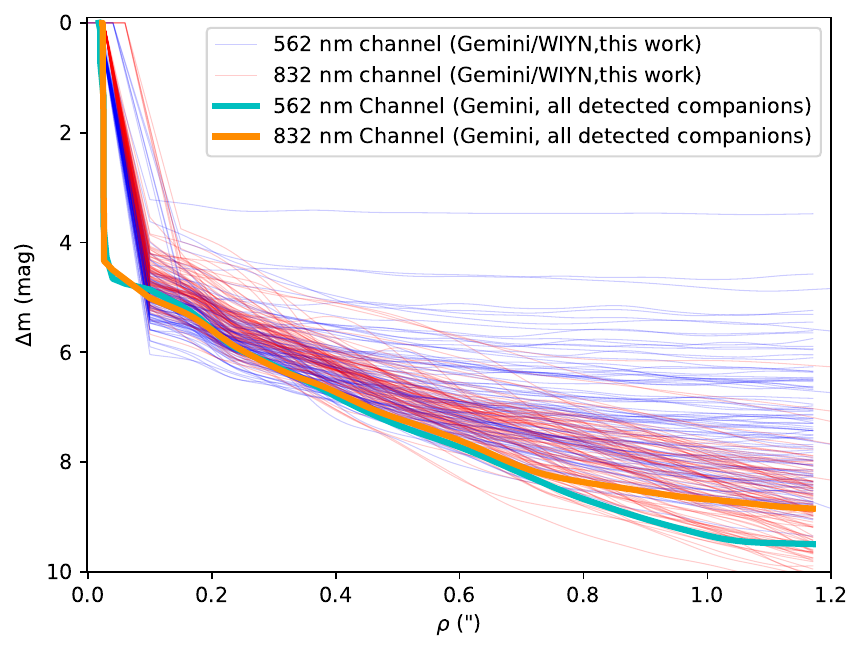}}
\subfigure{
\includegraphics[width=0.49\textwidth]{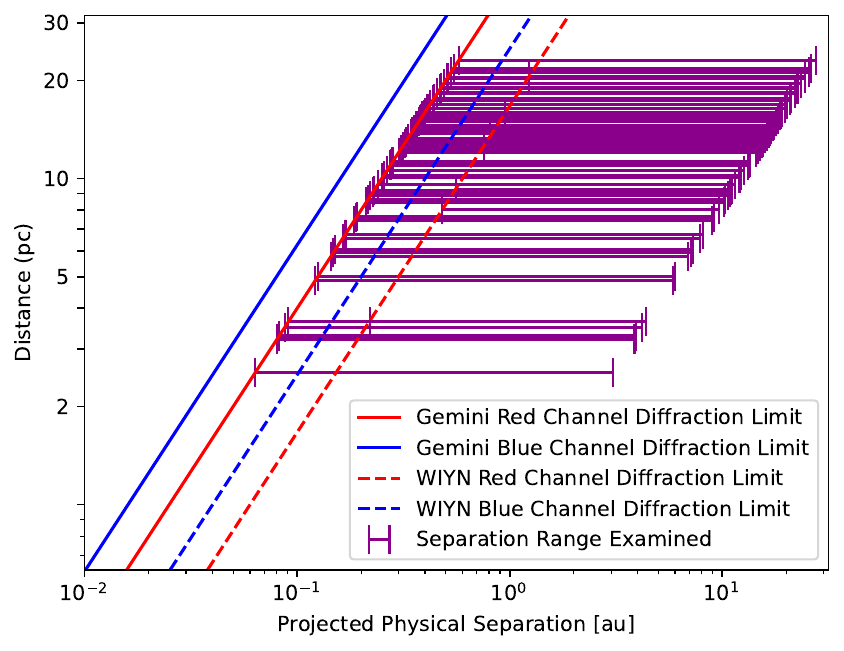}}
\caption{\label{fig:contrast} \textit{Left:} Contrast curves for the 80 HWO targets surveyed in this work. The 562\,nm and 832\,nm contrast curves are shown as blue and red lines, respectively. The solid cyan and orange lines represent a fit to the envelope of binaries detected by `Alopeke and Zorro and listed on ExoFOP. These contrast curves show that we detect stellar companions that are fainter than their primaries by 4-5 magnitudes at the diffraction limits of the telescopes. \textit{Right:} The projected physical separation range surveyed by our speckle observations for each target, shown as magenta bars. Solid and dashed lines represent the diffraction limits for the 562\,nm (blue) and 832\, nm (red) filters at Gemini (`Alopeke and Zorro) and WIYN (NESSI). This figure shows that speckle imaging surveys a necessary parameter space in the search for potential companions.}
\end{figure}

We also investigate the physical separation space probed by these observations. 
The left panel of Figure \ref{fig:contrast} shows the $5-\sigma$ contrast curves for each of the stars examined in this work as blue (562\,nm) and red (832\,nm) lines for both Gemini and WIYN.
We also provide two additional curves (cyan and orange for the 562\,nm and 832\,nm filters, respectively), which highlight our `Alopeke's and Zorro's sensitivity using previous companion detections over the past decade. 
Critically, these curves show that our observations could detect binaries down to the diffraction limit of Gemini with $\Delta m \sim4-5$ mag.
The right panel of Figure \ref{fig:contrast} highlights the range of projected physical separations, $\rho=$ angular separation $*$ distance, that our speckle observations are sensitive to as a function of distance to the target. 
The magenta bars show the limits for each target.
We set the lower limit of the physical separation range to be the diffraction limit of the red channel at Gemini or WIYN: 0.025\arcsec{} and 0.06\arcsec{}, respectively.
The upper limit is 1.2\arcsec{}, above which the speckles start to de-correlate and the sensitivity to stellar companions decreases. 
The diagonal lines represent the diffraction limits for the 562\,nm (blue) and 832\,nm (red) filters.

\subsection{Gaia Search}

We also cross-matched our sample of 80 stars with the Gaia Catalogue of Nearby Stars \citep[GCNS; ][]{2021gcns} and \citet{2021elbadry} to identify pairs of stars close on-sky with similar proper motions and parallaxes using the astrometric information provided by Gaia early Data Release 3 \citep{2016gaiamission,2021edr3}.  
Since these are nearby targets, substantial visual motion may occur between the components of these systems over the coming years; as such, identifying these common-proper-motion (CPM) pairs is critical.

Gaia Data Release 3 (DR3) also contains useful indicators of stellar multiplicity even when a companion is unresolved \citep{2023gaiamultiples}. 
We therefore cross-matched our sample of 80 targets with Gaia DR3 and retrieved known binary indicators \citep{2023gaiadr3}.
These indicators include the Reduced Unit Weight Error (RUWE), IPD\_gof\_harmonic\_amplitude (IPD\_gof), rv\_chisq\_pvalue (RV\_chi), rv\_renormalized\_gof (RV\_gof), radial\_velocity\_error, IPD\_frac\_multi\_peak (IPD\_fmp), rv\_nb\_transits (RV\_nb), and the Non\_Single\_Star flag (NSS, \citealt{2023gaiamultiples}). 
RUWE is an indicator of how good the Gaia astrometric fit is to a single star model, ipd\_gof\_harmonic\_amplitude is a measure of how much the IPD goodness-of-fit changes as a function of scan position angle, IPD\_fmp is the fraction of Gaia windows where a second peak is detected, RV\_chi gives the p-value of the measured RVs from Gaia, RV\_gof highlights the goodness-of-fit of the radial velocities, RV\_nb is the number of transits used to obtain the radial velocity, radial\_velocity\_error is the error in the measured radial velocity, and the NSS flag shows which targets have available information in the Gaia NSS tables. 
The NSS tables contain binary information on the astrometric, acceleration, and spectroscopic binaries that the Gaia mission has identified and characterized \citep{2023gaiamultiples}.
Table \ref{tab:gaiaresults} provides the TIC ID, Gaia DR3 ID, R.V. Error, RUWE, IPD\_fmp, whether the star is flagged as a non-single star, IPD\_gof, RV\_gof, RV\_chi, RV\_nb, and the source, Gaia ID, separation, and magnitude differences in Gaia $G$ of any companion(s) found in our wide binary search.

To flag a star as a potential multiple, we start with the criteria outlined by \citet{2025cifuentes}.
We then modified the criteria in two ways. 
First, we did not check the duplicated\_source parameter, as \citet{2025cifuentes} indicated that this is a secondary parameter rather than a primary parameter. 
This parameter flags sources that may have issues with astrometric quality, lack some data, or where Gaia may have detected multiple sources \citep{2025cifuentes}.
As this covers a wide range of possible issues, we do not check it as the other criteria are better at identifying stellar multiples.
Second, we lowered the IPD\_fmp cut to 2 following \citet{2023tokovinin}. 
Our modified criteria are:
\begin{itemize}
    \item RUWE $>2$
    \item ipd\_gof\_harmonic\_amplitude $>0.1$ \& RUWE $>1.4$
    \item ipd\_frac\_multi\_peak $>2$
    \item rv\_chisq\_pvalue $<0.01$ \& rv\_renormalized\_gof $>4$ \& rv\_nb\_transits $\geq 10$
    \item radial\_velocity\_error $\geq 10\SI{}{\kilo\metre\per\second}$
    \item non\_single\_star flagged
\end{itemize}

\begin{figure}
\centering
\includegraphics[width=\textwidth]{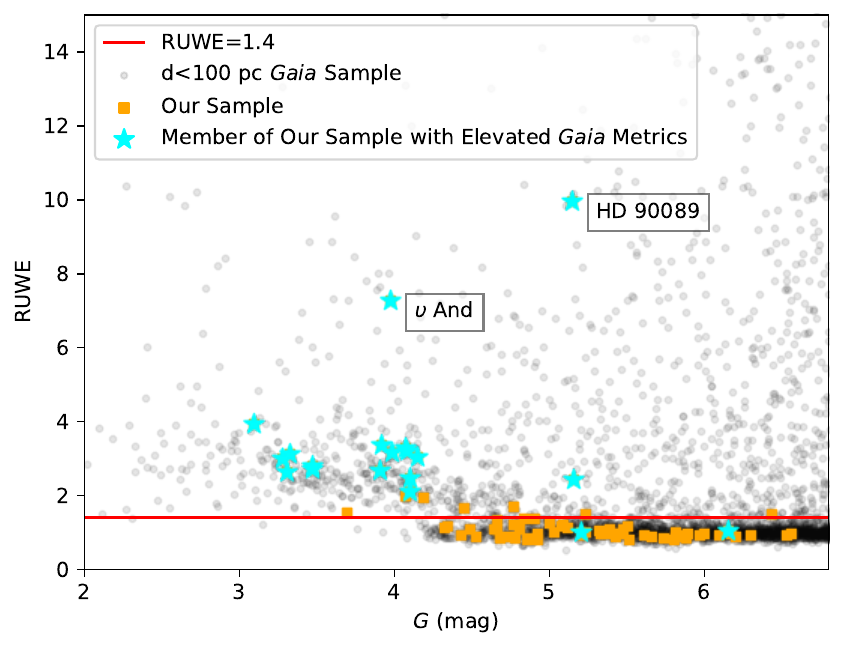}
\caption{\label{fig:ruwe} RUWE values of the 80 HWO targets surveyed in this work as a function of Gaia $G$ magnitude. Cyan stars indicate the 19 stars that were flagged as potential binaries by our Gaia metric criteria, while the remaining 61 stars are shown as orange squares. The red line indicates a RUWE value of 1.4, which is the value typically used to distinguish single and non-single stars. The black points represent a sample of stars within 100 pc from Gaia DR3. We label the two stars with RUWE$>7$. As one goes to brighter magnitudes, the RUWE values rise, which could indicate the presence of unresolved companions. However, it is more likely that the brightness of these targets is increasing the astrometric errors. As such, the RUWE value may not be the best indicator of stellar multiplicity for stars brighter than $G\sim4.5$\,mag.}
\end{figure}

Figure \ref{fig:ruwe} shows the RUWE values for the 80 targets in this study as a function of their Gaia $G$ magnitude, with the 19 targets that were flagged by the Gaia multiplicity criteria described above shown as cyan stars.
We note that at $G\sim4.5$\,mag, the RUWE values begin to rise.
While this could indicate the presence of an unresolved companion, it is more likely that the brightness of these targets increases the astrometric errors, leading to larger RUWE values. 
To demonstrate this effect, we also plot a sample of stars within 100 pc from Gaia.
Therefore, the RUWE value may be unreliable for stars brighter than $G\sim4.5$\,mag. 
As such, unless the target is flagged as a NSS by Gaia or a companion is found via other techniques, we consider the star to be single.
However, we do mark stars with these elevated Gaia metrics in Table \ref{tab:overall}, meaning one or some of the criteria above are met.

\subsection{Literature Search}

We searched several other sources for additional stellar companions, and to confirm companions found in our speckle images or in Gaia. 
The WDS was used as a starting point.
In most cases, the WDS entry is confirmed; however, in some cases notes in the WDS indicated that the companion may not be bound, or the detection is spurious. 
In particular, we disregard systems where components included a U, S, X and/or L notes.
From the WDS, U means that the pair is non-physical due to mismatched proper motions, S indicates that the components in the system have statistically different parallaxes and proper motions, X highlights a system that is a ``Dubious Double,'' and L indicates a linear solution.
 
We also examined \citet{2023golovin}, \citet{2024kirkpatrick}, \citet{2024gonzalez}, and \citet{2025gonzalezpayo} to identify any additional companions. \citet{2023golovin} and \citet{2024kirkpatrick} were both wide censuses of nearby stars with \citet{2023golovin} going to 25 pc and \citet{2024kirkpatrick} going to 20 pc. \citet{2024gonzalez} and \citet{2025gonzalezpayo} were targeted multiplicity searches with \citet{2024gonzalez} focused on exoplanet-hosting stars within 100 pc and \citet{2025gonzalezpayo} examining the multiplicity of stars within 10 pc. 

\section{Results} \label{sec:results}

\begin{figure}
\centering
\includegraphics[width=\textwidth]{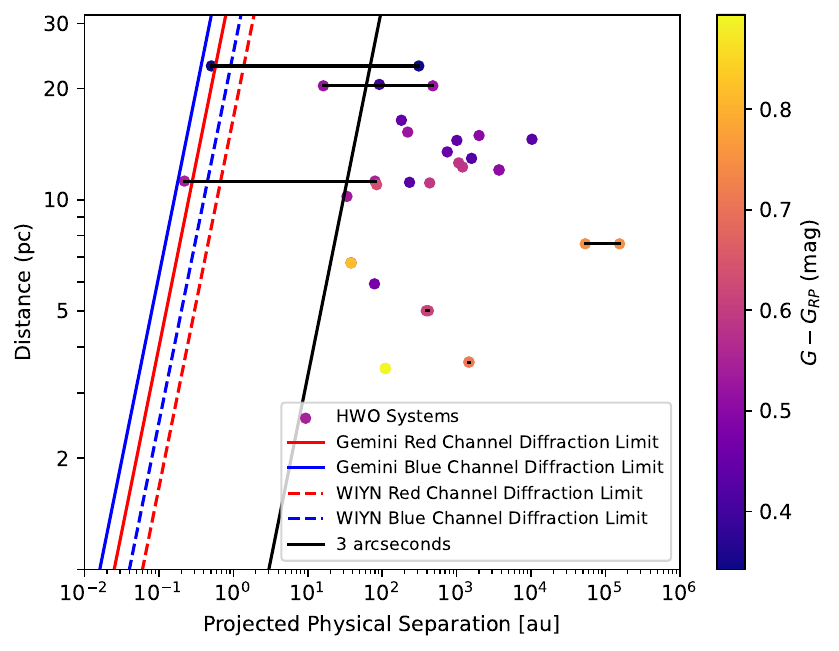}
\caption{\label{fig:hwophysep} 
Distances to the 27 multi-star systems identified in this study as a function of projected physical separation. The solid red and blue lines represent the nominal diffraction limits for the 562 nm (blue) and 832 nm (red) filters installed in `Alopeke and Zorro. The dashed lines represent the same for NESSI. The black line is the 3\arcsec{} cut used in the original sample selection for the ExEP provisional star list for HWO.
The color of the points highlights the $G-G_{RP}$ values of the stars.
If the system is triple, then it has two points representing the AB and AC components and we connect these systems by the black horizontal lines.
We assume the primary's distance as the distance for the system.
The majority of the stellar companions lie beyond 3\arcsec{}, which shows that the selection criteria from \citet{2024mamajek} removed most close-in multiples from the sample.
However, some systems are still found with separations less than 3\arcsec{}, which highlights the need for a uniform multiplicity study of all the potential HWO target stars.}
\end{figure}

Table \ref{tab:overall} summarizes the results from our speckle observations, the Gaia analysis, and the literature search. 
We provide the TIC ID, WDS ID, whether the star is part of a Gaia CPM pair, whether the star is flagged as a possible close binary by Gaia indicators, whether the star is binary in our speckle observations, the literature source(s) primarily used to make the multiplicity determination in addition to the speckle observations and Gaia data, the multiplicity status of the star after our analysis, and any notes on the system.
Figure \ref{fig:hwophysep} shows the projected physical separations of the multi-star systems that were identified in our analysis, with comparisons to the diffraction limits of the instruments and the selection criteria used in the ExEP provisional star list for HWO.

Overall, 27 of the 80 stars examined in this study were found to be members of multi-star systems. 
We identified a new stellar companion to HD 90089 (Section \ref{subsec:hip51502}), and we obtained an additional speckle observation of the known companion to HD 212330 (Section \ref{subsec:hip110649}).
In total, we find 33 possible companions with 24 common proper motion companions, two Gaia non-single stars, including one confirmed by our speckle observations, one potential speckle detection of a known companion, and five other companions listed in the literature. 
We discuss known and potentially new triple systems in Section \ref{subsec:triples}, and assess potentially missed companions in Section \ref{subsec:missingcompanions}.

\subsection{HD 90089/TIC 367631379} \label{subsec:hip51502}

HD 90089 is a F4V star and is $\sim23$ pc away. 
An examination of the literature data on Simbad shows few adaptive optics or speckle imaging observations of this star.
The most recent was \citet{1989mcalister} using speckle imaging on the 4-m Mayall Telescope at Kitt Peak National Observatory.
No companion was detected in those observations.

We detect a companion to HD 90089 at 832\,nm, with a separation of 0.022\arcsec, a position angle of 2.4\degree, and a magnitude difference of 0.45.
The companion is not detected at 562\,nm.
This system was already known to be multiple, with a companion at 13.5\arcsec{} noted in the WDS \citep{2016gatewood} and \citet{2021elbadry}.
We note that the 0.022\arcsec{} separation is below the nominal diffraction limit of the speckle instruments at Gemini Observatory; however, speckle imaging can go beyond this limit \citep{2011Horchb}.
This target also has elevated Gaia multiplicity metrics (e.g., RUWE=9.9) that provides more evidence towards the detected companion being real.

HD 90089 is also marked as a Gaia non-single star, and has an astrometric orbit listed in Gaia DR3.
To examine if our detection and the Gaia detection are one in the same, we used the orbital parameters from the Gaia orbit and the NSS Tools programs\footnote{\url{https://gitlab.obspm.fr/gaia/nsstools}} to determine where in the orbit the stellar companion would have been at the time of our observations \citep{2023halbwachs}.
This analysis put the companion at a separation of 0.024\arcsec{} at the time of our speckle observations, which is within the errors of our measurement.
However, our analysis predicted the companion would be at a position angle of 260.9\degree, which is different than our result. 

In speckle imaging, there is the potential for a 180\degree{} ambiguity in position angle due to the phase not being well constrained. 
This is caused by either the faintness of the companion resulting in low S/N, and thus a lack of phase information, or a close-in companion resulting in few fringes to fit during the analysis. 
As such, we re-ran this analysis with a 180\degree{} position angle offset from our measurement (i.e., 182.4\degree). 
We also note that the position angle from the Gaia orbit is not well-constrained, which could account for the remaining $\sim80$\degree difference between our speckle measurement and the Gaia prediction. 
To calculate an error estimate for the position angle, we bootstrapped the orbital calculation 100,000 times with the orbital elements drawn from a Gaussian distribution based on the calculated orbital values and errors and taking the standard deviation of the resulting values.
The error we derive on the Gaia position angle is $\sim50$\degree. 
While the position angles between the speckle and Gaia results are not in complete agreement, we believe our speckle detection and the Gaia companion are one in the same.

\subsection{HD 212330/TIC 259291108}
\label{subsec:hip110649}

HD 212330 is a member of a known triple system.
The star itself is G2 dwarf located $\sim20.3$ pc away.
A CPM companion at 23.957\arcsec{} is listed in the WDS and in \citet{2021elbadry}.
A third inner component was identified in a combined radial velocity and speckle imaging study \citep{2019kane}, which observed HD 212330 using DSSI on Gemini South on 2017, June 6. 
They found a companion at a separation of $\sim0.77$\arcsec{} with a magnitude difference of 6.7 at 692\,nm and of 5.5 at 880\,nm and a position angle of $\sim62\degree$.
Five years later, \citet{2023tokovinin} observed the system using HRCam on the Southern Astrophysical Research telescope and resolved the companion at a separation of 0.8796\arcsec{}, position angle of 244.6\degree{} and $\Delta m$ of 5\,mag in the $I$ filter.

\begin{figure}
\centering
\subfigure{
\includegraphics[width=0.49\textwidth]
{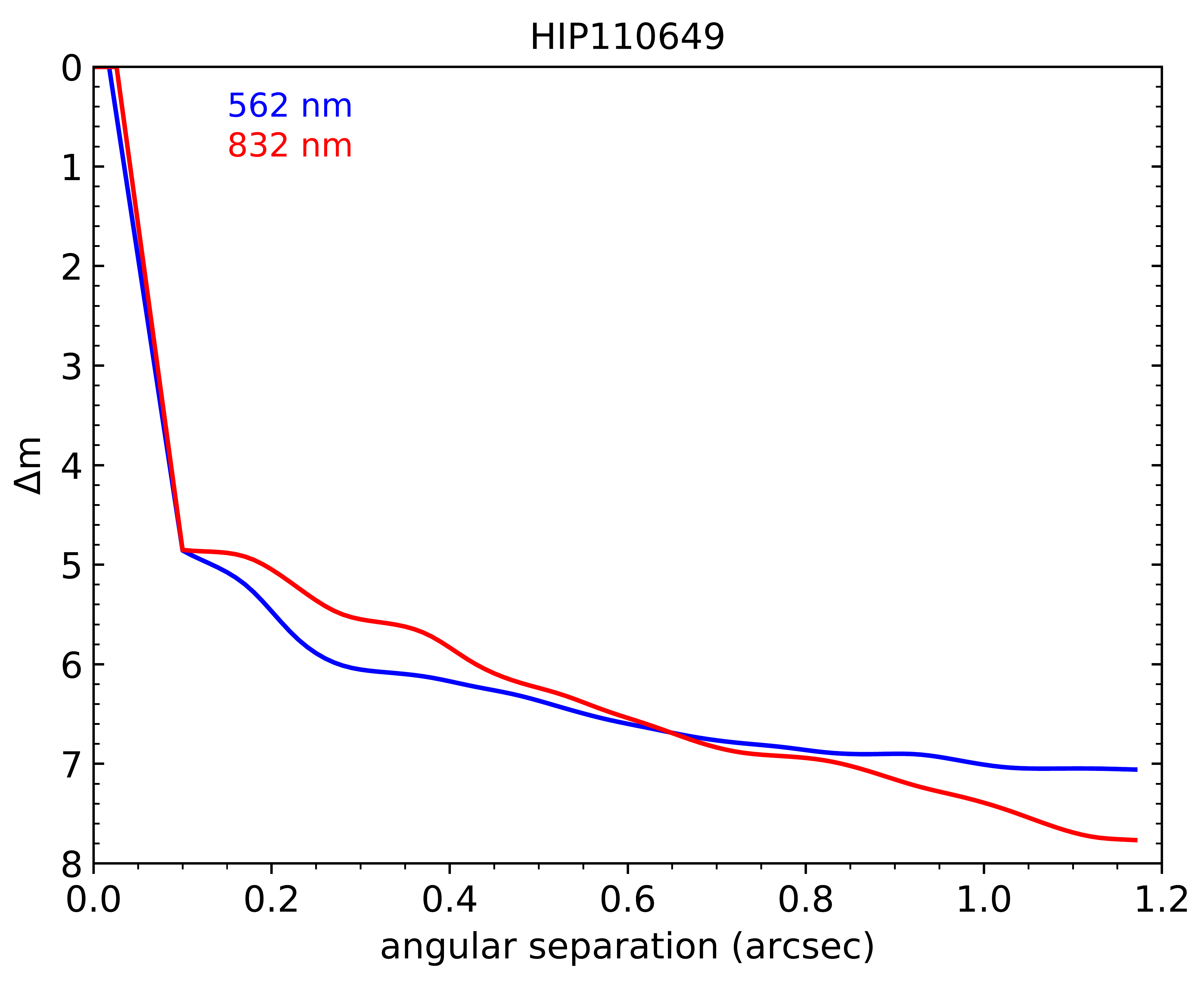}}
\subfigure{
\includegraphics[width=0.49\textwidth]{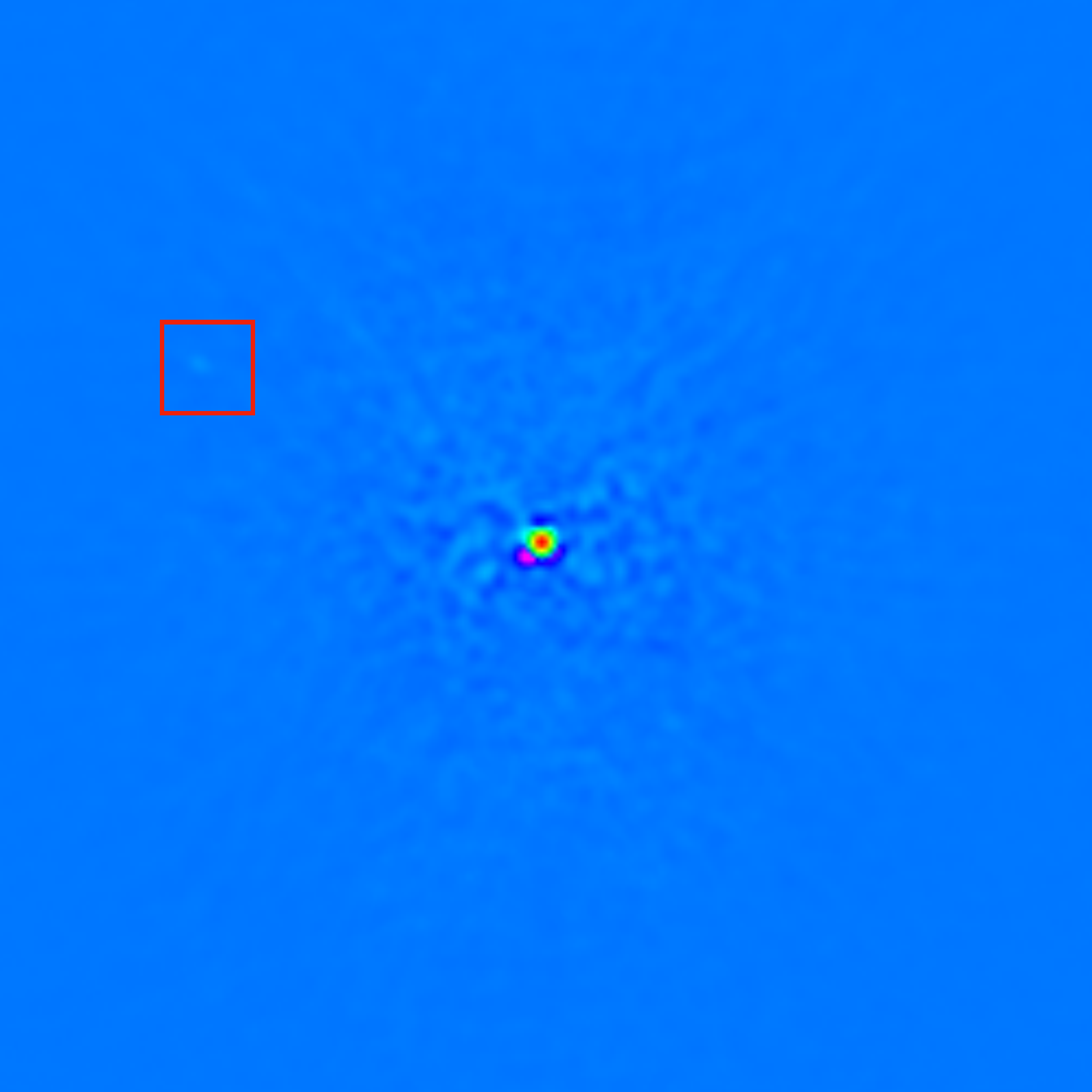}}
\caption{\label{fig:kane} Left panel: Contrast curve from the 2020 Zorro Observations of HD 212330/TIC 259291108 from the ExoFOP archive. Right panel: Reconstructed image from the 832\,nm filter. \citet{2019kane} reported a companion at 0.77\arcsec{} with a position angle of 62\degree{} and a magnitude difference of 5.5 at 880\,nm using DSSI on Gemini South in 2017. The red box highlights our possible detection of the same companion in this reconstructed image from 2020.}
\end{figure}

The star was also imaged in 2020 using Zorro on the Gemini South telescope.
Figure \ref{fig:kane} shows the Zorro observations of this target. 
The left panel shows the contrast curve and the right panel shows the reconstructed image from the 832 nm filter.
The power spectrum from this observation shows no indication of a companion.
However, looking at the right panel of Figure \ref{fig:kane}, there is a bright speckle at roughly the expected separation and position angle of the time-evolved companion from \citet{2019kane}.
This is highlighted by the red square.
Forcing a binary fit to the power spectrum yields a separation of 0.88\arcsec{} and a position angle of 63\degree, which is consistent with several of the orbit predictions.
However, the fit yields a $\Delta m = 7.5$ compared to the $\sim5$ found preivously in the DSSI and HRCam observations.

To examine the discrepancy between the Zorro speckle observations and those from DSSI and HRCam, we fit a number of orbits for this system and predicted the corresponding positions of the companion at the time of the 2020 observations. 
The HRCam position angle was roughly 180\degree{} offset from the DSSI and Zorro observations likely due to the 180\degree{} ambiguity mentioned in Section \ref{subsec:hip51502}, as the companion is quite faint.
As such, we subtracted 180\degree{} from that position angle to align with the other observations.
For the orbit fit, we used Gaussian distributions for the period, epoch of periastron, eccentricity, and the longitude of periastron of the secondary star based on the the spectroscopic orbital solution from \citet{2019kane}. 
The inclination, angular semi-major axis, and longitude of the ascending node of the secondary star were free parameters with uniform distributions. 
We tested $10^7$ combinations of these orbital parameters and calculated the $\chi^2$ statistic of each solution. We removed all solutions with $\chi^2$ > $\chi^2$\_min + 10.  
Our best-fit inclination was $85.55\degree{} \pm 1.73\degree{}$, our best-fit angular semi-major axis was $720.5$ mas $\pm$ $23.7$ mas, and our best-fit longitude of ascending node of the secondary was $244.1\degree{} \pm 0.8\degree{}$.
Because of the 180\degree{} ambiguity in the \citet{2023tokovinin} results, the longitude also has an 180\degree{} ambiguity, meaning it could be $64.1\degree{} \pm 0.8\degree{}$.
This would change the inclination to 94.45\degree{}.
However, this would only change the orientation and direction of the orbit on-sky, not the physical properties of the system. 

\begin{figure}
\centering
\includegraphics[width=\textwidth]{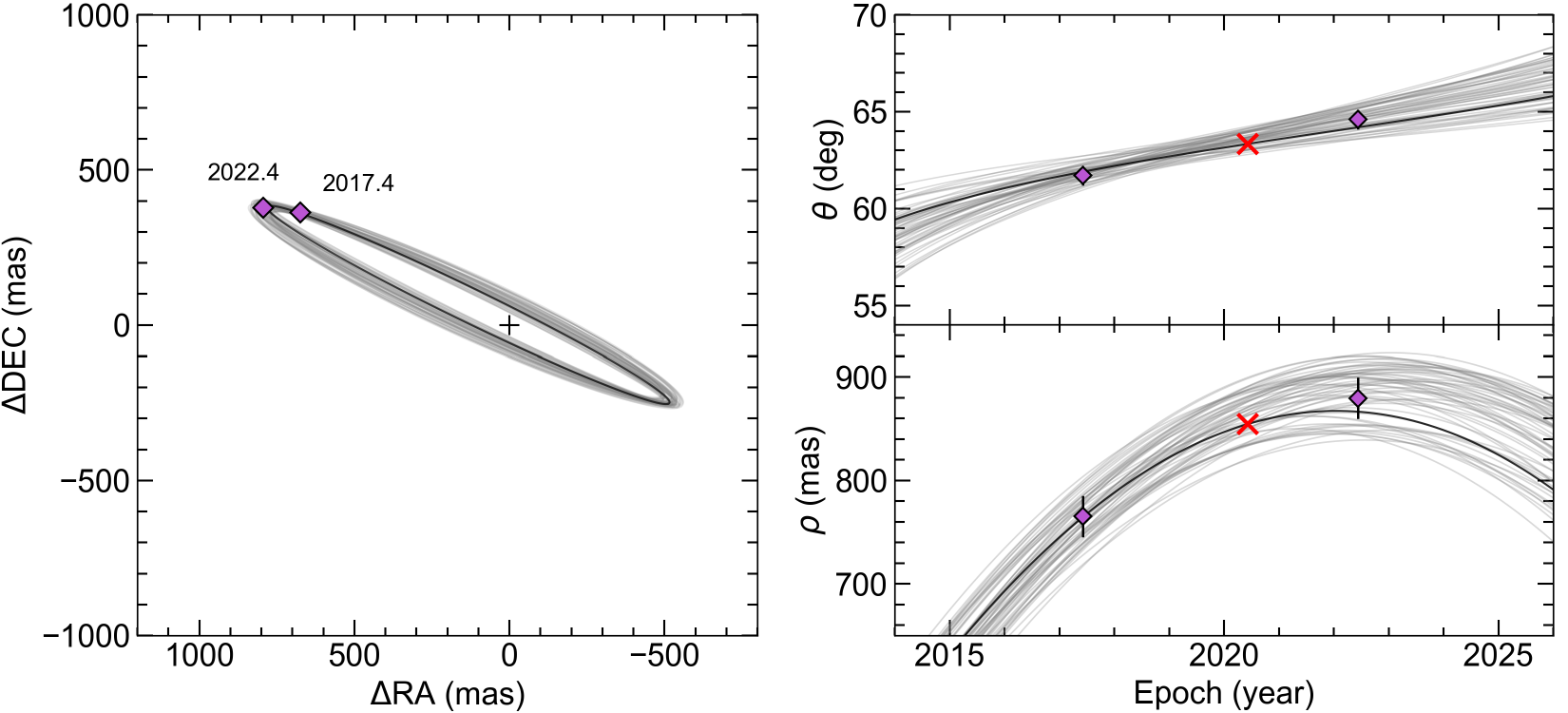}
\caption{\label{fig:kane2} Orbit fit for HD 212330/TIC 259291108 using the astrometric data from \citet{2019kane} at epoch 2017.4 and \citet{2023tokovinin} at epoch 2022.4, which are shown as purple diamonds, and the radial velocity fit from \citet{2019kane}. The black line represents the best fit, while the gray lines represent possible solutions within 2$\sigma$. The right panels show the separation and position angle of the orbit as a function of epoch. The $\times$ symbol shows the predicted position of the stellar companion at the time of the 2020 Zorro observations, which is consistent with our fit to the data.}
\end{figure}

The visual orbits for these solutions and predicted positions of the companion at the time of the 2020 observation are plotted in Figure \ref{fig:kane2}. 
Using the best-fit orbit, we predict the position of the companion during the 2020 Zorro observations. 
We find a predicted separation of 0.855\arcsec{} and a predicted position angle of 63.34\degree, which are consistent with the results from our fit to the data.
As such, we find that this is the same stellar companion that was detected in the DSSI and HRCam observations, but at a much lower significance.
This could explain why our $\Delta m$ value is off from the other two observations.
An examination of the night log finds that heavy clouds were present during the 2020 observations, which could explain the low significance.

\subsection{Potential New Triple Systems}
\label{subsec:triples}

Six of the 27 targets found to be in multi-star systems are members of known triple systems: 11 LMi (TIC 8915802), $o^2$ Eri (TIC 67772871), TW PsA (TIC 206686962), $\epsilon$ Ind (TIC 231698181), HD 212330 (TIC 259291108) and HD 90089 (TIC 367631379). Three of the 27 multiples are potentially new triple systems: 20 LMi (TIC 172954294) , $\theta$ Per (TIC 302158903) and $\pi^3$ Ori (TIC 399665349). 

20 LMi was listed by \citet{2024kirkpatrick} as a triple due to its over-luminosity. 
However, the companion has not been observationally detected yet.
While over-luminosity is a sign of multiplicity, it could also be caused by other factors, such as a star evolving from the main sequence.
Because of this, we do not consider 20 LMi to be a triple system.

$\theta$ Per is a known multiple with a wide companion at 20.932\arcsec{} as noted in the WDS.
\citet{2010tanner} reported the potential detection of an additional brown dwarf companion that was unconfirmed in their study, but \citet{2023golovin} lists only two components for this system.
As such, we consider $\theta$ Per a binary rather than a triple system.

Finally, $\pi^3$ Ori currently has no confirmed companions. 
However, \citet{2007lafreniere} notes that there were two point sources detected around $\pi^3$ Ori, but they did not obtain a second epoch of observation for this target.
Furthermore, \citet{2025gonzalezpayo}, \citet{2024kirkpatrick}, and \citet{2023golovin} consider this star to be single. 
As such, we consider this star single as well.

\subsection{Assessing Potentially Missed Companions}\label{subsec:missingcompanions}

One of the key goals of this study was to quantify the sensitivity of our existing speckle observations and determine what kinds of lower-mass stellar companions could remain undetected around these targets.
To do this, we used a code developed to simulate possible stellar companions to our targets and assess our sensitivity to them \citep{2020lund,2022clark, Clark2024AJ....167...56C, Clark2024AJ....167..174C}.
We provide a brief description of the code below.

First, the mass, effective temperature, surface gravity and metallicity of the target star were retrieved from the TIC. 
For ten targets, the TIC did not have values or errors for one or more of the four parameters used.
To correct this, we obtained these values from the ExEP target list, except for errors on the mass which were not given in the list.
We took the average of the errors on the masses for the 70 other stars which had errors as an estimate for the errors on these values.
In addition to this, six of the 80 stars were found to have directly measured masses listed by \citet{2024kirkpatrick}.
We adopted those masses and errors on the mass for this analysis, with one exception.
Taking these stellar parameters, a best-fit stellar isochrone was selected from the Dartmouth isochrones \citep{2008dotter}.
The errors on the masses are relatively large compared to the errors on the effective temperature and surface gravity, which can lead to a discrepancy between the input mass and the best-fit mass from the isochrone.
However, all of our best-fit masses are within the input mass errors.

100,000 potential companions were then simulated following mass ratio and period distributions as described below.
For our targets, we used two mass ratio and period distributions depending on the mass of the target. 
If the mass was between 0.6 $M_{\odot}$ to 1.42 $M_{\odot}$, we adopted the mass ratio and period distributions from \citet{2010raghavan}.
We used the left panel of Figure 16 in \citet{2010raghavan} to create a piecewise function where mass ratios in the intervals of 0-0.2, 0.2-0.95, and 0.95 - 1.0 have set values, and we normalized the resulting distribution. 
Below 0.6 $M_{\odot}$, we used the distributions from \citet{2013duchene}, who adopted a relatively flat power law distribution for the mass ratios and a normal distribution for the periods. 
In both cases, the minimum mass used for the simulations came from the best-fit isochrone and was roughly $0.1 M_{\odot}$ or spectral type M6.
The minimum mass was divided by the mass of the target star to determine the minimum mass ratio used by the simulation.
Beyond this minimum mass, there were no more points for the interpolation and, for our shown results, no companions with lower masses were simulated.

\begin{figure}
\centering
\includegraphics[width=\textwidth]{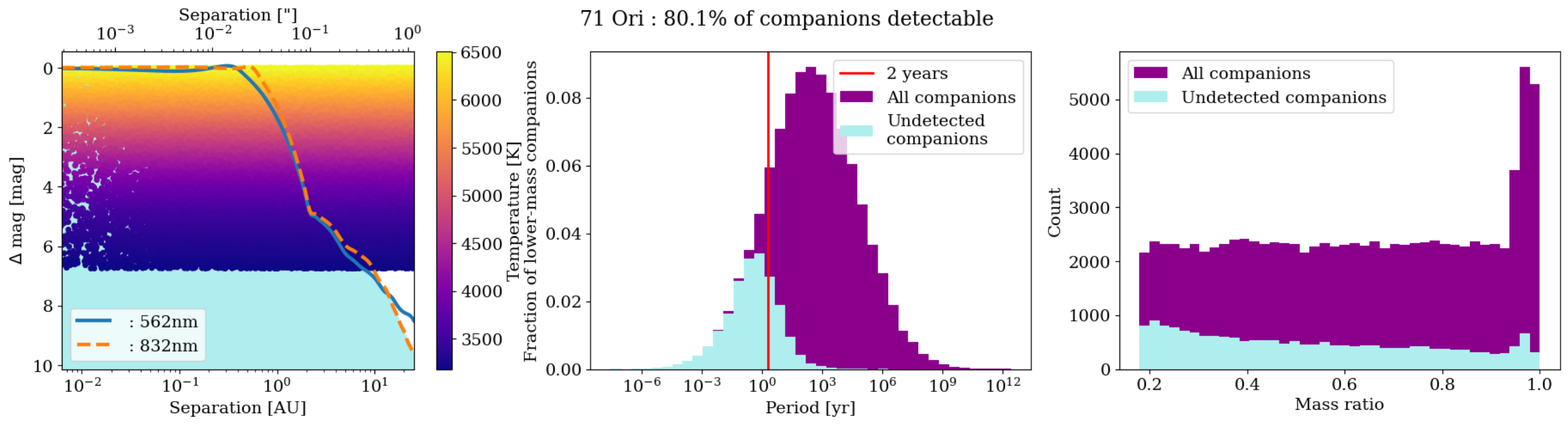}
\caption{\label{fig:71orimissingcomps} Results of the Missing Companions Analysis for 71 Ori. \textit{Left:} Simulated stellar companions as a function of magnitude difference and separation. The color of the points indicates the effective temperatures of the simulated companions. The solid and dashed red lines show the contrast curves from our speckle observations. Above these lines, we would expect to detect the companion. The region of $\Delta m$ vs. separation space that is shaded cyan represents the area where companions could reside and would not be detected by our speckle observations. The plot extends out to 1.17\arcsec{} as beyond this the correlation between the speckles breaks down and we lose detection sensitivity. For 71 Ori, this corresponds to a separation of 25.48\,au. \textit{Middle:} Period distributions of the simulated companions. The magenta histogram is the distribution of all 100,000 simulated companions, while the light blue histogram highlights what would not be detectable in our speckle observations. \textit{Right:} Similar to the middle panel, but for mass ratio. We find that 80.1\% of the simulated companions to this target are detected. Please see online Journal version for figures on the rest of the sample.} 
\end{figure}

Figure \ref{fig:71orimissingcomps} shows the results of this analysis for 71 Ori.
The left panel shows $\Delta m$ of the target star and the simulated companions as a function of physical separation. 
To determine this separation from our simulated period, we assumed circular orbits and use Kepler's Third law.
We then randomized the location of the simulated companion within a sphere the size of the max separation and converted this back to a separation for use in the analysis.
The red lines indicate where our speckle contrast curves fall on this plot.
In some cases, we have multiple epochs of data for a target.
All epochs are included in these plots and the best contrasts achieved overall are used.
The contrast curves were then used to assess the sensitivity of each observation to these simulated companions. 
It is assumed that wide-field, seeing-limited imaging or Gaia would detect companions outside the field-of-view of the speckle instruments. 
The light blue shaded areas highlight the region where our speckle observations would not be sensitive to companions. 
As mentioned previously, our simulated companions only go to companions of spectral type M6 due to the isochones being used. 
This is why there are no points below $\Delta m$ $\sim 6.5$ in Figure \ref{fig:71orimissingcomps}.
However, to test the potential effect of including lower mass companions, we extrapolated our interpolation to include masses down to 0.05 $M_{\odot}$.
We found a change of less than $2\%$ when these systems are included. 
The percentage change is dependent on the mass of the primary star as this changes the mass-ratio distribution for the lowest mass targets in our sample.

The middle and right panels of Figure \ref{fig:71orimissingcomps} show the period and mass ratio distributions for the simulated companions.
The purple distribution represents all simulated companions, while the cyan distribution highlights those that would be undetected by our speckle observations.
As seen, the majority of undetected companions are close-in systems and/or tend towards lower mass ratios.

Overall, our simulations show that when speckle imaging is combined with wide-field imaging, $\sim75\% - 85\%$ of simulated stellar companions should be detectable for our sample.
This value depends on a number of factors, including the distance to the target star.
Additionally, we treat each star as a single star for the purposes of assessing missing companions.
A detected companion would decrease the area where a third component would exist. 
We do note that the code will eliminate some simulated stellar companions if those companions suggest existing planet orbits would fall within the Hill sphere of the companion. 
This is only done for target stars that have TOIs on ExoFOP.
Among our 80 targets, only three are affected $\pi^1$ Men, HD 219134, and $\rho^1$ Cnc.
In these cases, the planet's semi-major axes were calculated from the orbital period listed in the TOI catalog using Kepler's Third Law.
The fact that our analysis only takes into account TOIs does bias our results for stars with planets detected through other methods. 
However, the magnitude of this bias depends on the location of the planet.
For close-in planets, there may be no change in our detectable percentage as any removed binaries would be below our detection limits.
Furthermore, we know that we are missing simulated companions below $0.1 M_{\odot}$ which impacts the percentage detected.
However, this change is relatively small, $<2\%$ drop in the percentage of detectable companions.


\section{Discussion}  \label{sec:Discussion}

\subsection{Comparing 20-ms and 60-ms Observations} \label{sec:exposures}

Currently, the recommended exposure time for `Alopeke and Zorro observations is 60-ms based on an analysis of bright stars. 
However, for many nearby, bright stars with $G<5-7$ -- which are prime targets for future space missions -- this exposure time results in counts that are close to the saturation limits of `Alopeke and Zorro on the 8-m Gemini telescopes.
This motivated the use of a 20-ms exposure time for our speckle observations of HWO targets.
However, some of the stars in our sample have observations using both 60-ms and 20-ms exposure times, allowing us to compare and contrast.

\begin{figure}
\centering
\subfigure{
\includegraphics[width=0.48\textwidth]{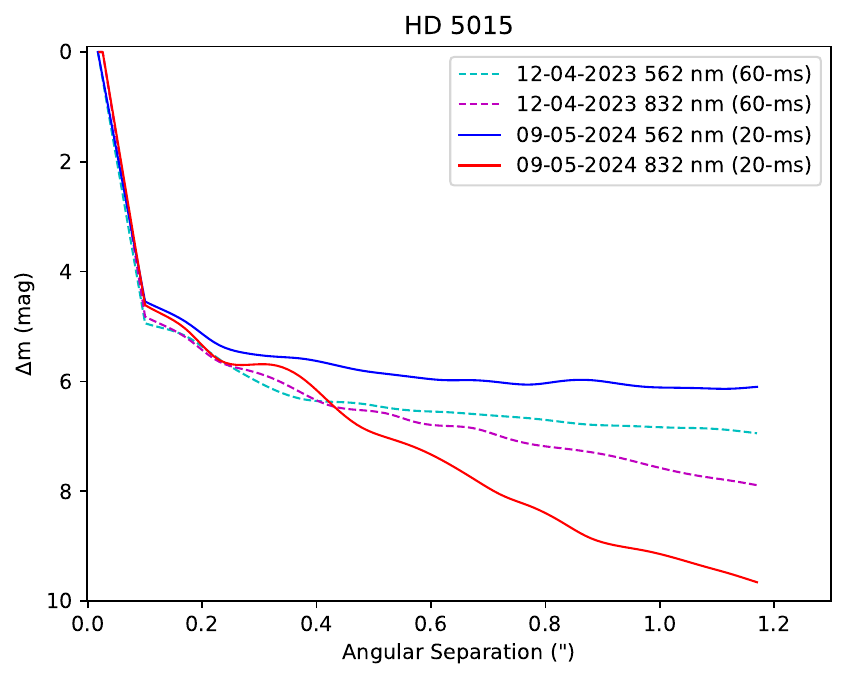}}
\subfigure{
\includegraphics[width=0.48\textwidth]{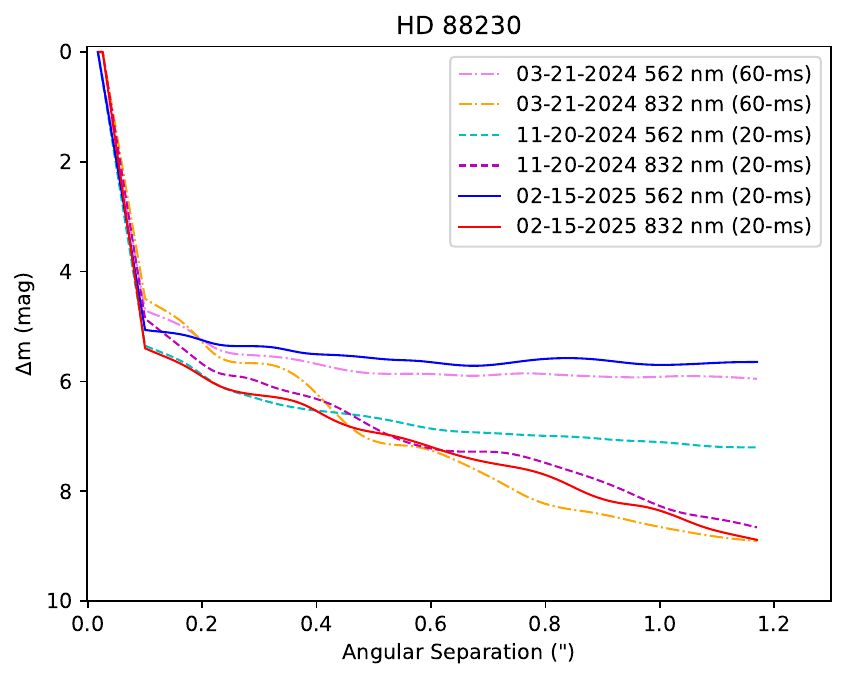}}
\subfigure{
\includegraphics[width=0.48\textwidth]{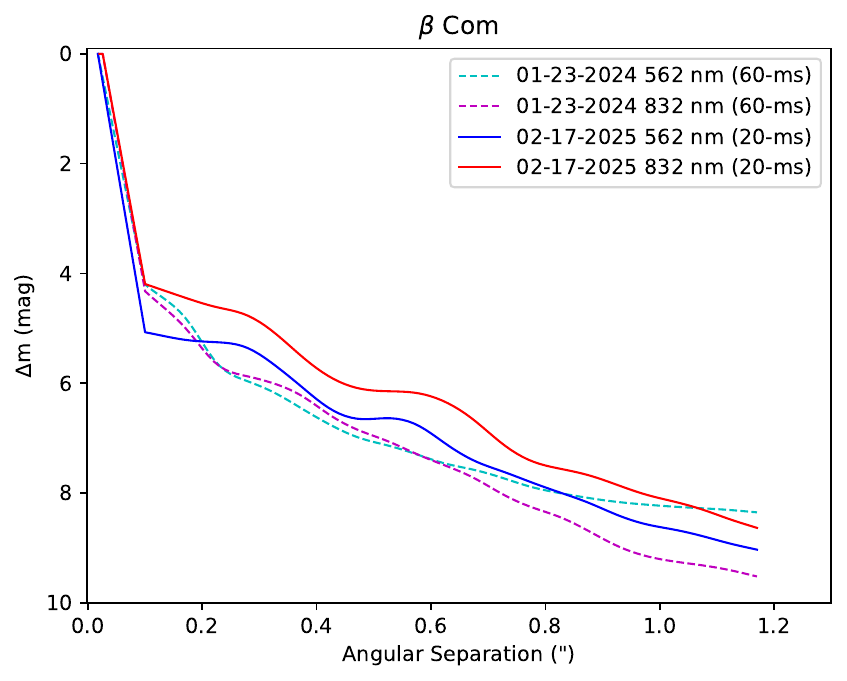}}
\subfigure{
\includegraphics[width=0.48\textwidth]{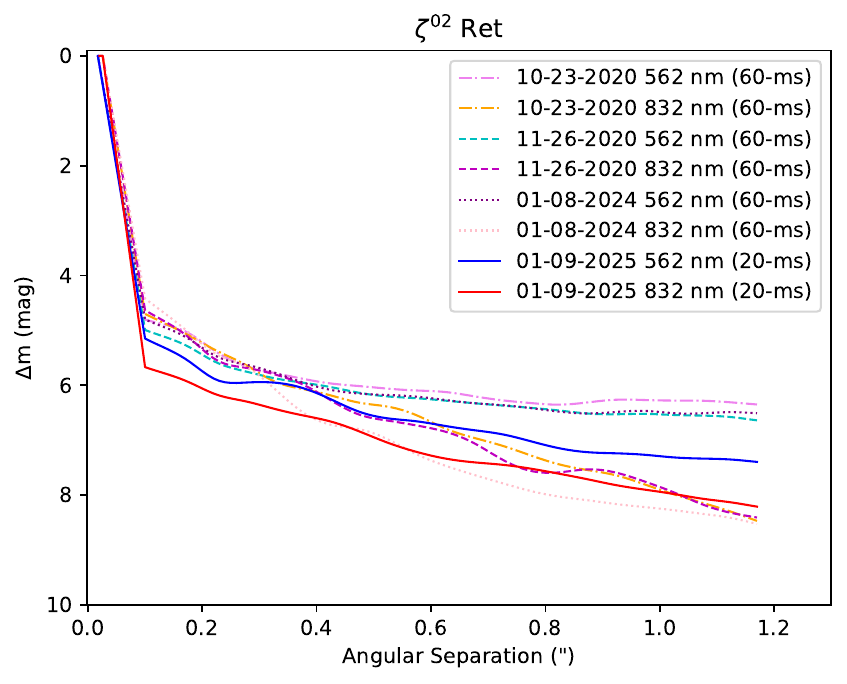}}
\caption{\label{fig:contrastcurvecomp} 20-ms and 60-ms contrast curves for HD 5015, HD 88230, $\beta$ Com, and $\zeta^{02}$ Ret. We find that similar results are produced for both exposure times.} 
\end{figure}

We show four examples for HD 5015, HD 88230, $\beta$ Com, and $\zeta^{02}$ Ret in Figure \ref{fig:contrastcurvecomp}. 
For each star, we show the contrast curves obtained from both the 20-ms and 60-ms observations.
In all four cases, no companion was detected.
With the exception of HD 5015 and the November 2024 observation of HD 88230, the time spent on-target for the 20-ms observation was double what is normally spent on-target (six minutes instead of three). 
In all cases, the 20-ms and 60-ms observations produce similar results if no saturation occurs, with some slight improvements for observations which observed the target for longer.

\subsection{Applications to Target Selection for HWO}

Finding and characterizing stellar companions to potential HWO target stars is key to the success of the mission, both from a technical and a scientific perspective.
As part of their effort to construct the ExEP provisional star list for HWO, \citet{2024mamajek} consulted the WDS and the SB9 to remove close-in stellar companions with separations less than 3\arcsec{} that would likely interfere with HWO coronagraphic observations.
Figure \ref{fig:hwophysep} shows that this effort was largely successful, as the majority of the companions identified in this work have separations larger than 3\arcsec.
However, our results also show that close-in stellar companions are still being identified around stars in this ``cleaned'' sample.

We have shown that Gaia can help in this effort.
Its contributions include the identification of CPM companions at 100s to 1000s of arcseconds, as well as useful indicators to highlight stars that host a currently unresolved companion.
In particular, the ability of Gaia to flag close-in companions as non-single stars enables targeted follow-up to confirm and assess the parameters of the companion.
However, for the bright stars on the HWO target list, our results indicate that the Gaia indicators may not be reliable, in particular, the RUWE parameter (Figure \ref{fig:ruwe}).
Therefore, caution needs to be exercised in using the RUWE values and other Gaia metrics to identify multiples amongst the brightest stars of the HWO target star list.

Overall, this work highlights the need for a uniform stellar multiplicity survey of all potential HWO targets using the best available facilities and a variety of techniques.
Our missing companion analysis shows that, although speckle imaging and wide field seeing limited imaging may detect the majority of companions, there is still the potential for a close-in or faint stellar companion around these target stars.
To examine the region of separation and contrast space that these companions occupy, other methods such long-term spectroscopic observations, space-based techniques, or the use of high-resolution imaging on the Extremely Large Telescope needs to be utilized.

\section{Conclusions} \label{sec:Conclusions}

Much like the previous Great Observatories, the Habitable Worlds Observatory will be an enormous undertaking for the astronomical community. 
In order to ensure the success of one of its key missions, the direct-imaging of 25 habitable-zone exoplanets, we must thoroughly vet any star that may be on the final target list.
This characterization includes an assessment of their abundances, ages, radii, masses, activity cycles, and nearby companions, both bound and unbound, stellar and planetary.

In this paper, we present results from new and archival speckle observations of 80 potential HWO targets stars drawn from the ExEP provisional star list for HWO \citep{2024mamajek}. 
We present one new definitive detection, and an ambiguous detection of a known companion.
After a search of Gaia and the literature, we find that 27 out of the 80 stars observed are in multi-star systems.

We note that 19 targets have elevated Gaia metrics indicating the potential presence of an unresolved companion. 
Two of these targets are flagged as NSS. 
However, we caution that bright targets may have higher astrometric errors than normal \citep{2023gaiadr3}, and as such, these Gaia indicators may not be as effective for bright targets. 
We find that only three of these 19 targets are shown to be multiples via other techniques.
Two of those three are flagged as Gaia Non-single stars, with one confirmed by our speckle observations. 
The third target (TIC 259291108) has an elevated IPD\_fmp value and is discuss in Section \ref{subsec:hip110649}.

For each target that was observed with speckle imaging, we simulate potential stellar companions and assess our sensitivity to them using the contrast curves from our speckle observations.
We find that the majority (75-85\%) of simulated companions are recovered, although we caution that companions with masses lower than $\sim0.1 M_{\odot}$ are not included in this analysis.
Furthermore, known companions and known planets other than TOIs are not included in the analysis.
Future work will be to include these known stellar and substellar companions into the analysis.
The undetected companions are very faint or very close-in.
For these companions, we will need to implement long-term spectroscopic monitoring, use space-based techniques, or observe them using the high-angular resolution imagers that will be available on the ELTs.

We also find that 20-ms observations yield similar results to the typical 60-ms observations from `Alopeke and Zorro.
This enables their use on bright targets that might prove difficult at other facilities.

Before the launch of HWO, many more potential targets will need to be vetted for stellar companions.
By identifying which targets are part of multi-star systems and the parameters of the companions in those systems, we can enable the future architects of the final target list to make informed decisions on whether to include the examined stars as HWO targets.
However, this work should to be started now rather than in five to ten years.
This is because exoplanet yield simulations are being run now in order to help design and guide both the science design requirements and the optomechanical design of the mission.
These simulations must be as realistic as possible and one of the key inputs into these simulations is the multiplicity of the potential target.
High-resolution imaging, particularly speckle imaging, provides a time-efficient method for revealing companions at close separations and moderate contrasts.
We are currently planning on observing the remainder of the ExEP list and more potential HWO targets with high resolution imaging as a next step in paving the road to the Habitable Worlds Observatory.

\section{acknowledgments}

We would like to thank the anonymous referee for their insightful comments. We would also like to thank Ruslan Belikov, Michael Bottom, Thierry Forveille, Madison LeBlanc, Tim Johns, and Dan Sirbu for their comments on a draft of this manuscript.

This research has made use of the Exoplanet Follow-up Observation Program (ExoFOP; DOI: 10.26134/ExoFOP5) website, which is operated by the California Institute of Technology, under contract with the National Aeronautics and Space Administration under the Exoplanet Exploration Program.

The International Gemini Observatory, a program of NSF NOIRLab, is managed by the Association of Universities for Research in Astronomy (AURA) under a cooperative agreement with the U.S. National Science Foundation on behalf of the Gemini partnership: the U.S. National Science Foundation (United States), the National Research Council (Canada), Agencia Nacional de Investigación y Desarrollo (Chile), Ministerio de Ciencia, Tecnología e Innovación (Argentina), Ministério da Ciência, Tecnologia, Inovações e Comunicações (Brazil), and Korea Astronomy and Space Science Institute (Republic of Korea).

Observations in the paper made use of the High-Resolution Imaging instruments `Alopeke and Zorro. `Alopeke and Zorro were funded by the NASA Exoplanet Exploration Program and built at the NASA Ames Research Center by Steve B. Howell, Nic Scott, Elliott P. Horch, and Emmett Quigley. `Alopeke and Zorro were mounted on the Gemini North and South telescope of the international Gemini Observatory, a program of NSF NOIRLab, which is managed by the Association of Universities for Research in Astronomy (AURA) under a cooperative agreement with the U.S. National Science Foundation. on behalf of the Gemini partnership: the U.S. National Science Foundation (United States), National Research Council (Canada), Agencia Nacional de Investigación y Desarrollo (Chile), Ministerio de Ciencia, Tecnología e Innovación (Argentina), Ministério da Ciência, Tecnologia, Inovações e Comunicações (Brazil), and Korea Astronomy and Space Science Institute (Republic of Korea).

Based on observations at NSF Kitt Peak National Observatory, NSF NOIRLab (Prop:2017A-0006, 2023A-807455 PI: S.B. Howell), which is managed by the Association of Universities for Research in Astronomy (AURA) under a cooperative agreement with the U.S. National Science Foundation. The authors are honored to be permitted to conduct astronomical research on I'oligam Du’ag (Kitt Peak), a mountain with particular significance to the Tohono O’odham.

Some of the observations in the paper made use of the NN-EXPLORE Exoplanet and Stellar Speckle Imager (NESSI). NESSI was funded by the NASA Exoplanet Exploration Program and the NASA Ames Research Center. NESSI was built at the Ames Research Center by Steve B. Howell, Nic Scott, Elliott P. Horch, and Emmett Quigley.

This research has made use of the Washington Double Star Catalog maintained at the U.S. Naval Observatory.

We acknowledge financial support from the Agencia Estatal de Investigaci\'on (AEI/10.13039/501100011033) of the Ministerio de Ciencia e Innovaci\'on and the ERDF ``A way of making Europe'' through project PID2022-137241NB-C42.

Part of this research was carried out at the Jet Propulsion Laboratory, California Institute of Technology, under a contract with the National Aeronautics and Space Administration (80NM0018D0004).


\bibliography{sample631}{}

@ARTICLE{Clark2024AJ....167..174C,
       author = {{Clark}, Catherine A. and {van Belle}, Gerard T. and {Horch}, Elliott P. and {Ciardi}, David R. and {von Braun}, Kaspar and {Skiff}, Brian A. and {Winters}, Jennifer G. and {Lund}, Michael B. and {Everett}, Mark E. and {Hartman}, Zachary D. and {Llama}, Joe},
        title = "{The POKEMON Speckle Survey of Nearby M Dwarfs. III. The Stellar Multiplicity Rate of M Dwarfs within 15 pc}",
      journal = {\aj},
         year = 2024,
        month = apr,
       volume = {167},
       number = {4},
          eid = {174},
        pages = {174},
          doi = {10.3847/1538-3881/ad267d},
archivePrefix = {arXiv},
       eprint = {2401.14703},
 primaryClass = {astro-ph.SR},
       adsurl = {https://ui.adsabs.harvard.edu/abs/2024AJ....167..174C}
}

@ARTICLE{Clark2024AJ....167...56C,
       author = {{Clark}, Catherine A. and {van Belle}, Gerard T. and {Horch}, Elliott P. and {Lund}, Michael B. and {Ciardi}, David R. and {von Braun}, Kaspar and {Winters}, Jennifer G. and {Everett}, Mark E. and {Hartman}, Zachary D. and {Llama}, Joe},
        title = "{The POKEMON Speckle Survey of Nearby M Dwarfs. II. Observations of 1125 Targets}",
      journal = {\aj},
         year = 2024,
        month = feb,
       volume = {167},
       number = {2},
          eid = {56},
        pages = {56},
          doi = {10.3847/1538-3881/ad0bfd},
archivePrefix = {arXiv},
       eprint = {2311.08489},
 primaryClass = {astro-ph.SR},
       adsurl = {https://ui.adsabs.harvard.edu/abs/2024AJ....167...56C}
}

@ARTICLE{RafikovSilsbee2015ApJ...798...70R,
       author = {{Rafikov}, Roman R. and {Silsbee}, Kedron},
        title = "{Planet Formation in Stellar Binaries. II. Overcoming the Fragmentation Barrier in {\ensuremath{\alpha}} Centauri and {\ensuremath{\gamma}} Cephei-like Systems}",
      journal = {\apj},
         year = 2015,
        month = jan,
       volume = {798},
       number = {2},
          eid = {70},
        pages = {70},
          doi = {10.1088/0004-637X/798/2/70},
archivePrefix = {arXiv},
       eprint = {1408.4819},
 primaryClass = {astro-ph.EP},
       adsurl = {https://ui.adsabs.harvard.edu/abs/2015ApJ...798...70R}
}

@ARTICLE{RafikovSilsbee2015ApJ...798...69R,
       author = {{Rafikov}, Roman R. and {Silsbee}, Kedron},
        title = "{Planet Formation in Stellar Binaries. I. Planetesimal Dynamics in Massive Protoplanetary Disks}",
      journal = {\apj},
         year = 2015,
        month = jan,
       volume = {798},
       number = {2},
          eid = {69},
        pages = {69},
          doi = {10.1088/0004-637X/798/2/69},
archivePrefix = {arXiv},
       eprint = {1405.7054},
 primaryClass = {astro-ph.EP},
       adsurl = {https://ui.adsabs.harvard.edu/abs/2015ApJ...798...69R}
}

@ARTICLE{Jang-Condell2015ApJ...799..147J,
       author = {{Jang-Condell}, Hannah},
        title = "{On the Likelihood of Planet Formation in Close Binaries}",
      journal = {\apj},
         year = 2015,
        month = feb,
       volume = {799},
       number = {2},
          eid = {147},
        pages = {147},
          doi = {10.1088/0004-637X/799/2/147},
archivePrefix = {arXiv},
       eprint = {1501.00617},
 primaryClass = {astro-ph.EP},
       adsurl = {https://ui.adsabs.harvard.edu/abs/2015ApJ...799..147J}
}

@ARTICLE{HaghighipourRaymond2007ApJ...666..436H,
       author = {{Haghighipour}, Nader and {Raymond}, Sean N.},
        title = "{Habitable Planet Formation in Binary Planetary Systems}",
      journal = {\apj},
         year = 2007,
        month = sep,
       volume = {666},
       number = {1},
        pages = {436-446},
          doi = {10.1086/520501},
archivePrefix = {arXiv},
       eprint = {astro-ph/0702706},
 primaryClass = {astro-ph},
       adsurl = {https://ui.adsabs.harvard.edu/abs/2007ApJ...666..436H}
}

@INPROCEEDINGS{Scowen2025AAS...24530504S,
       author = {{Scowen}, Paul and {Bolcar}, Matthew and {Zhao}, Feng},
        title = "{HWO Technology Roadmapping - High Sensitivity UV and Visible Band Technologies for Instrumentation}",
    booktitle = {American Astronomical Society Meeting Abstracts},
         year = 2025,
       series = {American Astronomical Society Meeting Abstracts},
       volume = {245},
        month = jan,
          eid = {305.04},
        pages = {305.04},
       adsurl = {https://ui.adsabs.harvard.edu/abs/2025AAS...24530504S}
}

@ARTICLE{2021paegert,
       author = {{Paegert}, Martin and {Stassun}, Keivan G. and {Collins}, Karen A. and {Pepper}, Joshua and {Torres}, Guillermo and {Jenkins}, Jon and {Twicken}, Joseph D. and {Latham}, David W.},
        title = "{TESS Input Catalog versions 8.1 and 8.2: Phantoms in the 8.0 Catalog and How to Handle Them}",
      journal = {arXiv e-prints},
         year = 2021,
        month = aug,
          eid = {arXiv:2108.04778},
        pages = {arXiv:2108.04778},
          doi = {10.48550/arXiv.2108.04778},
archivePrefix = {arXiv},
       eprint = {2108.04778},
 primaryClass = {astro-ph.EP},
       adsurl = {https://ui.adsabs.harvard.edu/abs/2021arXiv210804778P}
}

@ARTICLE{2019stassun,
       author = {{Stassun}, Keivan G. and {Oelkers}, Ryan J. and {Paegert}, Martin and {Torres}, Guillermo and {Pepper}, Joshua and {De Lee}, Nathan and {Collins}, Kevin and {Latham}, David W. and {Muirhead}, Philip S. and {Chittidi}, Jay and {Rojas-Ayala}, B{\'a}rbara and {Fleming}, Scott W. and {Rose}, Mark E. and {Tenenbaum}, Peter and {Ting}, Eric B. and {Kane}, Stephen R. and {Barclay}, Thomas and {Bean}, Jacob L. and {Brassuer}, C.~E. and {Charbonneau}, David and {Ge}, Jian and {Lissauer}, Jack J. and {Mann}, Andrew W. and {McLean}, Brian and {Mullally}, Susan and {Narita}, Norio and {Plavchan}, Peter and {Ricker}, George R. and {Sasselov}, Dimitar and {Seager}, S. and {Sharma}, Sanjib and {Shiao}, Bernie and {Sozzetti}, Alessandro and {Stello}, Dennis and {Vanderspek}, Roland and {Wallace}, Geoff and {Winn}, Joshua N.},
        title = "{The Revised TESS Input Catalog and Candidate Target List}",
      journal = {\aj},
         year = 2019,
        month = oct,
       volume = {158},
       number = {4},
          eid = {138},
        pages = {138},
          doi = {10.3847/1538-3881/ab3467},
archivePrefix = {arXiv},
       eprint = {1905.10694},
 primaryClass = {astro-ph.SR},
       adsurl = {https://ui.adsabs.harvard.edu/abs/2019AJ....158..138S}
}

@ARTICLE{2025matson,
       author = {{Matson}, Rachel A. and {Gore}, Rebecca and {Howell}, Steve B. and {Ciardi}, David R. and {Christiansen}, Jessie L. and {Clark}, Catherine A. and {Crossfield}, Ian J.~M. and {Fajardo-Acosta}, Sergio B. and {Fernandes}, Rachel B. and {Furlan}, Elise and {Gilbert}, Emily A. and {Gonzales}, Erica and {Lester}, Kathryn V. and {Lund}, Michael B. and {Matthews}, Elisabeth C. and {Polanski}, Alex S. and {Schlieder}, Joshua E. and {Ziegler}, Carl},
        title = "{Demographics of M Dwarf Binary Exoplanet Hosts Discovered by TESS}",
      journal = {\aj},
         year = 2025,
        month = feb,
       volume = {169},
       number = {2},
          eid = {76},
        pages = {76},
          doi = {10.3847/1538-3881/ad9923},
archivePrefix = {arXiv},
       eprint = {2412.08465},
 primaryClass = {astro-ph.EP},
       adsurl = {https://ui.adsabs.harvard.edu/abs/2025AJ....169...76M}
}

@ARTICLE{2021lester,
       author = {{Lester}, Kathryn V. and {Matson}, Rachel A. and {Howell}, Steve B. and {Furlan}, Elise and {Gnilka}, Crystal L. and {Scott}, Nicholas J. and {Ciardi}, David R. and {Everett}, Mark E. and {Hartman}, Zachary D. and {Hirsch}, Lea A.},
        title = "{Speckle Observations of TESS Exoplanet Host Stars. II. Stellar Companions at 1-1000 au and Implications for Small Planet Detection}",
      journal = {\aj},
         year = 2021,
        month = aug,
       volume = {162},
       number = {2},
          eid = {75},
        pages = {75},
          doi = {10.3847/1538-3881/ac0d06},
archivePrefix = {arXiv},
       eprint = {2106.13354},
 primaryClass = {astro-ph.SR},
       adsurl = {https://ui.adsabs.harvard.edu/abs/2021AJ....162...75L}
}

@ARTICLE{2021elbadry,
       author = {{El-Badry}, Kareem and {Rix}, Hans-Walter and {Heintz}, Tyler M.},
        title = "{A million binaries from Gaia eDR3: sample selection and validation of Gaia parallax uncertainties}",
      journal = {\mnras},
         year = 2021,
        month = sep,
       volume = {506},
       number = {2},
        pages = {2269-2295},
          doi = {10.1093/mnras/stab323},
archivePrefix = {arXiv},
       eprint = {2101.05282},
 primaryClass = {astro-ph.SR},
       adsurl = {https://ui.adsabs.harvard.edu/abs/2021MNRAS.506.2269E}
}

@ARTICLE{1983Lohmann,
       author = {{Lohmann}, A.~W. and {Weigelt}, G. and {Wirnitzer}, B.},
        title = "{Speckle masking in astronomy: triple correlation theory and applications.}",
      journal = {\ao},
         year = 1983,
        month = dec,
       volume = {22},
        pages = {4028-4037},
          doi = {10.1364/AO.22.004028},
       adsurl = {https://ui.adsabs.harvard.edu/abs/1983ApOpt..22.4028L}
}

@ARTICLE{2019hutter,
       author = {{Hutter}, D.~J. and {Tycner}, C. and {Zavala}, R.~T. and {Benson}, J.~A. and {Hummel}, C.~A. and {Sanborn}, J.},
        title = "{Surveying the Bright Stars by Optical Interferometry. II. A Volume-limited Multiplicity Survey of Main-sequence F Stars}",
      journal = {\apjs},
         year = 2019,
        month = aug,
       volume = {243},
       number = {2},
          eid = {32},
        pages = {32},
          doi = {10.3847/1538-4365/ab32e1},
       adsurl = {https://ui.adsabs.harvard.edu/abs/2019ApJS..243...32H}
}

@ARTICLE{2021moe,
       author = {{Moe}, Maxwell and {Kratter}, Kaitlin M.},
        title = "{Impact of binary stars on planet statistics - I. Planet occurrence rates and trends with stellar mass}",
      journal = {\mnras},
         year = 2021,
        month = nov,
       volume = {507},
       number = {3},
        pages = {3593-3611},
          doi = {10.1093/mnras/stab2328},
archivePrefix = {arXiv},
       eprint = {1912.01699},
 primaryClass = {astro-ph.EP},
       adsurl = {https://ui.adsabs.harvard.edu/abs/2021MNRAS.507.3593M}
}

@ARTICLE{2021hirsch,
       author = {{Hirsch}, Lea A. and {Rosenthal}, Lee and {Fulton}, Benjamin J. and {Howard}, Andrew W. and {Ciardi}, David R. and {Marcy}, Geoffrey W. and {Nielsen}, Eric and {Petigura}, Erik A. and {de Rosa}, Robert J. and {Isaacson}, Howard and {Weiss}, Lauren M. and {Sinukoff}, Evan and {Macintosh}, Bruce},
        title = "{Understanding the Impacts of Stellar Companions on Planet Formation and Evolution: A Survey of Stellar and Planetary Companions within 25 pc}",
      journal = {\aj},
         year = 2021,
        month = mar,
       volume = {161},
       number = {3},
          eid = {134},
        pages = {134},
          doi = {10.3847/1538-3881/abd639},
archivePrefix = {arXiv},
       eprint = {2012.09190},
 primaryClass = {astro-ph.EP},
       adsurl = {https://ui.adsabs.harvard.edu/abs/2021AJ....161..134H}
}

@ARTICLE{2022clark,
       author = {{Clark}, Catherine A. and {van Belle}, Gerard T. and {Ciardi}, David R. and {Lund}, Michael B. and {Howell}, Steve B. and {Everett}, Mark E. and {Beichman}, Charles A. and {Winters}, Jennifer G.},
        title = "{A Dearth of Close-in Stellar Companions to M-dwarf TESS Objects of Interest}",
      journal = {\aj},
         year = 2022,
        month = may,
       volume = {163},
       number = {5},
          eid = {232},
        pages = {232},
          doi = {10.3847/1538-3881/ac6101},
archivePrefix = {arXiv},
       eprint = {2203.12795},
 primaryClass = {astro-ph.SR},
       adsurl = {https://ui.adsabs.harvard.edu/abs/2022AJ....163..232C}
}

@ARTICLE{2014tokovinin,
       author = {{Tokovinin}, Andrei},
        title = "{From Binaries to Multiples. II. Hierarchical Multiplicity of F and G Dwarfs}",
      journal = {\aj},
         year = 2014,
        month = apr,
       volume = {147},
       number = {4},
          eid = {87},
        pages = {87},
          doi = {10.1088/0004-6256/147/4/87},
archivePrefix = {arXiv},
       eprint = {1401.6827},
 primaryClass = {astro-ph.SR},
       adsurl = {https://ui.adsabs.harvard.edu/abs/2014AJ....147...87T}
}

@INCOLLECTION{2015thebault,
       author = {{Thebault}, P. and {Haghighipour}, N.},
        title = "{Planet Formation in Binaries}",
    booktitle = {Planetary Exploration and Science: Recent Results and Advances},
         year = 2015,
       editor = {{Jin}, Shuanggen and {Haghighipour}, Nader and {Ip}, Wing-Huen},
        pages = {309-340},
          doi = {10.1007/978-3-662-45052-9_13},
       adsurl = {https://ui.adsabs.harvard.edu/abs/2015pes..book..309T},
      publisher = {Springer}
}

@ARTICLE{2020hawkins,
       author = {{Hawkins}, Keith and {Lucey}, Madeline and {Ting}, Yuan-Sen and {Ji}, Alexander and {Katzberg}, Dustin and {Thompson}, Megan and {El-Badry}, Kareem and {Teske}, Johanna and {Nelson}, Tyler and {Carrillo}, Andreia},
        title = "{Identical or fraternal twins? The chemical homogeneity of wide binaries from Gaia DR2}",
      journal = {\mnras},
         year = 2020,
        month = feb,
       volume = {492},
       number = {1},
        pages = {1164-1179},
          doi = {10.1093/mnras/stz3132},
archivePrefix = {arXiv},
       eprint = {1912.08895},
 primaryClass = {astro-ph.SR},
       adsurl = {https://ui.adsabs.harvard.edu/abs/2020MNRAS.492.1164H}
}

@ARTICLE{2016gatewood,
       author = {{Gatewood}, George and {Gatewood}, Carolyn},
        title = "{Near IR Measures II, 9-12hr and Reference Stars}",
      journal = {Journal of Double Star Observations},
         year = 2016,
        month = oct,
       volume = {12},
       number = {6},
        pages = {580-584},
       adsurl = {https://ui.adsabs.harvard.edu/abs/2016JDSO...12..580G}
}

@ARTICLE{1989mcalister,
       author = {{McAlister}, Harold A. and {Hartkopf}, William I. and {Sowell}, James R. and {Dombrowski}, Edmund G. and {Franz}, Otto G.},
        title = "{ICCD Speckle Observations of Binary Stars. IV. Measurements During 1986-1988 from the Kitt Peak 4-m Telescope}",
      journal = {\aj},
         year = 1989,
        month = feb,
       volume = {97},
        pages = {510},
          doi = {10.1086/115001},
       adsurl = {https://ui.adsabs.harvard.edu/abs/1989AJ.....97..510M}
}

@ARTICLE{2018baroch,
       author = {{Baroch}, D. and {Morales}, J.~C. and {Ribas}, I. and {Tal-Or}, L. and {Zechmeister}, M. and {Reiners}, A. and {Caballero}, J.~A. and {Quirrenbach}, A. and {Amado}, P.~J. and {Dreizler}, S. and {Lalitha}, S. and {Jeffers}, S.~V. and {Lafarga}, M. and {B{\'e}jar}, V.~J.~S. and {Colom{\'e}}, J. and {Cort{\'e}s-Contreras}, M. and {D{\'\i}ez-Alonso}, E. and {Galad{\'\i}-Enr{\'\i}quez}, D. and {Guenther}, E.~W. and {Hagen}, H. -J. and {Henning}, T. and {Herrero}, E. and {K{\"u}rster}, M. and {Montes}, D. and {Nagel}, E. and {Passegger}, V.~M. and {Perger}, M. and {Rosich}, A. and {Schweitzer}, A. and {Seifert}, W.},
        title = "{The CARMENES search for exoplanets around M dwarfs. Nine new double-line spectroscopic binary stars}",
      journal = {\aap},
         year = 2018,
        month = nov,
       volume = {619},
          eid = {A32},
        pages = {A32},
          doi = {10.1051/0004-6361/201833440},
archivePrefix = {arXiv},
       eprint = {1808.06895},
 primaryClass = {astro-ph.SR},
       adsurl = {https://ui.adsabs.harvard.edu/abs/2018A&A...619A..32B}
}

@ARTICLE{2017tokovinin,
       author = {{Tokovinin}, Andrei},
        title = "{Formation of wide binary stars from adjacent cores}",
      journal = {\mnras},
         year = 2017,
        month = jul,
       volume = {468},
       number = {3},
        pages = {3461-3467},
          doi = {10.1093/mnras/stx707},
archivePrefix = {arXiv},
       eprint = {1703.06794},
 primaryClass = {astro-ph.SR},
       adsurl = {https://ui.adsabs.harvard.edu/abs/2017MNRAS.468.3461T}
}

@ARTICLE{2024heintz,
       author = {{Heintz}, Tyler M. and {Hermes}, J.~J. and {Tremblay}, P. -E. and {Ould Rouis}, Lou Baya and {Reding}, Joshua S. and {Kaiser}, B.~C. and {van Saders}, Jennifer L.},
        title = "{A Test of Spectroscopic Age Estimates of White Dwarfs Using Wide WD+WD Binaries}",
      journal = {\apj},
         year = 2024,
        month = jul,
       volume = {969},
       number = {1},
          eid = {68},
        pages = {68},
          doi = {10.3847/1538-4357/ad479b},
archivePrefix = {arXiv},
       eprint = {2405.02423},
 primaryClass = {astro-ph.SR},
       adsurl = {https://ui.adsabs.harvard.edu/abs/2024ApJ...969...68H}
}

@ARTICLE{2022feng,
       author = {{Feng}, Fabo and {Butler}, R. Paul and {Vogt}, Steven S. and {Clement}, Matthew S. and {Tinney}, C.~G. and {Cui}, Kaiming and {Aizawa}, Masataka and {Jones}, Hugh R.~A. and {Bailey}, J. and {Burt}, Jennifer and {Carter}, B.~D. and {Crane}, Jeffrey D. and {Flammini Dotti}, Francesco and {Holden}, Bradford and {Ma}, Bo and {Ogihara}, Masahiro and {Oppenheimer}, Rebecca and {O'Toole}, S.~J. and {Shectman}, Stephen A. and {Wittenmyer}, Robert A. and {Wang}, Sharon X. and {Wright}, D.~J. and {Xuan}, Yifan},
        title = "{3D Selection of 167 Substellar Companions to Nearby Stars}",
      journal = {\apjs},
         year = 2022,
        month = sep,
       volume = {262},
       number = {1},
          eid = {21},
        pages = {21},
          doi = {10.3847/1538-4365/ac7e57},
archivePrefix = {arXiv},
       eprint = {2208.12720},
 primaryClass = {astro-ph.EP},
       adsurl = {https://ui.adsabs.harvard.edu/abs/2022ApJS..262...21F}
}

@ARTICLE{2006quintana,
       author = {{Quintana}, Elisa V. and {Lissauer}, Jack J.},
        title = "{Terrestrial planet formation surrounding close binary stars}",
      journal = {\icarus},
         year = 2006,
        month = nov,
       volume = {185},
       number = {1},
        pages = {1-20},
          doi = {10.1016/j.icarus.2006.06.016},
archivePrefix = {arXiv},
       eprint = {astro-ph/0607222},
 primaryClass = {astro-ph},
       adsurl = {https://ui.adsabs.harvard.edu/abs/2006Icar..185....1Q}
}

@ARTICLE{2021hamers,
       author = {{Hamers}, Adrian S. and {Rantala}, Antti and {Neunteufel}, Patrick and {Preece}, Holly and {Vynatheya}, Pavan},
        title = "{Multiple Stellar Evolution: a population synthesis algorithm to model the stellar, binary, and dynamical evolution of multiple-star systems}",
      journal = {\mnras},
         year = 2021,
        month = apr,
       volume = {502},
       number = {3},
        pages = {4479-4512},
          doi = {10.1093/mnras/stab287},
archivePrefix = {arXiv},
       eprint = {2011.04513},
 primaryClass = {astro-ph.SR},
       adsurl = {https://ui.adsabs.harvard.edu/abs/2021MNRAS.502.4479H}
}

@ARTICLE{1991duquennoy,
       author = {{Duquennoy}, A. and {Mayor}, M.},
        title = "{Multiplicity among Solar Type Stars in the Solar Neighbourhood - Part Two - Distribution of the Orbital Elements in an Unbiased Sample}",
      journal = {\aap},
         year = 1991,
        month = aug,
       volume = {248},
        pages = {485},
       adsurl = {https://ui.adsabs.harvard.edu/abs/1991A&A...248..485D}
}

@ARTICLE{2008dotter,
       author = {{Dotter}, Aaron and {Chaboyer}, Brian and {Jevremovi{\'c}}, Darko and {Kostov}, Veselin and {Baron}, E. and {Ferguson}, Jason W.},
        title = "{The Dartmouth Stellar Evolution Database}",
      journal = {\apjs},
         year = 2008,
        month = sep,
       volume = {178},
       number = {1},
        pages = {89-101},
          doi = {10.1086/589654},
archivePrefix = {arXiv},
       eprint = {0804.4473},
 primaryClass = {astro-ph},
       adsurl = {https://ui.adsabs.harvard.edu/abs/2008ApJS..178...89D}
}

@ARTICLE{2020bohn,
       author = {{Bohn}, A.~J. and {Southworth}, J. and {Ginski}, C. and {Kenworthy}, M.~A. and {Maxted}, P.~F.~L. and {Evans}, D.~F.},
        title = "{A multiplicity study of transiting exoplanet host stars. I. High-contrast imaging with VLT/SPHERE}",
      journal = {\aap},
         year = 2020,
        month = mar,
       volume = {635},
          eid = {A73},
        pages = {A73},
          doi = {10.1051/0004-6361/201937127},
archivePrefix = {arXiv},
       eprint = {2001.08224},
 primaryClass = {astro-ph.EP},
       adsurl = {https://ui.adsabs.harvard.edu/abs/2020A&A...635A..73B}
}

@ARTICLE{2004zucker,
       author = {{Zucker}, S. and {Mazeh}, T. and {Santos}, N.~C. and {Udry}, S. and {Mayor}, M.},
        title = "{Multi-order TODCOR: Application to observations taken with the CORALIE echelle spectrograph. II. A planet in the system <ASTROBJ>HD 41004</ASTROBJ>}",
      journal = {\aap},
         year = 2004,
        month = nov,
       volume = {426},
        pages = {695-698},
          doi = {10.1051/0004-6361:20040384},
       adsurl = {https://ui.adsabs.harvard.edu/abs/2004A&A...426..695Z}
}

@ARTICLE{2019matson,
       author = {{Matson}, Rachel A. and {Howell}, Steve B. and {Ciardi}, David R.},
        title = "{Detecting Unresolved Binaries in TESS Data with Speckle Imaging}",
      journal = {\aj},
         year = 2019,
        month = may,
       volume = {157},
       number = {5},
          eid = {211},
        pages = {211},
          doi = {10.3847/1538-3881/ab1755},
archivePrefix = {arXiv},
       eprint = {1811.02108},
 primaryClass = {astro-ph.EP},
       adsurl = {https://ui.adsabs.harvard.edu/abs/2019AJ....157..211M}
}

@ARTICLE{2023tokovinin,
       author = {{Tokovinin}, Andrei},
        title = "{Exploring Thousands of Nearby Hierarchical Systems with Gaia and Speckle Interferometry}",
      journal = {\aj},
         year = 2023,
        month = apr,
       volume = {165},
       number = {4},
          eid = {180},
        pages = {180},
          doi = {10.3847/1538-3881/acc464},
       adsurl = {https://ui.adsabs.harvard.edu/abs/2023AJ....165..180T}
}

@ARTICLE{2021scott,
       author = {{Scott}, Nicholas J. and {Howell}, Steve B. and {Gnilka}, Crystal L. and {Stephens}, Andrew W. and {Salinas}, Ricardo and {Matson}, Rachel A. and {Furlan}, Elise and {Horch}, Elliott P. and {Everett}, Mark E. and {Ciardi}, David R. and {Mills}, Dave and {Quigley}, Emmett A.},
        title = "{Twin High-resolution, High-speed Imagers for the Gemini Telescopes: Instrument description and science verification results}",
      journal = {Frontiers in Astronomy and Space Sciences},
         year = 2021,
        month = sep,
       volume = {8},
          eid = {138},
        pages = {138},
          doi = {10.3389/fspas.2021.716560},
       adsurl = {https://ui.adsabs.harvard.edu/abs/2021FrASS...8..138S}
}

@ARTICLE{2021edr3,
       author = {{Gaia Collaboration} and {Brown}, A.~G.~A. and {Vallenari}, A. and {Prusti}, T. and {de Bruijne}, J.~H.~J. and {Babusiaux}, C. and {Biermann}, M. and {Creevey}, O.~L. and {Evans}, D.~W. and {Eyer}, L. and {Hutton}, A. and {Jansen}, F. and {Jordi}, C. and {Klioner}, S.~A. and {Lammers}, U. and {Lindegren}, L. and {Luri}, X. and {Mignard}, F. and {Panem}, C. and {Pourbaix}, D. and {Randich}, S. and {Sartoretti}, P. and {Soubiran}, C. and {Walton}, N.~A. and {Arenou}, F. and {Bailer-Jones}, C.~A.~L. and {Bastian}, U. and {Cropper}, M. and {Drimmel}, R. and {Katz}, D. and {Lattanzi}, M.~G. and {van Leeuwen}, F. and {Bakker}, J. and {Cacciari}, C. and {Casta{\~n}eda}, J. and {De Angeli}, F. and {Ducourant}, C. and {Fabricius}, C. and {Fouesneau}, M. and {Fr{\'e}mat}, Y. and {Guerra}, R. and {Guerrier}, A. and {Guiraud}, J. and {Jean-Antoine Piccolo}, A. and {Masana}, E. and {Messineo}, R. and {Mowlavi}, N. and {Nicolas}, C. and {Nienartowicz}, K. and {Pailler}, F. and {Panuzzo}, P. and {Riclet}, F. and {Roux}, W. and {Seabroke}, G.~M. and {Sordo}, R. and {Tanga}, P. and {Th{\'e}venin}, F. and {Gracia-Abril}, G. and {Portell}, J. and {Teyssier}, D. and {Altmann}, M. and {Andrae}, R. and {Bellas-Velidis}, I. and {Benson}, K. and {Berthier}, J. and {Blomme}, R. and {Brugaletta}, E. and {Burgess}, P.~W. and {Busso}, G. and {Carry}, B. and {Cellino}, A. and {Cheek}, N. and {Clementini}, G. and {Damerdji}, Y. and {Davidson}, M. and {Delchambre}, L. and {Dell'Oro}, A. and {Fern{\'a}ndez-Hern{\'a}ndez}, J. and {Galluccio}, L. and {Garc{\'\i}a-Lario}, P. and {Garcia-Reinaldos}, M. and {Gonz{\'a}lez-N{\'u}{\~n}ez}, J. and {Gosset}, E. and {Haigron}, R. and {Halbwachs}, J. -L. and {Hambly}, N.~C. and {Harrison}, D.~L. and {Hatzidimitriou}, D. and {Heiter}, U. and {Hern{\'a}ndez}, J. and {Hestroffer}, D. and {Hodgkin}, S.~T. and {Holl}, B. and {Jan{\ss}en}, K. and {Jevardat de Fombelle}, G. and {Jordan}, S. and {Krone-Martins}, A. and {Lanzafame}, A.~C. and {L{\"o}ffler}, W. and {Lorca}, A. and {Manteiga}, M. and {Marchal}, O. and {Marrese}, P.~M. and {Moitinho}, A. and {Mora}, A. and {Muinonen}, K. and {Osborne}, P. and {Pancino}, E. and {Pauwels}, T. and {Petit}, J. -M. and {Recio-Blanco}, A. and {Richards}, P.~J. and {Riello}, M. and {Rimoldini}, L. and {Robin}, A.~C. and {Roegiers}, T. and {Rybizki}, J. and {Sarro}, L.~M. and {Siopis}, C. and {Smith}, M. and {Sozzetti}, A. and {Ulla}, A. and {Utrilla}, E. and {van Leeuwen}, M. and {van Reeven}, W. and {Abbas}, U. and {Abreu Aramburu}, A. and {Accart}, S. and {Aerts}, C. and {Aguado}, J.~J. and {Ajaj}, M. and {Altavilla}, G. and {{\'A}lvarez}, M.~A. and {{\'A}lvarez Cid-Fuentes}, J. and {Alves}, J. and {Anderson}, R.~I. and {Anglada Varela}, E. and {Antoja}, T. and {Audard}, M. and {Baines}, D. and {Baker}, S.~G. and {Balaguer-N{\'u}{\~n}ez}, L. and {Balbinot}, E. and {Balog}, Z. and {Barache}, C. and {Barbato}, D. and {Barros}, M. and {Barstow}, M.~A. and {Bartolom{\'e}}, S. and {Bassilana}, J. -L. and {Bauchet}, N. and {Baudesson-Stella}, A. and {Becciani}, U. and {Bellazzini}, M. and {Bernet}, M. and {Bertone}, S. and {Bianchi}, L. and {Blanco-Cuaresma}, S. and {Boch}, T. and {Bombrun}, A. and {Bossini}, D. and {Bouquillon}, S. and {Bragaglia}, A. and {Bramante}, L. and {Breedt}, E. and {Bressan}, A. and {Brouillet}, N. and {Bucciarelli}, B. and {Burlacu}, A. and {Busonero}, D. and {Butkevich}, A.~G. and {Buzzi}, R. and {Caffau}, E. and {Cancelliere}, R. and {C{\'a}novas}, H. and {Cantat-Gaudin}, T. and {Carballo}, R. and {Carlucci}, T. and {Carnerero}, M.~I. and {Carrasco}, J.~M. and {Casamiquela}, L. and {Castellani}, M. and {Castro-Ginard}, A. and {Castro Sampol}, P. and {Chaoul}, L. and {Charlot}, P. and {Chemin}, L. and {Chiavassa}, A. and {Cioni}, M. -R.~L. and {Comoretto}, G. and {Cooper}, W.~J. and {Cornez}, T. and {Cowell}, S. and {Crifo}, F. and {Crosta}, M. and {Crowley}, C. and {Dafonte}, C. and {Dapergolas}, A. and {David}, M. and {David}, P.},
        title = "{Gaia Early Data Release 3. Summary of the contents and survey properties}",
      journal = {\aap},
         year = 2021,
        month = may,
       volume = {649},
          eid = {A1},
        pages = {A1},
          doi = {10.1051/0004-6361/202039657},
archivePrefix = {arXiv},
       eprint = {2012.01533},
 primaryClass = {astro-ph.GA},
       adsurl = {https://ui.adsabs.harvard.edu/abs/2021A&A...649A...1G}
}

@ARTICLE{2024sullivan,
       author = {{Sullivan}, Kendall and {Kraus}, Adam L. and {Berger}, Travis A. and {Dupuy}, Trent J. and {Evans}, Elise and {Gaidos}, Eric and {Huber}, Daniel and {Ireland}, Michael J. and {Mann}, Andrew W. and {Petigura}, Erik A. and {Thao}, Pa Chia and {Wood}, Mackenna L. and {Zhang}, Jingwen},
        title = "{Revising Properties of Planet{\textendash}Host Binary Systems. IV. The Radius Distribution of Small Planets in Binary Star Systems Is Dependent on Stellar Separation}",
      journal = {\aj},
         year = 2024,
        month = sep,
       volume = {168},
       number = {3},
          eid = {129},
        pages = {129},
          doi = {10.3847/1538-3881/ad6310},
archivePrefix = {arXiv},
       eprint = {2406.17648},
 primaryClass = {astro-ph.EP},
       adsurl = {https://ui.adsabs.harvard.edu/abs/2024AJ....168..129S}
}

@INPROCEEDINGS{2023offner,
       author = {{Offner}, S.~S.~R. and {Moe}, M. and {Kratter}, K.~M. and {Sadavoy}, S.~I. and {Jensen}, E.~L.~N. and {Tobin}, J.~J.},
        title = "{The Origin and Evolution of Multiple Star Systems}",
    booktitle = {Protostars and Planets VII},
         year = 2023,
       editor = {{Inutsuka}, S. and {Aikawa}, Y. and {Muto}, T. and {Tomida}, K. and {Tamura}, M.},
       series = {Astronomical Society of the Pacific Conference Series},
       volume = {534},
        month = jul,
        pages = {275},
          doi = {10.48550/arXiv.2203.10066},
archivePrefix = {arXiv},
       eprint = {2203.10066},
 primaryClass = {astro-ph.SR},
       adsurl = {https://ui.adsabs.harvard.edu/abs/2023ASPC..534..275O}
}

@ARTICLE{2016kraus,
       author = {{Kraus}, Adam L. and {Ireland}, Michael J. and {Huber}, Daniel and {Mann}, Andrew W. and {Dupuy}, Trent J.},
        title = "{The Impact of Stellar Multiplicity on Planetary Systems. I. The Ruinous Influence of Close Binary Companions}",
      journal = {\aj},
         year = 2016,
        month = jul,
       volume = {152},
       number = {1},
          eid = {8},
        pages = {8},
          doi = {10.3847/0004-6256/152/1/8},
archivePrefix = {arXiv},
       eprint = {1604.05744},
 primaryClass = {astro-ph.EP},
       adsurl = {https://ui.adsabs.harvard.edu/abs/2016AJ....152....8K}
}

@ARTICLE{2010tanner,
       author = {{Tanner}, Angelle M. and {Gelino}, Christopher R. and {Law}, Nicholas M.},
        title = "{A High-Contrast Imaging Survey of SIM Lite Planet Search Targets}",
      journal = {\pasp},
         year = 2010,
        month = oct,
       volume = {122},
       number = {896},
        pages = {1195},
          doi = {10.1086/656481},
archivePrefix = {arXiv},
       eprint = {1007.4315},
 primaryClass = {astro-ph.SR},
       adsurl = {https://ui.adsabs.harvard.edu/abs/2010PASP..122.1195T}
}

@ARTICLE{2025gonzalezpayo,
       author = {{Gonz{\'a}lez-Payo}, J. and {Caballero}, J.~A. and {Cifuentes}, C. and {Cort{\'e}s-Contreras},M. and {Rica},F.M. and {Deveny}, S. and {Hartman}, Z. and {Howell}, S. and {Littlefield}, C.},
        title = "{Characterisation of all known multiple stellar systems within 10 pc}",
      journal = {\aap},
         year = 2025,
        month = jan,
       volume = {Submitted},
       number = {},
        pages = {},
          doi = {},
       adsurl = {}
}

@ARTICLE{2023golovin,
       author = {{Golovin}, Alex and {Reffert}, Sabine and {Just}, Andreas and {Jordan}, Stefan and {Vani}, Akash and {Jahrei{\ss}}, Hartmut},
        title = "{The Fifth Catalogue of Nearby Stars (CNS5)}",
      journal = {\aap},
         year = 2023,
        month = feb,
       volume = {670},
          eid = {A19},
        pages = {A19},
          doi = {10.1051/0004-6361/202244250},
archivePrefix = {arXiv},
       eprint = {2211.01449},
 primaryClass = {astro-ph.SR},
       adsurl = {https://ui.adsabs.harvard.edu/abs/2023A&A...670A..19G}
}

@ARTICLE{2025cifuentes,
       author = {{Cifuentes}, C. and {Caballero}, J.~A. and {Gonz{\'a}lez-Payo}, J. and {Amado}, P.~J. and {B{\'e}jar}, V.~J.~S. and {Burgasser}, A.~J. and {Cort{\'e}s-Contreras}, M. and {Lodieu}, N. and {Montes}, D. and {Quirrenbach}, A. and {Reiners}, A. and {Ribas}, I. and {Sanz-Forcada}, J. and {Seifert}, W. and {Zapatero Osorio}, M.~R.},
        title = "{CARMENES input catalogue of M dwarfs: IX. Multiplicity from close spectroscopic binaries to ultra-wide systems}",
      journal = {\aap},
         year = 2025,
        month = jan,
       volume = {693},
          eid = {A228},
        pages = {A228},
          doi = {10.1051/0004-6361/202452527},
archivePrefix = {arXiv},
       eprint = {2412.12264},
 primaryClass = {astro-ph.SR},
       adsurl = {https://ui.adsabs.harvard.edu/abs/2025A&A...693A.228C}
}

@INPROCEEDINGS{2018scott,
       author = {{Scott}, Nicholas J. and {Howell}, Steve B.},
        title = "{NESSI and 'Alopeke: two new dual-channel speckle imaging instruments}",
    booktitle = {Optical and Infrared Interferometry and Imaging VI},
         year = 2018,
       editor = {{Creech-Eakman}, Michelle J. and {Tuthill}, Peter G. and {M{\'e}rand}, Antoine},
       series = {Society of Photo-Optical Instrumentation Engineers (SPIE) Conference Series},
       volume = {10701},
        month = jul,
          eid = {107010G},
        pages = {107010G},
          doi = {10.1117/12.2311539},
       adsurl = {https://ui.adsabs.harvard.edu/abs/2018SPIE10701E..0GS}
}

@ARTICLE{2015wardduong,
       author = {{Ward-Duong}, K. and {Patience}, J. and {De Rosa}, R.~J. and {Bulger}, J. and {Rajan}, A. and {Goodwin}, S.~P. and {Parker}, Richard J. and {McCarthy}, D.~W. and {Kulesa}, C.},
        title = "{The M-dwarfs in Multiples (MINMS) survey - I. Stellar multiplicity among low-mass stars within 15 pc}",
      journal = {\mnras},
         year = 2015,
        month = may,
       volume = {449},
       number = {3},
        pages = {2618-2637},
          doi = {10.1093/mnras/stv384},
archivePrefix = {arXiv},
       eprint = {1503.00724},
 primaryClass = {astro-ph.SR},
       adsurl = {https://ui.adsabs.harvard.edu/abs/2015MNRAS.449.2618W}
}

@ARTICLE{2007lafreniere,
       author = {{Lafreni{\`e}re}, David and {Doyon}, Ren{\'e} and {Marois}, Christian and {Nadeau}, Daniel and {Oppenheimer}, Ben R. and {Roche}, Patrick F. and {Rigaut}, Fran{\c{c}}ois and {Graham}, James R. and {Jayawardhana}, Ray and {Johnstone}, Doug and {Kalas}, Paul G. and {Macintosh}, Bruce and {Racine}, Ren{\'e}},
        title = "{The Gemini Deep Planet Survey}",
      journal = {\apj},
         year = 2007,
        month = dec,
       volume = {670},
       number = {2},
        pages = {1367-1390},
          doi = {10.1086/522826},
archivePrefix = {arXiv},
       eprint = {0705.4290},
 primaryClass = {astro-ph},
       adsurl = {https://ui.adsabs.harvard.edu/abs/2007ApJ...670.1367L}
}

@ARTICLE{2024gonzalez,
       author = {{Gonz{\'a}lez-Payo}, J. and {Caballero}, J.~A. and {Gorgas}, J. and {Cort{\'e}s-Contreras}, M. and {G{\'a}lvez-Ortiz}, M. -C. and {Cifuentes}, C.},
        title = "{Multiplicity of stars with planets in the solar neighbourhood}",
      journal = {\aap},
         year = 2024,
        month = sep,
       volume = {689},
          eid = {A302},
        pages = {A302},
          doi = {10.1051/0004-6361/202450048},
archivePrefix = {arXiv},
       eprint = {2407.20138},
 primaryClass = {astro-ph.SR},
       adsurl = {https://ui.adsabs.harvard.edu/abs/2024A&A...689A.302G}
}

@ARTICLE{2024kirkpatrick,
       author = {{Kirkpatrick}, J. Davy and {Marocco}, Federico and {Gelino}, Christopher R. and {Raghu}, Yadukrishna and {Faherty}, Jacqueline K. and {Bardalez Gagliuffi}, Daniella C. and {Schurr}, Steven D. and {Apps}, Kevin and {Schneider}, Adam C. and {Meisner}, Aaron M. and {Kuchner}, Marc J. and {Caselden}, Dan and {Smart}, R.~L. and {Casewell}, S.~L. and {Raddi}, Roberto and {Kesseli}, Aurora and {Stevnbak Andersen}, Nikolaj and {Antonini}, Edoardo and {Beaulieu}, Paul and {Bickle}, Thomas P. and {Bilsing}, Martin and {Chieng}, Raymond and {Colin}, Guillaume and {Deen}, Sam and {Dereveanco}, Alexandru and {Doll}, Katharina and {Durantini Luca}, Hugo A. and {Frazer}, Anya and {Gantier}, Jean Marc and {Gramaize}, L{\'e}opold and {Grant}, Kristin and {Hamlet}, Leslie K. and {Higashimura}, Hiro and {Hyogo}, Michiharu and {Ja{\l}owiczor}, Peter A. and {Jonkeren}, Alexander and {Kabatnik}, Martin and {Kiwy}, Frank and {Martin}, David W. and {Michaels}, Marianne N. and {Pendrill}, William and {Pessanha Machado}, Celso and {Pumphrey}, Benjamin and {Rothermich}, Austin and {Russwurm}, Rebekah and {Sainio}, Arttu and {Sanchez}, John and {Sapelkin-Tambling}, Fyodor Theo and {Sch{\"u}mann}, J{\"o}rg and {Selg-Mann}, Karl and {Singh}, Harshdeep and {Stenner}, Andres and {Sun}, Guoyou and {Tanner}, Christopher and {Th{\'e}venot}, Melina and {Ventura}, Maurizio and {Voloshin}, Nikita V. and {Walla}, Jim and {W{\k{e}}dracki}, Zbigniew and {Adorno}, Jose I. and {Aganze}, Christian and {Allers}, Katelyn N. and {Brooks}, Hunter and {Burgasser}, Adam J. and {Calamari}, Emily and {Connor}, Thomas and {Costa}, Edgardo and {Eisenhardt}, Peter R. and {Gagn{\'e}}, Jonathan and {Gerasimov}, Roman and {Gonzales}, Eileen C. and {Hsu}, Chih-Chun and {Kiman}, Rocio and {Li}, Guodong and {Low}, Ryan and {Mamajek}, Eric and {Pantoja}, Blake M. and {Popinchalk}, Mark and {Rees}, Jon M. and {Stern}, Daniel and {Su{\'a}rez}, Genaro and {Theissen}, Christopher and {Tsai}, Chao-Wei and {Vos}, Johanna M. and {Zurek}, David and {The Backyard Worlds: Planet 9 Collaboration}},
        title = "{The Initial Mass Function Based on the Full-sky 20 pc Census of {\ensuremath{\sim}}3600 Stars and Brown Dwarfs}",
      journal = {\apjs},
         year = 2024,
        month = apr,
       volume = {271},
       number = {2},
          eid = {55},
        pages = {55},
          doi = {10.3847/1538-4365/ad24e2},
archivePrefix = {arXiv},
       eprint = {2312.03639},
 primaryClass = {astro-ph.SR},
       adsurl = {https://ui.adsabs.harvard.edu/abs/2024ApJS..271...55K}
}

@ARTICLE{2019kane,
       author = {{Kane}, Stephen R. and {Dalba}, Paul A. and {Li}, Zhexing and {Horch}, Elliott P. and {Hirsch}, Lea A. and {Horner}, Jonathan and {Wittenmyer}, Robert A. and {Howell}, Steve B. and {Everett}, Mark E. and {Butler}, R. Paul and {Tinney}, Christopher G. and {Carter}, Brad D. and {Wright}, Duncan J. and {Jones}, Hugh R.~A. and {Bailey}, Jeremy and {O'Toole}, Simon J.},
        title = "{Detection of Planetary and Stellar Companions to Neighboring Stars via a Combination of Radial Velocity and Direct Imaging Techniques}",
      journal = {\aj},
         year = 2019,
        month = jun,
       volume = {157},
       number = {6},
          eid = {252},
        pages = {252},
          doi = {10.3847/1538-3881/ab1ddf},
archivePrefix = {arXiv},
       eprint = {1904.12931},
 primaryClass = {astro-ph.EP},
       adsurl = {https://ui.adsabs.harvard.edu/abs/2019AJ....157..252K}
}

@INPROCEEDINGS{2020lund,
       author = {{Lund}, M.~B. and {Ciardi}, D.},
        title = "{Reducing Errors in Derived Planetary Radii Caused by Undetected Stellar Companions}",
    booktitle = {American Astronomical Society Meeting Abstracts \#235},
         year = 2020,
       series = {American Astronomical Society Meeting Abstracts},
       volume = {235},
        month = jan,
          eid = {249.06},
        pages = {249.06},
       adsurl = {https://ui.adsabs.harvard.edu/abs/2020AAS...23524906L}
}

@INPROCEEDINGS{2023sirbu,
       author = {{Sirbu}, Dan and {Belikov}, Ruslan and {Bendek}, Eduardo and {Marx}, David and {Riggs}, A.~J. Eldorado and {Mejia Prada}, Camilo and {Seo}, Byoung-Joon and {Zhou}, Hanying},
        title = "{Multi-star wavefront control at the occulting mask coronagraph testbed: monochromatic laboratory demonstration for the Roman Coronagraph instrument}",
    booktitle = {Society of Photo-Optical Instrumentation Engineers (SPIE) Conference Series},
         year = 2023,
       series = {Society of Photo-Optical Instrumentation Engineers (SPIE) Conference Series},
       volume = {12680},
        month = oct,
          eid = {126800Y},
        pages = {126800Y},
          doi = {10.1117/12.2677778},
       adsurl = {https://ui.adsabs.harvard.edu/abs/2023SPIE12680E..0YS}
}

@INPROCEEDINGS{2021bendek,
       author = {{Bendek}, Eduardo A. and {Belikov}, Ruslan and {Sirbu}, Dan and {Ruane}, Garreth and {Riggs}, A.~J. Eldorado and {Balasubramanian}, Bala and {Mejia Prada}, Camilo and {Pluzhnyk}, Eugene},
        title = "{Enabling binary stars high-contrast imaging on the Roman Space Telescope coronagraph instrument}",
    booktitle = {Techniques and Instrumentation for Detection of Exoplanets X},
         year = 2021,
       editor = {{Shaklan}, Stuart B. and {Ruane}, Garreth J.},
       series = {Society of Photo-Optical Instrumentation Engineers (SPIE) Conference Series},
       volume = {11823},
        month = sep,
          eid = {1182311},
        pages = {1182311},
          doi = {10.1117/12.2594992},
       adsurl = {https://ui.adsabs.harvard.edu/abs/2021SPIE11823E..11B}
}

@INPROCEEDINGS{2015belikov,
       author = {{Belikov}, Ruslan and {Bendek}, Eduardo and {Thomas}, Sandrine and {Males}, Jared and {Lozi}, Julien},
        title = "{How to directly image a habitable planet around Alpha Centauri with a \raisebox{-0.5ex}\textasciitilde30-45cm space telescope}",
    booktitle = {Techniques and Instrumentation for Detection of Exoplanets VII},
         year = 2015,
       editor = {{Shaklan}, Stuart},
       series = {Society of Photo-Optical Instrumentation Engineers (SPIE) Conference Series},
       volume = {9605},
        month = sep,
          eid = {960517},
        pages = {960517},
          doi = {10.1117/12.2188732},
archivePrefix = {arXiv},
       eprint = {1510.02479},
 primaryClass = {astro-ph.IM},
       adsurl = {https://ui.adsabs.harvard.edu/abs/2015SPIE.9605E..17B}
}

@ARTICLE{2017sirbu,
       author = {{Sirbu}, D. and {Thomas}, S. and {Belikov}, R. and {Bendek}, E.},
        title = "{Techniques for High-contrast Imaging in Multi-star Systems. II. Multi-star Wavefront Control}",
      journal = {\apj},
         year = 2017,
        month = nov,
       volume = {849},
       number = {2},
          eid = {142},
        pages = {142},
          doi = {10.3847/1538-4357/aa8e02},
archivePrefix = {arXiv},
       eprint = {1704.05441},
 primaryClass = {astro-ph.IM},
       adsurl = {https://ui.adsabs.harvard.edu/abs/2017ApJ...849..142S}
}

@ARTICLE{2015thomas,
       author = {{Thomas}, S. and {Belikov}, R. and {Bendek}, E.},
        title = "{Techniques for High-contrast Imaging in Multi-star Systems. I. Super-Nyquist Wavefront Control}",
      journal = {\apj},
         year = 2015,
        month = sep,
       volume = {810},
       number = {1},
          eid = {81},
        pages = {81},
          doi = {10.1088/0004-637X/810/1/81},
archivePrefix = {arXiv},
       eprint = {1501.01583},
 primaryClass = {astro-ph.IM},
       adsurl = {https://ui.adsabs.harvard.edu/abs/2015ApJ...810...81T}
}

@ARTICLE{2001hartkopforb6,
       author = {{Hartkopf}, William I. and {Mason}, Brian D. and {Worley}, Charles E.},
        title = "{The 2001 US Naval Observatory Double Star CD-ROM. II. The Fifth Catalog of Orbits of Visual Binary Stars}",
      journal = {\aj},
         year = 2001,
        month = dec,
       volume = {122},
       number = {6},
        pages = {3472-3479},
          doi = {10.1086/323921},
       adsurl = {https://ui.adsabs.harvard.edu/abs/2001AJ....122.3472H}
}

@ARTICLE{2010raghavan,
       author = {{Raghavan}, Deepak and {McAlister}, Harold A. and {Henry}, Todd J. and {Latham}, David W. and {Marcy}, Geoffrey W. and {Mason}, Brian D. and {Gies}, Douglas R. and {White}, Russel J. and {ten Brummelaar}, Theo A.},
        title = "{A Survey of Stellar Families: Multiplicity of Solar-type Stars}",
      journal = {\apjs},
         year = 2010,
        month = sep,
       volume = {190},
       number = {1},
        pages = {1-42},
          doi = {10.1088/0067-0049/190/1/1},
archivePrefix = {arXiv},
       eprint = {1007.0414},
 primaryClass = {astro-ph.SR},
       adsurl = {https://ui.adsabs.harvard.edu/abs/2010ApJS..190....1R}
}

@ARTICLE{2013duchene,
       author = {{Duch{\^e}ne}, Gaspard and {Kraus}, Adam},
        title = "{Stellar Multiplicity}",
      journal = {\araa},
         year = 2013,
        month = aug,
       volume = {51},
       number = {1},
        pages = {269-310},
          doi = {10.1146/annurev-astro-081710-102602},
archivePrefix = {arXiv},
       eprint = {1303.3028},
 primaryClass = {astro-ph.SR},
       adsurl = {https://ui.adsabs.harvard.edu/abs/2013ARA&A..51..269D}
}

@ARTICLE{2023gaiamultiples,
       author = {{Gaia Collaboration} and {Arenou}, F. and {Babusiaux}, C. and {Barstow}, M.~A. and {Faigler}, S. and {Jorissen}, A. and {Kervella}, P. and {Mazeh}, T. and {Mowlavi}, N. and {Panuzzo}, P. and {Sahlmann}, J. and {Shahaf}, S. and {Sozzetti}, A. and {Bauchet}, N. and {Damerdji}, Y. and {Gavras}, P. and {Giacobbe}, P. and {Gosset}, E. and {Halbwachs}, J. -L. and {Holl}, B. and {Lattanzi}, M.~G. and {Leclerc}, N. and {Morel}, T. and {Pourbaix}, D. and {Re Fiorentin}, P. and {Sadowski}, G. and {S{\'e}gransan}, D. and {Siopis}, C. and {Teyssier}, D. and {Zwitter}, T. and {Planquart}, L. and {Brown}, A.~G.~A. and {Vallenari}, A. and {Prusti}, T. and {de Bruijne}, J.~H.~J. and {Biermann}, M. and {Creevey}, O.~L. and {Ducourant}, C. and {Evans}, D.~W. and {Eyer}, L. and {Guerra}, R. and {Hutton}, A. and {Jordi}, C. and {Klioner}, S.~A. and {Lammers}, U.~L. and {Lindegren}, L. and {Luri}, X. and {Mignard}, F. and {Panem}, C. and {Randich}, S. and {Sartoretti}, P. and {Soubiran}, C. and {Tanga}, P. and {Walton}, N.~A. and {Bailer-Jones}, C.~A.~L. and {Bastian}, U. and {Drimmel}, R. and {Jansen}, F. and {Katz}, D. and {van Leeuwen}, F. and {Bakker}, J. and {Cacciari}, C. and {Casta{\~n}eda}, J. and {De Angeli}, F. and {Fabricius}, C. and {Fouesneau}, M. and {Fr{\'e}mat}, Y. and {Galluccio}, L. and {Guerrier}, A. and {Heiter}, U. and {Masana}, E. and {Messineo}, R. and {Nicolas}, C. and {Nienartowicz}, K. and {Pailler}, F. and {Riclet}, F. and {Roux}, W. and {Seabroke}, G.~M. and {Sordo}, R. and {Th{\'e}venin}, F. and {Gracia-Abril}, G. and {Portell}, J. and {Altmann}, M. and {Andrae}, R. and {Audard}, M. and {Bellas-Velidis}, I. and {Benson}, K. and {Berthier}, J. and {Blomme}, R. and {Burgess}, P.~W. and {Busonero}, D. and {Busso}, G. and {C{\'a}novas}, H. and {Carry}, B. and {Cellino}, A. and {Cheek}, N. and {Clementini}, G. and {Davidson}, M. and {de Teodoro}, P. and {Nu{\~n}ez Campos}, M. and {Delchambre}, L. and {Dell'Oro}, A. and {Esquej}, P. and {Fern{\'a}ndez-Hern{\'a}ndez}, J. and {Fraile}, E. and {Garabato}, D. and {Garc{\'\i}a-Lario}, P. and {Haigron}, R. and {Hambly}, N.~C. and {Harrison}, D.~L. and {Hern{\'a}ndez}, J. and {Hestroffer}, D. and {Hodgkin}, S.~T. and {Jan{\ss}en}, K. and {Jevardat de Fombelle}, G. and {Jordan}, S. and {Krone-Martins}, A. and {Lanzafame}, A.~C. and {L{\"o}ffler}, W. and {Marchal}, O. and {Marrese}, P.~M. and {Moitinho}, A. and {Muinonen}, K. and {Osborne}, P. and {Pancino}, E. and {Pauwels}, T. and {Recio-Blanco}, A. and {Reyl{\'e}}, C. and {Riello}, M. and {Rimoldini}, L. and {Roegiers}, T. and {Rybizki}, J. and {Sarro}, L.~M. and {Smith}, M. and {Utrilla}, E. and {van Leeuwen}, M. and {Abbas}, U. and {{\'A}brah{\'a}m}, P. and {Abreu Aramburu}, A. and {Aerts}, C. and {Aguado}, J.~J. and {Ajaj}, M. and {Aldea-Montero}, F. and {Altavilla}, G. and {{\'A}lvarez}, M.~A. and {Alves}, J. and {Anders}, F. and {Anderson}, R.~I. and {Anglada Varela}, E. and {Antoja}, T. and {Baines}, D. and {Baker}, S.~G. and {Balaguer-N{\'u}{\~n}ez}, L. and {Balbinot}, E. and {Balog}, Z. and {Barache}, C. and {Barbato}, D. and {Barros}, M. and {Bartolom{\'e}}, S. and {Bassilana}, J. -L. and {Becciani}, U. and {Bellazzini}, M. and {Berihuete}, A. and {Bernet}, M. and {Bertone}, S. and {Bianchi}, L. and {Binnenfeld}, A. and {Blanco-Cuaresma}, S. and {Blazere}, A. and {Boch}, T. and {Bombrun}, A. and {Bossini}, D. and {Bouquillon}, S. and {Bragaglia}, A. and {Bramante}, L. and {Breedt}, E. and {Bressan}, A. and {Brouillet}, N. and {Brugaletta}, E. and {Bucciarelli}, B. and {Burlacu}, A. and {Butkevich}, A.~G. and {Buzzi}, R. and {Caffau}, E. and {Cancelliere}, R. and {Cantat-Gaudin}, T. and {Carballo}, R. and {Carlucci}, T. and {Carnerero}, M.~I. and {Carrasco}, J.~M. and {Casamiquela}, L. and {Castellani}, M. and {Castro-Ginard}, A. and {Chaoul}, L. and {Charlot}, P. and {Chemin}, L. and {Chiaramida}, V. and {Chiavassa}, A. and {Chornay}, N. and {Comoretto}, G.},
        title = "{Gaia Data Release 3. Stellar multiplicity, a teaser for the hidden treasure}",
      journal = {\aap},
         year = 2023,
        month = jun,
       volume = {674},
          eid = {A34},
        pages = {A34},
          doi = {10.1051/0004-6361/202243782},
archivePrefix = {arXiv},
       eprint = {2206.05595},
 primaryClass = {astro-ph.SR},
       adsurl = {https://ui.adsabs.harvard.edu/abs/2023A&A...674A..34G}
}

@ARTICLE{2015teske,
       author = {{Teske}, Johanna K. and {Everett}, Mark E. and {Hirsch}, Lea and {Furlan}, Elise and {Horch}, Elliott P. and {Howell}, Steve B. and {Ciardi}, David R. and {Gonzales}, Erica and {Crepp}, Justin R.},
        title = "{A Comparison of Spectroscopic versus Imaging Techniques for Detecting Close Companions to Kepler Objects of Interest}",
      journal = {\aj},
         year = 2015,
        month = nov,
       volume = {150},
       number = {5},
          eid = {144},
        pages = {144},
          doi = {10.1088/0004-6256/150/5/144},
archivePrefix = {arXiv},
       eprint = {1508.06502},
 primaryClass = {astro-ph.SR},
       adsurl = {https://ui.adsabs.harvard.edu/abs/2015AJ....150..144T}
}

@ARTICLE{2023gaiadr3,
       author = {{Gaia Collaboration} and {Vallenari}, A. and {Brown}, A.~G.~A. and {Prusti}, T. and {de Bruijne}, J.~H.~J. and {Arenou}, F. and {Babusiaux}, C. and {Biermann}, M. and {Creevey}, O.~L. and {Ducourant}, C. and {Evans}, D.~W. and {Eyer}, L. and {Guerra}, R. and {Hutton}, A. and {Jordi}, C. and {Klioner}, S.~A. and {Lammers}, U.~L. and {Lindegren}, L. and {Luri}, X. and {Mignard}, F. and {Panem}, C. and {Pourbaix}, D. and {Randich}, S. and {Sartoretti}, P. and {Soubiran}, C. and {Tanga}, P. and {Walton}, N.~A. and {Bailer-Jones}, C.~A.~L. and {Bastian}, U. and {Drimmel}, R. and {Jansen}, F. and {Katz}, D. and {Lattanzi}, M.~G. and {van Leeuwen}, F. and {Bakker}, J. and {Cacciari}, C. and {Casta{\~n}eda}, J. and {De Angeli}, F. and {Fabricius}, C. and {Fouesneau}, M. and {Fr{\'e}mat}, Y. and {Galluccio}, L. and {Guerrier}, A. and {Heiter}, U. and {Masana}, E. and {Messineo}, R. and {Mowlavi}, N. and {Nicolas}, C. and {Nienartowicz}, K. and {Pailler}, F. and {Panuzzo}, P. and {Riclet}, F. and {Roux}, W. and {Seabroke}, G.~M. and {Sordo}, R. and {Th{\'e}venin}, F. and {Gracia-Abril}, G. and {Portell}, J. and {Teyssier}, D. and {Altmann}, M. and {Andrae}, R. and {Audard}, M. and {Bellas-Velidis}, I. and {Benson}, K. and {Berthier}, J. and {Blomme}, R. and {Burgess}, P.~W. and {Busonero}, D. and {Busso}, G. and {C{\'a}novas}, H. and {Carry}, B. and {Cellino}, A. and {Cheek}, N. and {Clementini}, G. and {Damerdji}, Y. and {Davidson}, M. and {de Teodoro}, P. and {Nu{\~n}ez Campos}, M. and {Delchambre}, L. and {Dell'Oro}, A. and {Esquej}, P. and {Fern{\'a}ndez-Hern{\'a}ndez}, J. and {Fraile}, E. and {Garabato}, D. and {Garc{\'\i}a-Lario}, P. and {Gosset}, E. and {Haigron}, R. and {Halbwachs}, J. -L. and {Hambly}, N.~C. and {Harrison}, D.~L. and {Hern{\'a}ndez}, J. and {Hestroffer}, D. and {Hodgkin}, S.~T. and {Holl}, B. and {Jan{\ss}en}, K. and {Jevardat de Fombelle}, G. and {Jordan}, S. and {Krone-Martins}, A. and {Lanzafame}, A.~C. and {L{\"o}ffler}, W. and {Marchal}, O. and {Marrese}, P.~M. and {Moitinho}, A. and {Muinonen}, K. and {Osborne}, P. and {Pancino}, E. and {Pauwels}, T. and {Recio-Blanco}, A. and {Reyl{\'e}}, C. and {Riello}, M. and {Rimoldini}, L. and {Roegiers}, T. and {Rybizki}, J. and {Sarro}, L.~M. and {Siopis}, C. and {Smith}, M. and {Sozzetti}, A. and {Utrilla}, E. and {van Leeuwen}, M. and {Abbas}, U. and {{\'A}brah{\'a}m}, P. and {Abreu Aramburu}, A. and {Aerts}, C. and {Aguado}, J.~J. and {Ajaj}, M. and {Aldea-Montero}, F. and {Altavilla}, G. and {{\'A}lvarez}, M.~A. and {Alves}, J. and {Anders}, F. and {Anderson}, R.~I. and {Anglada Varela}, E. and {Antoja}, T. and {Baines}, D. and {Baker}, S.~G. and {Balaguer-N{\'u}{\~n}ez}, L. and {Balbinot}, E. and {Balog}, Z. and {Barache}, C. and {Barbato}, D. and {Barros}, M. and {Barstow}, M.~A. and {Bartolom{\'e}}, S. and {Bassilana}, J. -L. and {Bauchet}, N. and {Becciani}, U. and {Bellazzini}, M. and {Berihuete}, A. and {Bernet}, M. and {Bertone}, S. and {Bianchi}, L. and {Binnenfeld}, A. and {Blanco-Cuaresma}, S. and {Blazere}, A. and {Boch}, T. and {Bombrun}, A. and {Bossini}, D. and {Bouquillon}, S. and {Bragaglia}, A. and {Bramante}, L. and {Breedt}, E. and {Bressan}, A. and {Brouillet}, N. and {Brugaletta}, E. and {Bucciarelli}, B. and {Burlacu}, A. and {Butkevich}, A.~G. and {Buzzi}, R. and {Caffau}, E. and {Cancelliere}, R. and {Cantat-Gaudin}, T. and {Carballo}, R. and {Carlucci}, T. and {Carnerero}, M.~I. and {Carrasco}, J.~M. and {Casamiquela}, L. and {Castellani}, M. and {Castro-Ginard}, A. and {Chaoul}, L. and {Charlot}, P. and {Chemin}, L. and {Chiaramida}, V. and {Chiavassa}, A. and {Chornay}, N. and {Comoretto}, G. and {Contursi}, G. and {Cooper}, W.~J. and {Cornez}, T. and {Cowell}, S. and {Crifo}, F. and {Cropper}, M. and {Crosta}, M. and {Crowley}, C. and {Dafonte}, C. and {Dapergolas}, A. and {David}, M. and {David}, P. and {de Laverny}, P. and {De Luise}, F. and {De March}, R.},
        title = "{Gaia Data Release 3. Summary of the content and survey properties}",
      journal = {\aap},
         year = 2023,
        month = jun,
       volume = {674},
          eid = {A1},
        pages = {A1},
          doi = {10.1051/0004-6361/202243940},
archivePrefix = {arXiv},
       eprint = {2208.00211},
 primaryClass = {astro-ph.GA},
       adsurl = {https://ui.adsabs.harvard.edu/abs/2023A&A...674A...1G}
}

@ARTICLE{2016gaiamission,
       author = {{Gaia Collaboration} and {Prusti}, T. and {de Bruijne}, J.~H.~J. and {Brown}, A.~G.~A. and {Vallenari}, A. and {Babusiaux}, C. and {Bailer-Jones}, C.~A.~L. and {Bastian}, U. and {Biermann}, M. and {Evans}, D.~W. and {Eyer}, L. and {Jansen}, F. and {Jordi}, C. and {Klioner}, S.~A. and {Lammers}, U. and {Lindegren}, L. and {Luri}, X. and {Mignard}, F. and {Milligan}, D.~J. and {Panem}, C. and {Poinsignon}, V. and {Pourbaix}, D. and {Randich}, S. and {Sarri}, G. and {Sartoretti}, P. and {Siddiqui}, H.~I. and {Soubiran}, C. and {Valette}, V. and {van Leeuwen}, F. and {Walton}, N.~A. and {Aerts}, C. and {Arenou}, F. and {Cropper}, M. and {Drimmel}, R. and {H{\o}g}, E. and {Katz}, D. and {Lattanzi}, M.~G. and {O'Mullane}, W. and {Grebel}, E.~K. and {Holland}, A.~D. and {Huc}, C. and {Passot}, X. and {Bramante}, L. and {Cacciari}, C. and {Casta{\~n}eda}, J. and {Chaoul}, L. and {Cheek}, N. and {De Angeli}, F. and {Fabricius}, C. and {Guerra}, R. and {Hern{\'a}ndez}, J. and {Jean-Antoine-Piccolo}, A. and {Masana}, E. and {Messineo}, R. and {Mowlavi}, N. and {Nienartowicz}, K. and {Ord{\'o}{\~n}ez-Blanco}, D. and {Panuzzo}, P. and {Portell}, J. and {Richards}, P.~J. and {Riello}, M. and {Seabroke}, G.~M. and {Tanga}, P. and {Th{\'e}venin}, F. and {Torra}, J. and {Els}, S.~G. and {Gracia-Abril}, G. and {Comoretto}, G. and {Garcia-Reinaldos}, M. and {Lock}, T. and {Mercier}, E. and {Altmann}, M. and {Andrae}, R. and {Astraatmadja}, T.~L. and {Bellas-Velidis}, I. and {Benson}, K. and {Berthier}, J. and {Blomme}, R. and {Busso}, G. and {Carry}, B. and {Cellino}, A. and {Clementini}, G. and {Cowell}, S. and {Creevey}, O. and {Cuypers}, J. and {Davidson}, M. and {De Ridder}, J. and {de Torres}, A. and {Delchambre}, L. and {Dell'Oro}, A. and {Ducourant}, C. and {Fr{\'e}mat}, Y. and {Garc{\'\i}a-Torres}, M. and {Gosset}, E. and {Halbwachs}, J. -L. and {Hambly}, N.~C. and {Harrison}, D.~L. and {Hauser}, M. and {Hestroffer}, D. and {Hodgkin}, S.~T. and {Huckle}, H.~E. and {Hutton}, A. and {Jasniewicz}, G. and {Jordan}, S. and {Kontizas}, M. and {Korn}, A.~J. and {Lanzafame}, A.~C. and {Manteiga}, M. and {Moitinho}, A. and {Muinonen}, K. and {Osinde}, J. and {Pancino}, E. and {Pauwels}, T. and {Petit}, J. -M. and {Recio-Blanco}, A. and {Robin}, A.~C. and {Sarro}, L.~M. and {Siopis}, C. and {Smith}, M. and {Smith}, K.~W. and {Sozzetti}, A. and {Thuillot}, W. and {van Reeven}, W. and {Viala}, Y. and {Abbas}, U. and {Abreu Aramburu}, A. and {Accart}, S. and {Aguado}, J.~J. and {Allan}, P.~M. and {Allasia}, W. and {Altavilla}, G. and {{\'A}lvarez}, M.~A. and {Alves}, J. and {Anderson}, R.~I. and {Andrei}, A.~H. and {Anglada Varela}, E. and {Antiche}, E. and {Antoja}, T. and {Ant{\'o}n}, S. and {Arcay}, B. and {Atzei}, A. and {Ayache}, L. and {Bach}, N. and {Baker}, S.~G. and {Balaguer-N{\'u}{\~n}ez}, L. and {Barache}, C. and {Barata}, C. and {Barbier}, A. and {Barblan}, F. and {Baroni}, M. and {Barrado y Navascu{\'e}s}, D. and {Barros}, M. and {Barstow}, M.~A. and {Becciani}, U. and {Bellazzini}, M. and {Bellei}, G. and {Bello Garc{\'\i}a}, A. and {Belokurov}, V. and {Bendjoya}, P. and {Berihuete}, A. and {Bianchi}, L. and {Bienaym{\'e}}, O. and {Billebaud}, F. and {Blagorodnova}, N. and {Blanco-Cuaresma}, S. and {Boch}, T. and {Bombrun}, A. and {Borrachero}, R. and {Bouquillon}, S. and {Bourda}, G. and {Bouy}, H. and {Bragaglia}, A. and {Breddels}, M.~A. and {Brouillet}, N. and {Br{\"u}semeister}, T. and {Bucciarelli}, B. and {Budnik}, F. and {Burgess}, P. and {Burgon}, R. and {Burlacu}, A. and {Busonero}, D. and {Buzzi}, R. and {Caffau}, E. and {Cambras}, J. and {Campbell}, H. and {Cancelliere}, R. and {Cantat-Gaudin}, T. and {Carlucci}, T. and {Carrasco}, J.~M. and {Castellani}, M. and {Charlot}, P. and {Charnas}, J. and {Charvet}, P. and {Chassat}, F. and {Chiavassa}, A. and {Clotet}, M. and {Cocozza}, G. and {Collins}, R.~S. and {Collins}, P. and {Costigan}, G.},
        title = "{The Gaia mission}",
      journal = {\aap},
         year = 2016,
        month = nov,
       volume = {595},
          eid = {A1},
        pages = {A1},
          doi = {10.1051/0004-6361/201629272},
archivePrefix = {arXiv},
       eprint = {1609.04153},
 primaryClass = {astro-ph.IM},
       adsurl = {https://ui.adsabs.harvard.edu/abs/2016A&A...595A...1G}
}

@ARTICLE{2023halbwachs,
       author = {{Halbwachs}, Jean-Louis and {Pourbaix}, Dimitri and {Arenou}, Fr{\'e}d{\'e}ric and {Galluccio}, Laurent and {Guillout}, Patrick and {Bauchet}, Nathalie and {Marchal}, Olivier and {Sadowski}, Gilles and {Teyssier}, David},
        title = "{Gaia Data Release 3. Astrometric binary star processing}",
      journal = {\aap},
         year = 2023,
        month = jun,
       volume = {674},
          eid = {A9},
        pages = {A9},
          doi = {10.1051/0004-6361/202243969},
archivePrefix = {arXiv},
       eprint = {2206.05726},
 primaryClass = {astro-ph.SR},
       adsurl = {https://ui.adsabs.harvard.edu/abs/2023A&A...674A...9H}
}

@ARTICLE{2021gcns,
       author = {{Gaia Collaboration} and {Smart}, R.~L. and {Sarro}, L.~M. and {Rybizki}, J. and {Reyl{\'e}}, C. and {Robin}, A.~C. and {Hambly}, N.~C. and {Abbas}, U. and {Barstow}, M.~A. and {de Bruijne}, J.~H.~J. and {Bucciarelli}, B. and {Carrasco}, J.~M. and {Cooper}, W.~J. and {Hodgkin}, S.~T. and {Masana}, E. and {Michalik}, D. and {Sahlmann}, J. and {Sozzetti}, A. and {Brown}, A.~G.~A. and {Vallenari}, A. and {Prusti}, T. and {Babusiaux}, C. and {Biermann}, M. and {Creevey}, O.~L. and {Evans}, D.~W. and {Eyer}, L. and {Hutton}, A. and {Jansen}, F. and {Jordi}, C. and {Klioner}, S.~A. and {Lammers}, U. and {Lindegren}, L. and {Luri}, X. and {Mignard}, F. and {Panem}, C. and {Pourbaix}, D. and {Randich}, S. and {Sartoretti}, P. and {Soubiran}, C. and {Walton}, N.~A. and {Arenou}, F. and {Bailer-Jones}, C.~A.~L. and {Bastian}, U. and {Cropper}, M. and {Drimmel}, R. and {Katz}, D. and {Lattanzi}, M.~G. and {van Leeuwen}, F. and {Bakker}, J. and {Casta{\~n}eda}, J. and {De Angeli}, F. and {Ducourant}, C. and {Fabricius}, C. and {Fouesneau}, M. and {Fr{\'e}mat}, Y. and {Guerra}, R. and {Guerrier}, A. and {Guiraud}, J. and {Jean-Antoine Piccolo}, A. and {Messineo}, R. and {Mowlavi}, N. and {Nicolas}, C. and {Nienartowicz}, K. and {Pailler}, F. and {Panuzzo}, P. and {Riclet}, F. and {Roux}, W. and {Seabroke}, G.~M. and {Sordo}, R. and {Tanga}, P. and {Th{\'e}venin}, F. and {Gracia-Abril}, G. and {Portell}, J. and {Teyssier}, D. and {Altmann}, M. and {Andrae}, R. and {Bellas-Velidis}, I. and {Benson}, K. and {Berthier}, J. and {Blomme}, R. and {Brugaletta}, E. and {Burgess}, P.~W. and {Busso}, G. and {Carry}, B. and {Cellino}, A. and {Cheek}, N. and {Clementini}, G. and {Damerdji}, Y. and {Davidson}, M. and {Delchambre}, L. and {Dell'Oro}, A. and {Fern{\'a}ndez-Hern{\'a}ndez}, J. and {Galluccio}, L. and {Garc{\'\i}a-Lario}, P. and {Garcia-Reinaldos}, M. and {Gonz{\'a}lez-N{\'u}{\~n}ez}, J. and {Gosset}, E. and {Haigron}, R. and {Halbwachs}, J. -L. and {Harrison}, D.~L. and {Hatzidimitriou}, D. and {Heiter}, U. and {Hern{\'a}ndez}, J. and {Hestroffer}, D. and {Holl}, B. and {Jan{\ss}en}, K. and {Jevardat de Fombelle}, G. and {Jordan}, S. and {Krone-Martins}, A. and {Lanzafame}, A.~C. and {L{\"o}ffler}, W. and {Lorca}, A. and {Manteiga}, M. and {Marchal}, O. and {Marrese}, P.~M. and {Moitinho}, A. and {Mora}, A. and {Muinonen}, K. and {Osborne}, P. and {Pancino}, E. and {Pauwels}, T. and {Recio-Blanco}, A. and {Richards}, P.~J. and {Riello}, M. and {Rimoldini}, L. and {Roegiers}, T. and {Siopis}, C. and {Smith}, M. and {Ulla}, A. and {Utrilla}, E. and {van Leeuwen}, M. and {van Reeven}, W. and {Abreu Aramburu}, A. and {Accart}, S. and {Aerts}, C. and {Aguado}, J.~J. and {Ajaj}, M. and {Altavilla}, G. and {{\'A}lvarez}, M.~A. and {{\'A}lvarez Cid-Fuentes}, J. and {Alves}, J. and {Anderson}, R.~I. and {Anglada Varela}, E. and {Antoja}, T. and {Audard}, M. and {Baines}, D. and {Baker}, S.~G. and {Balaguer-N{\'u}{\~n}ez}, L. and {Balbinot}, E. and {Balog}, Z. and {Barache}, C. and {Barbato}, D. and {Barros}, M. and {Bartolom{\'e}}, S. and {Bassilana}, J. -L. and {Bauchet}, N. and {Baudesson-Stella}, A. and {Becciani}, U. and {Bellazzini}, M. and {Bernet}, M. and {Bertone}, S. and {Bianchi}, L. and {Blanco-Cuaresma}, S. and {Boch}, T. and {Bombrun}, A. and {Bossini}, D. and {Bouquillon}, S. and {Bragaglia}, A. and {Bramante}, L. and {Breedt}, E. and {Bressan}, A. and {Brouillet}, N. and {Burlacu}, A. and {Busonero}, D. and {Butkevich}, A.~G. and {Buzzi}, R. and {Caffau}, E. and {Cancelliere}, R. and {C{\'a}novas}, H. and {Cantat-Gaudin}, T. and {Carballo}, R. and {Carlucci}, T. and {Carnerero}, M.~I. and {Casamiquela}, L. and {Castellani}, M. and {Castro-Ginard}, A. and {Castro Sampol}, P. and {Chaoul}, L. and {Charlot}, P. and {Chemin}, L. and {Chiavassa}, A. and {Cioni}, M. -R.~L. and {Comoretto}, G. and {Cornez}, T. and {Cowell}, S. and {Crifo}, F. and {Crosta}, M. and {Crowley}, C. and {Dafonte}, C. and {Dapergolas}, A.},
        title = "{Gaia Early Data Release 3. The Gaia Catalogue of Nearby Stars}",
      journal = {\aap},
         year = 2021,
        month = may,
       volume = {649},
          eid = {A6},
        pages = {A6},
          doi = {10.1051/0004-6361/202039498},
archivePrefix = {arXiv},
       eprint = {2012.02061},
 primaryClass = {astro-ph.SR},
       adsurl = {https://ui.adsabs.harvard.edu/abs/2021A&A...649A...6G}
}

@book{2020decadal,
  title     = "Pathways to Discovery in Astronomy and Astrophysics for the 2020s",
  author    = "National Academies of Sciences Engineering and Medicine",
  year      = 2021,
  publisher = "National Academies Press",
  address   = "Washington D.C."
}

@ARTICLE{2011Howell,
       author = {{Howell}, Steve B. and {Everett}, Mark E. and {Sherry}, William and {Horch}, Elliott and {Ciardi}, David R.},
        title = "{Speckle Camera Observations for the NASA Kepler Mission Follow-up Program}",
      journal = {\aj},
         year = 2011,
        month = jul,
       volume = {142},
       number = {1},
          eid = {19},
        pages = {19},
          doi = {10.1088/0004-6256/142/1/19},
       adsurl = {https://ui.adsabs.harvard.edu/abs/2011AJ....142...19H}
}

@ARTICLE{2011Horcha,
       author = {{Horch}, Elliott P. and {Gomez}, Shamilia C. and {Sherry}, William H. and {Howell}, Steve B. and {Ciardi}, David R. and {Anderson}, Lisa M. and {van Altena}, William F.},
        title = "{Observations of Binary Stars with the Differential Speckle Survey Instrument. II. Hipparcos Stars Observed in 2010 January and June}",
      journal = {\aj},
         year = 2011,
        month = feb,
       volume = {141},
       number = {2},
          eid = {45},
        pages = {45},
          doi = {10.1088/0004-6256/141/2/45},
       adsurl = {https://ui.adsabs.harvard.edu/abs/2011AJ....141...45H}
}

@ARTICLE{2011Horchb,
       author = {{Horch}, Elliott P. and {van Altena}, William F. and {Howell}, Steve B. and {Sherry}, William H. and {Ciardi}, David R.},
        title = "{Observations of Binary Stars with the Differential Speckle Survey Instrument. III. Measures below the Diffraction Limit of the WIYN Telescope}",
      journal = {\aj},
         year = 2011,
        month = jun,
       volume = {141},
       number = {6},
          eid = {180},
        pages = {180},
          doi = {10.1088/0004-6256/141/6/180},
       adsurl = {https://ui.adsabs.harvard.edu/abs/2011AJ....141..180H}
}

@ARTICLE{2009Horch,
       author = {{Horch}, Elliott P. and {Veillette}, Daniel R. and {Baena Gall{\'e}}, Roberto and {Shah}, Sagar C. and {O'Rielly}, Grant V. and {van Altena}, William F.},
        title = "{Observations of Binary Stars with the Differential Speckle Survey Instrument. I. Instrument Description and First Results}",
      journal = {\aj},
         year = 2009,
        month = jun,
       volume = {137},
       number = {6},
        pages = {5057-5067},
          doi = {10.1088/0004-6256/137/6/5057},
       adsurl = {https://ui.adsabs.harvard.edu/abs/2009AJ....137.5057H}
}

@ARTICLE{2004A&A...424..727P,
       author = {{Pourbaix}, D. and {Tokovinin}, A.~A. and {Batten}, A.~H. and {Fekel}, F.~C. and {Hartkopf}, W.~I. and {Levato}, H. and {Morrell}, N.~I. and {Torres}, G. and {Udry}, S.},
        title = "{S$_{B$^{9}$}$: The ninth catalogue of spectroscopic binary orbits}",
      journal = {\aap},
         year = 2004,
        month = sep,
       volume = {424},
        pages = {727-732},
          doi = {10.1051/0004-6361:20041213},
archivePrefix = {arXiv},
       eprint = {astro-ph/0406573},
 primaryClass = {astro-ph},
       adsurl = {https://ui.adsabs.harvard.edu/abs/2004A&A...424..727P}
}

@ARTICLE{2024mamajek,
       author = {{Mamajek}, Eric and {Stapelfeldt}, Karl},
        title = "{NASA Exoplanet Exploration Program (ExEP) Mission Star List for the Habitable Worlds Observatory (2023)}",
      journal = {arXiv e-prints},
         year = 2024,
        month = feb,
          eid = {arXiv:2402.12414},
        pages = {arXiv:2402.12414},
          doi = {10.48550/arXiv.2402.12414},
archivePrefix = {arXiv},
       eprint = {2402.12414},
 primaryClass = {astro-ph.IM},
       adsurl = {https://ui.adsabs.harvard.edu/abs/2024arXiv240212414M}
}

@ARTICLE{2001mason,
       author = {{Mason}, Brian D. and {Wycoff}, Gary L. and {Hartkopf}, William I. and {Douglass}, Geoffrey G. and {Worley}, Charles E.},
        title = "{The 2001 US Naval Observatory Double Star CD-ROM. I. The Washington Double Star Catalog}",
      journal = {\aj},
         year = 2001,
        month = dec,
       volume = {122},
       number = {6},
        pages = {3466-3471},
          doi = {10.1086/323920},
       adsurl = {https://ui.adsabs.harvard.edu/abs/2001AJ....122.3466M}
}
\bibliographystyle{aasjournal}



\startlongtable
\begin{deluxetable}{ccccccccccccccc}
\tablewidth{700pt}
\tabletypesize{\scriptsize}
\tablehead{
\colhead{TIC ID} & \colhead{Gaia DR3 ID} & \colhead{Simbad Name} & \colhead{R.A.} & \colhead{decl.} & \colhead{$PM_{R.A.}$} & \colhead{$PM_{decl.}$} & \colhead{$\pi$}  & \colhead{$G$} & \colhead{$G-G_{RP}$}  \\
\colhead{-} & \colhead{-} & \colhead{-} & \colhead{[\degree]} & \colhead{[\degree]} & \colhead{[mas/yr]} & \colhead{[mas/yr]} & \colhead{[mas]}  & \colhead{[mag]} & \colhead{[mag]}}
\tablecaption{\label{tab:sample}Potential HWO Target Stars Observed}
\startdata
3962869 & 2473608009504466688 & $\phi^2$ Cet & 12.53059 & -10.64535 & -228.03115 & -229.5775 & 0.0628 & 5.02954 & 0.41624 \\
8915802 & 798068905726303232 & 11 LMi & 143.91061 & 35.80898 & -726.51388 & -259.05745 & 0.08901 & 5.20043 & 0.54685 \\
11310083 & 5663558453071724288 & HD 84117 & 145.55813 & -23.9144 & -399.49675 & 262.31931 & 0.06688 & 4.77215 & 0.419 \\
21535479 & 777254360337133312 & 47 UMa & 164.8647 & 40.4305 & -316.8499 & 55.1804 & 0.07201 & 4.86659 & 0.47652 \\
23969522 & 821455861647126656 & 15 LMi & 147.1488 & 46.0206 & 221.75005 & -92.35563 & 0.05313 & 4.91655 & 0.46953 \\
27136704 & 3407611182744289152 &  $\mu$ Tau  & 76.86503 & 18.64513 & 534.67392 & 18.44878 & 0.06283 & 4.73246 & 0.49986 \\
38511251 & 5164120762333028736 &  $\delta$ Eri  & 55.81166 & -9.76008 & -93.63406 & 744.36039 & 0.11003 & 3.27557 & 0.5888 \\
46125330 & 3374052988354372992 & 71 Ori & 93.71153 & 19.15564 & -97.60836 & -182.44758 & 0.04592 & 5.07295 & 0.35995 \\
47346402 & 3400292798990117888 & 111 Tau & 81.10726 & 17.3835 & 250.4829 & -7.15628 & 0.06859 & 4.85019 & 0.42812 \\
60716322 & 2719475542667772416 &  $\xi$ Peg  & 341.67431 & 12.17069 & 233.35408 & -493.13255 & 0.06092 & 4.06936 & 0.4435 \\
67772871 & 3195919528989223040 & $o^2$ Eri & 63.80795 & -7.66808 & -2240.08498 & -3421.80862 & 0.19961 & 4.17989 & 0.61014 \\
72748794 & 5117974602912370432 & HD 14412 & 34.74269 & -25.94371 & -217.73411 & 444.60889 & 0.07791 & 6.14786 & 0.5497 \\
80226651 & 3381727613874753536 & 37 Gem & 103.82759 & 25.3758 & -37.59497 & 24.23372 & 0.05746 & 5.60278 & 0.47347 \\
95431211 & 624684408880100992 & 40 Leo & 154.93294 & 19.46996 & -230.58856 & -214.06859 & 0.04713 & 4.65715 & 0.37559 \\
101641846 & 4025850731201819392 & 61 UMa & 175.2625 & 34.19994 & -12.52161 & -381.19587 & 0.10443 & 5.111 & 0.52984 \\
113710966 & 96331172942614528 & 107 Psc & 25.62259 & 20.26552 & -300.73692 & -673.5389 & 0.13082 & 4.99968 & 0.58235 \\
116988032 & 436648129327098496 &  $\iota$ Per  & 47.2754 & 49.61287 & 1262.19279 & -91.08282 & 0.09454 & 3.90264 & 0.47579 \\
117979951 & 2980492387558410112 & 58 Eri & 71.90182 & -16.9337 & 130.26356 & 169.33843 & 0.07553 & 5.31838 & 0.4909 \\
118572803 & 5164707970261890560 &  $\epsilon$ Eri  & 53.22829 & -9.45817 & -974.75815 & 20.87584 & 0.31058 & 3.46575 & 0.6236 \\
135656809 & 4345775217221821312 & 18 Sco & 243.90634 & -8.37164 & 232.23032 & -495.3782 & 0.07074 & 5.32765 & 0.4898 \\
141810080 & 5264749303462961280 &  $\alpha$ Men  & 92.56236 & -74.75399 & 121.59582 & -212.41088 & 0.09792 & 4.90014 & 0.53205 \\
147407292 & 6564091190988411520 & HD 207129 & 327.06671 & -47.30493 & 165.06854 & -295.55273 & 0.06427 & 5.41939 & 0.47192 \\
155315739 & 6553614253923452800 & HD 217987 & 346.50392 & -35.84716 & 6765.99514 & 1330.28527 & 0.30414 & 6.52203 & 1.03053 \\
156890613 & 974887555341533440 & HD 55575 & 108.95911 & 47.23914 & 30.07864 & -185.90471 & 0.05934 & 5.38816 & 0.47489 \\
157364190 & 1384195086792268800 &  $\chi$ Her  & 238.17156 & 42.45432 & 438.83869 & 629.54062 & 0.0629 & 4.42426 & 0.46506 \\
157966796 & 1284257623084916480 &  $\sigma$ Boo  & 218.67104 & 29.74571 & 188.57436 & 131.57967 & 0.06347 & 4.34196 & 0.33059 \\
165602000 & 1872046609345556480 & 61 Cyg A & 316.74848 & 38.76386 & 4164.20869 & 3249.61388 & 0.28599 & 4.76671 & 0.78951 \\
165602023 & 1872046574983497216 & 61 Cyg B & 316.75366 & 38.75607 & 4105.97643 & 3155.94164 & 0.28601 & 5.45064 & 0.89373 \\
166646191 & 762815470562110464 & HD 95735 & 165.83096 & 35.94865 & -580.05709 & -4776.58872 & 0.39275 & 6.55117 & 1.07566 \\
172954294 & 746545172372256384 & 20 LMi & 150.24997 & 31.92176 & -528.87061 & -429.37566 & 0.067 & 5.20496 & 0.50104 \\
176521059 & 3023711269067191296 & HD 38858 & 87.14586 & -4.09566 & 61.427 & -229.2912 & 0.06574 & 5.80588 & 0.49249 \\
179348425 & 6009538585839374336 & HD 140901 & 236.86891 & -37.91726 & -415.50658 & -213.99094 & 0.06559 & 5.83384 & 0.52322 \\
189576919 & 348020482735930112 &  $\upsilon$ And  & 24.19832 & 41.40376 & -171.89184 & -381.81516 & 0.07419 & 3.96613 & 0.45044 \\
202380743 & 1661568850770761600 & HD 122064 & 209.38328 & 61.49382 & -31.69308 & 216.28964 & 0.09933 & 6.16992 & 0.68668 \\
206686962 & 6604147121141267712 &  TW PsA  & 344.10194 & -31.56627 & 330.20338 & -158.60233 & 0.13155 & 6.09186 & 0.74553 \\
213041474 & 3211461469444773376 & HD 32147 & 75.20662 & -5.7586 & 549.30869 & -1108.24503 & 0.11307 & 5.88649 & 0.6996 \\
219709102 & 1068982214258617216 & $\sigma^2$ UMa & 137.59826 & 67.13363 & 13.1189 & -87.88146 & 0.04874 & 4.67962 & 0.39739 \\
229092427 & 4934923028038871296 &  $\nu$ Phe  & 18.80056 & -45.53087 & 665.08565 & 178.0698 & 0.06553 & 4.82792 & 0.44801 \\
229902025 & 1264630412816366720 &  c Boo  & 226.82618 & 24.86846 & 184.71405 & -164.27907 & 0.05119 & 4.8023 & 0.36277 \\
231698181 & 6412595290592307840 &  $\epsilon$ Ind  & 330.87241 & -56.79725 & 3966.66054 & -2536.19199 & 0.27484 & 4.3229 & 0.71235 \\
238432056 & 2860924621205256704 & HD 166 & 1.6552 & 29.02071 & 380.15925 & -177.7296 & 0.07264 & 5.87833 & 0.54521 \\
257393898 & 2552925644460225152 & HD 4628 & 12.09911 & 5.27554 & 755.8935 & -1141.7193 & 0.13449 & 5.46769 & 0.63892 \\
259237827 & 2261614264931057664 &  $\sigma$ Dra  & 293.0976 & 69.65345 & 597.38424 & -1738.2861 & 0.17349 & 4.44904 & 0.58387 \\
259291108 & 6408551802222368384 & HD 212330 & 336.23595 & -57.79902 & 129.65788 & -352.4248 & 0.04916 & 5.15285 & 0.52138 \\
261136679 & 4623036865373793408 &  $\pi$ Men  & 84.29954 & -80.46446 & 310.90883 & 1049.06049 & 0.05468 & 5.51158 & 0.46077 \\
279649049 & 4722135642226902656 & $\zeta^2$ Ret & 49.56623 & -62.50348 & 1331.02709 & 647.7254 & 0.08306 & 5.07961 & 0.48554 \\
279649057 & 4722111590409480064 & $\zeta^1$ Ret & 49.45525 & -62.57243 & 1337.53038 & 649.83975 & 0.08302 & 5.34462 & 0.50483 \\
283722336 & 2009481748875806976 & HD 219134 & 348.33773 & 57.16967 & 2074.41351 & 294.45198 & 0.15286 & 5.2319 & 0.68166 \\
285544488 & 427322415301550208 & HD 5015 & 13.26686 & 61.12473 & -68.29788 & 169.43506 & 0.0529 & 4.63591 & 0.43715 \\
289673491 & 2416611663182821120 & 6 Cet & 2.81569 & -15.46918 & -82.82753 & -269.54932 & 0.05295 & 4.76362 & 0.42554 \\
301051051 & 4847957293278177024 &  e Eri  & 50.00034 & -43.06655 & 3035.01732 & 726.96448 & 0.16552 & 4.06392 & 0.55562 \\
302158903 & 438829629114680704 &  $\theta$ Per  & 41.05222 & 49.22805 & 334.70816 & -89.25073 & 0.08968 & 3.983 & 0.42831 \\
309599261 & 4034171629042489088 & HD 103095 & 178.26735 & 37.69283 & 4002.65464 & -5817.80019 & 0.10903 & 6.1985 & 0.59273 \\
311063391 & 263916708025623680 & HD 37394 & 85.33475 & 53.47873 & 2.80354 & -523.47749 & 0.0815 & 5.99095 & 0.58834 \\
311092847 & 3263836568394170880 & 10 Tau & 54.21723 & 0.39952 & -232.56337 & -481.47228 & 0.07184 & 4.14109 & 0.46525 \\
328324648 & 976893923544104576 & 22 Lyn & 112.48392 & 49.67209 & 111.04698 & -82.88428 & 0.04901 & 5.22393 & 0.39558 \\
332064670 & 704967037090946688 & $\rho^1$ Cnc & 133.14676 & 28.32978 & -485.68081 & -233.51701 & 0.07945 & 5.73268 & 0.58567 \\
343813545 & 3269362645115584640 & $\kappa^1$ Cet & 49.8416 & 3.37062 & 269.29794 & 93.95652 & 0.1078 & 4.65649 & 0.50198 \\
355127594 & 714116137767540096 & HD 78366 & 137.21177 & 33.88171 & -191.41313 & -114.6302 & 0.05277 & 5.80906 & 0.46222 \\
366661076 & 3796442680948579328 &  $\beta$ Vir  & 177.67712 & 1.76352 & 740.74634 & -270.92723 & 0.09089 & 3.46856 & 0.46228 \\
367631379 & 1146337324038514176 & HD 90089 & 157.76666 & 82.55874 & -86.13273 & 19.83218 & 0.04344 & 5.14406 & 0.34225 \\
371520835 & 823773494718931968 & HD 88230 & 152.83293 & 49.45199 & -1363.28719 & -505.77036 & 0.20531 & 5.96117 & 0.89508 \\
373694425 & 512167948043650816 & HD 10780 & 26.94268 & 63.85141 & 581.68397 & -246.46227 & 0.09959 & 5.4104 & 0.56842 \\
373765355 & 3936909723803146368 & HD 115404 & 199.21567 & 17.016 & 636.28474 & -264.67829 & 0.09102 & 6.29768 & 0.63509 \\
377415363 & 1193030490492925824 &  $\gamma$ Ser  & 239.1147 & 15.65591 & 311.18255 & -1282.76704 & 0.08956 & 3.69443 & 0.40156 \\
389853353 & 1521686782461765888 & 10 CVn & 191.24546 & 39.27953 & -359.69875 & 139.0162 & 0.05696 & 5.80303 & 0.46462 \\
399665349 & 3288921720025503360 & $\pi^3$ Ori & 72.46212 & 6.96133 & 463.9544 & 11.93701 & 0.12462 & 3.08792 & 0.37977 \\
409104974 & 188771135583573504 &  $\lambda$ Aur  & 79.78834 & 40.0961 & 520.56891 & -664.48787 & 0.0796 & 4.52588 & 0.46843 \\
416519065 & 853819947756949120 & 36 UMa & 157.65517 & 55.98039 & -177.0448 & -32.63371 & 0.07725 & 4.67389 & 0.42207 \\
417762326 & 1092545710514654464 & $\pi^1$ UMa & 129.79848 & 65.0213 & -27.27447 & 87.88083 & 0.06926 & 5.49898 & 0.48628 \\
419015728 & 2452378776434477184 &  $\tau$ Cet  & 26.00906 & -15.93368 & -1721.72779 & 854.96316 & 0.27381 & 3.3004 & 0.5521 \\
425935521 & 4900108950849461248 &  $\zeta$ Tuc  & 5.03561 & -64.86962 & 1706.74686 & 1164.95944 & 0.11618 & 4.07351 & 0.47331 \\
434210589 & 2802397960855105920 & 54 Psc & 9.83865 & 21.24883 & -461.94829 & -369.62427 & 0.09002 & 5.65432 & 0.59171 \\
437886584 & 3331979901042416512 &  k Ori  & 94.11129 & 12.27299 & 82.77515 & 186.48042 & 0.05105 & 4.91641 & 0.36162 \\
441709021 & 1604859511344724864 &  $\theta$ Boo  & 216.29746 & 51.84897 & -235.96992 & -399.6957 & 0.06907 & 3.91723 & 0.43786 \\
445070560 & 1460229442589223424 &  $\beta$ Com  & 197.96428 & 27.88211 & -800.7201 & 882.30132 & 0.10873 & 4.09328 & 0.46766 \\
445258206 & 425040000962559616 &  $\eta$ Cas  & 12.28523 & 57.81273 & 1078.60917 & -551.13342 & 0.16883 & 3.32007 & 0.47145 \\
458445966 & 1534011998572555776 &  $\beta$ CVn  & 188.43143 & 41.35878 & -704.70212 & 292.15466 & 0.11803 & 4.09467 & 0.49284 \\
1101124558 & 1237090738916392704 &  $\xi$ Boo  & 222.84805 & 19.10027 & 127.46788 & -40.56865 & 0.14807 & 4.48048 & 0.53605 \\
1101124559 & 1237090738916392832 &  $\xi$ Boo B  & 222.84664 & 19.1011 & 133.37605 & -182.05918 & 0.14818 & 6.43009 & 0.81497 \\
\enddata
\tablecomments{Coordinates are on ICRS frame and in epoch 2016.0}
\end{deluxetable}

\startlongtable
\begin{deluxetable}{cccccccc}
\tablewidth{700pt}
\tabletypesize{\tiny}
\tablehead{
\colhead{TIC ID} & \colhead{Instrument} & \colhead{P.I.} & \colhead{Date} & \multicolumn{2}{c}{562\,nm} & \multicolumn{2}{c}{832\,nm} \\
\colhead{} & \colhead{} & \colhead{} & \colhead{} &  \colhead{Contrast @ 0.1$\arcsec{}$} & \colhead{Contrast @ 0.5$\arcsec{}$} & \colhead{Contrast @ 0.1\arcsec{}} & \colhead{Contrast @ 0.5\arcsec{}} \\
\colhead{-} & \colhead{-} & \colhead{-} & \colhead{yyyy/mm/dd} & \colhead{[mag]} & \colhead{[mag]} & \colhead{[mag]} & \colhead{[mag]} }
\tablecaption{\label{tab:speckleresults}Speckle Imaging Results}
\startdata
3962869 &  `Alopeke  &  Burt  & 2020/08/10 & 5.280 & 6.755 & 4.951 & 7.131 \\
8915802 &  `Alopeke  &  Hartman  & 2025/02/18 & 5.034 & 7.588 & 5.142 & 7.136 \\
11310083 &  Zorro  &  Howell  & 2020/12/31 & 5.395 & 6.609 & 5.234 & 6.677 \\
21535479 &  `Alopeke  &  Hartman  & 2025/02/17 & 4.803 & 6.090 & 4.360 & 6.874 \\
23969522 &  `Alopeke  &  Hartman  & 2025/02/17 & 4.888 & 7.014 & 4.070 & 6.835 \\
27136704 &  `Alopeke  &  Hartman  & 2024/09/26 & 4.818 & 7.159 & 3.862 & 6.856 \\
''  &  ''  &  ''  & 2025/02/18 & 5.219 & 7.093 & 4.622 & 6.799 \\
38511251 &  `Alopeke  &  Burt  & 2020/12/06 & 4.729 & 5.619 & 4.949 & 6.883 \\
46125330 &  `Alopeke  &  Hartman  & 2025/02/16 & 4.886 & 7.231 & 4.830 & 7.092 \\
47346402 &  `Alopeke  &  Hartman  & 2025/02/18 & 5.059 & 7.019 & 4.121 & 7.025 \\
60716322 &  `Alopeke  &  Burt  & 2024/05/25 & 4.650 & 6.665 & 4.545 & 6.850 \\
67772871 &  `Alopeke  &  Burt  & 2020/12/06 & 5.034 & 6.480 & 4.516 & 6.757 \\
72748794 &  `Alopeke  &  Burt  & 2020/11/24 & 5.474 & 7.159 & 4.763 & 6.443 \\
80226651 &  `Alopeke  &  Hartman  & 2025/02/17 & 5.046 & 7.304 & 4.206 & 7.307 \\
95431211 &  `Alopeke  &  Hartman  & 2025/02/17 & 4.850 & 6.954 & 4.131 & 6.471 \\
101641846 &  `Alopeke  &  Hartman  & 2025/02/17 & 4.886 & 7.372 & 4.201 & 6.972 \\
113710966 &  `Alopeke  &  Burt  & 2020/12/04 & 5.150 & 6.920 & 4.488 & 6.537 \\
113710966 &  `Alopeke  &  Howell  & 2024/09/25 & 5.069 & 7.168 & 4.185 & 6.895 \\
116988032 &  `Alopeke  &  Burt  & 2020/12/02 & 5.083 & 7.007 & 4.750 & 6.988 \\
''  &  ''  &  Howell  & 2024/09/27 & 5.029 & 6.943 & 4.782 & 6.745 \\
117979951 &  Zorro  &  Burt  & 2020/11/27 & 4.319 & 6.697 & 5.088 & 6.681 \\
118572803 &  `Alopeke  &  Burt  & 2020/12/06 & 5.089 & 5.977 & 4.919 & 6.800 \\
135656809 &  Zorro  &  Howell  & 2021/06/27 & 4.659 & 6.291 & 4.737 & 6.972 \\
''  &  ''  &  ''  & 2021/07/23 &  --  &  --  & 4.856 & 6.613 \\
141810080 &  Zorro  &  Howell  & 2019/12/22 & 3.995 & 6.329 & 4.542 & 6.224 \\
''  &  ''  &  Burt  & 2020/11/25 & 3.219 & 3.456 & 4.569 & 6.417 \\
147407292 &  Zorro  &  Burt  & 2020/10/23 & 5.238 & 6.531 & 5.211 & 6.646 \\
155315739 &  Zorro  &  Hartman  & 2025/01/09 & 4.503 & 5.721 & 4.843 & 6.732 \\
156890613 &  `Alopeke  &  Hartman  & 2025/02/17 & 4.700 & 6.424 & 4.386 & 7.216 \\
157364190 &  NESSI  &  Howell  & 2023/07/12 & 2.761 & 6.221 & 1.906 & 6.080 \\
157966796 &  `Alopeke  &  Hartman  & 2025/02/17 & 4.541 & 7.490 & 3.978 & 7.128 \\
165602000 &  `Alopeke  &  Burt  & 2020/08/10 & 5.319 & 7.237 & 4.690 & 6.989 \\
''  &  ''  &  Howell  & 2024/08/13 & 5.789 & 7.329 & 4.800 & 7.560 \\
165602023 &  `Alopeke  &  Burt  & 2020/08/10 & 5.369 & 6.959 & 4.496 & 6.984 \\
''  &  ''  &  Howell  & 2024/08/13 & 5.899 & 7.167 & 4.884 & 7.222 \\
166646191 &  `Alopeke  &  Hartman  & 2025/02/17 & 4.806 & 6.731 & 4.101 & 6.529 \\
172954294 &  `Alopeke  &  Hartman  & 2025/02/14 & 5.588 & 7.234 & 5.225 & 7.044 \\
176521059 &  `Alopeke  &  Burt  & 2020/12/06 & 5.187 & 6.910 & 4.460 & 6.702 \\
179348425 &  Zorro  &  Howell  & 2021/07/22 &  --  &  --  & 4.444 & 6.707 \\
189576919 &  NESSI  &  Howell  & 2018/11/18 & 2.601 & 6.711 & 1.671 & 5.342 \\
202380743 &  `Alopeke  &  Hartman  & 2025/02/17 & 5.033 & 6.135 & 4.868 & 6.657 \\
206686962 &  Zorro  &  Burt  & 2020/11/24 & 5.308 & 6.948 & 4.912 & 6.617 \\
213041474 &  `Alopeke  &  Burt  & 2020/12/07 & 5.146 & 6.770 & 4.683 & 6.794 \\
219709102 &  NESSI  &  Howell  & 2019/01/25 & 2.247 & 5.223 & 1.910 & 5.816 \\
229092427 &  Zorro  &  Burt  & 2020/11/24 & 5.408 & 7.032 & 4.644 & 6.683 \\
229902025 &  `Alopeke  &  Hartman  & 2025/02/18 & 6.042 & 7.259 & 5.191 & 6.716 \\
231698181 &  Zorro  &  Burt  & 2020/10/23 & 5.600 & 6.545 & 5.278 & 6.725 \\
238432056 &  `Alopeke  &  Burt  & 2020/12/06 & 5.034 & 6.944 & 5.244 & 6.635 \\
257393898 &  `Alopeke  &  Burt  & 2020/12/24 & 5.194 & 6.743 & 4.293 & 6.270 \\
''  &  ''  &  Howell  & 2024/09/26 & 5.408 & 7.111 & 5.147 & 6.968 \\
259237827 &  `Alopeke  &  Howell  & 2024/08/13 & 4.780 & 5.416 & 4.828 & 7.089 \\
259291108 &  Zorro  &  Burt  & 2020/01/02 & 4.857 & 6.366 & 4.852 & 6.238 \\
261136679 &  Zorro  &  Burt  & 2020/11/26 & 4.415 & 4.600 & 5.142 & 6.568 \\
279649049 &  Zorro  &  Howell  & 2020/10/23 & 4.803 & 6.040 & 4.701 & 6.355 \\
''  &  ''  &  Burt  & 2020/11/26 & 4.993 & 6.177 & 4.633 & 6.604 \\
''  &  ''  &  Hartman  & 2025/01/08 & 4.803 & 6.167 & 4.420 & 6.876 \\
''  &  ''  &  ''  & 2025/01/09 & 5.149 & 6.559 & 5.671 & 6.948 \\
279649057 &  Zorro  &  Howell  & 2020/10/23 & 5.019 & 6.638 & 4.820 & 6.499 \\
''  &  ''  &  Burt  & 2020/11/26 & 5.117 & 6.098 & 4.744 & 6.293 \\
283722336 &  `Alopeke  &  Burt  & 2020/08/04 & 4.583 & 5.917 & 4.089 & 6.025 \\
''  &  ''  &  Howell  & 2024/09/27 & 4.813 & 6.730 & 4.863 & 7.632 \\
285544488 &  `Alopeke  &  Howell  & 2023/12/04 & 4.943 & 6.441 & 4.821 & 6.546 \\
''  &  ''  &  Hartman  & 2024/09/25 & 4.546 & 5.836 & 4.611 & 6.943 \\
289673491 &  Zorro  &  Burt  & 2020/11/24 & 4.844 & 6.775 & 4.735 & 6.640 \\
301051051 &  Zorro  &  Burt  & 2020/11/25 & 4.996 & 6.727 & 4.666 & 6.460 \\
302158903 &  `Alopeke  &  Howell  & 2024/09/27 & 4.734 & 6.706 & 4.619 & 6.715 \\
309599261 &  `Alopeke  &  Hartman  & 2025/02/15 & 5.791 & 7.178 & 5.171 & 6.811 \\
311063391 &  `Alopeke  &  Hartman  & 2025/02/16 & 5.141 & 6.879 & 4.695 & 7.280 \\
311092847 &  `Alopeke  &  Burt  & 2020/12/07 & 4.666 & 7.090 & 4.452 & 6.963 \\
328324648 &  `Alopeke  &  Hartman  & 2025/02/17 & 4.883 & 6.435 & 4.244 & 7.260 \\
332064670 &  NESSI  &  Howell  & 2018/11/18 & 2.166 & 4.556 & 1.991 & 5.388 \\
343813545 &  NESSI  &  Howell  & 2019/01/25 & 2.490 & 5.283 & 2.009 & 6.215 \\
''  &  `Alopeke  &  Burt  & 2020/12/05 & 5.068 & 6.987 & 4.315 & 6.183 \\
''  &  ''  &  Howell  & 2023/12/02 & 5.583 & 6.954 & 4.455 & 6.498 \\
''  &  ''  &  ''  & 2024/09/26 & 5.431 & 7.069 & 4.778 & 7.217 \\
355127594 &  `Alopeke  &  Hartman  & 2025/02/18 & 5.466 & 7.124 & 4.76 & 6.867 \\
366661076 &  Zorro  &  Howell  & 2023/03/05 & 4.911 & 6.999 & 5.185 & 7.660 \\
367631379 &  `Alopeke  &  Hartman  & 2025/02/17 & 4.914 & 5.845 & 4.814 & 6.578 \\ 
371520835 &  `Alopeke  &  Howell  & 2024/03/21 & 4.707 & 5.855 & 4.494 & 7.085 \\
''  &  ''  &  ''  & 2024/11/20 & 5.350 & 6.659 & 4.589 & 6.845 \\
''  &  ''  &  Hartman  & 2025/02/15 & 5.064 & 5.580 & 5.398 & 6.929 \\
373694425 &  `Alopeke  &  Burt  & 2020/12/03 & 4.685 & 5.585 & 4.709 & 6.419 \\
''  &  ''  &  Howell  & 2024/08/15 & 4.812 & 6.239 & 4.515 & 7.322 \\
373765355 &  `Alopeke  &  Hartman  & 2025/02/17 & 5.365 & 7.222 & 4.267 & 6.927 \\
377415363 &  `Alopeke  &  Howell  & 2019/06/07 & 5.465 & 6.880 & 4.719 & 6.737 \\
389853353 &  `Alopeke  &  Hartman  & 2025/02/17 & 5.037 & 6.794 & 4.790 & 6.751 \\
399665349 &  NESSI  &  Howell  & 2019/01/23 & 2.594 & 6.313 & 2.022 & 5.827 \\
''  &  `Alopeke  &  ''  & 2024/11/21 & 5.089 & 7.396 & 4.543 & 7.045 \\
409104974 &  `Alopeke  &  Hartman  & 2025/02/18 & 4.718 & 6.777 & 3.844 & 6.851 \\
416519065 &  `Alopeke  &  Howell  & 05/23/2024 & 4.557 & 5.722 & 3.541 & 6.896 \\
''  &  ''  &  ''  & 2024/11/21 & 5.404 & 6.168 & 4.756 & 6.624 \\
''  &  ''  &  Hartman  & 2025/02/17 & 3.618 & 5.197 & 4.220 & 5.917 \\
417762326 &  `Alopeke  &  Hartman  & 2025/02/18 & 4.200 & 5.160 & 4.527 & 7.019 \\
419015728 &  NESSI  &  Howell  & 2018/11/20 & 2.921 & 7.273 & 1.768 & 6.607 \\
''  &  Zorro  &  Burt  & 2020/11/28 & 5.303 & 7.047 & 4.890 & 6.769 \\
425935521 &  Zorro  &  Howell  & 2020/10/29 & 4.309 & 6.555 & 4.431 & 6.357 \\
'' &  Zorro  &  Burt  & 2020/11/25 & 5.304 & 6.346 & 4.477 & 6.651 \\
434210589 &  `Alopeke  &  Burt  & 2020/12/06 & 5.007 & 6.850 & 4.613 & 6.572 \\
437886584 &  `Alopeke  &  Hartman  & 2025/02/18 & 5.436 & 7.336 & 4.516 & 7.187 \\
441709021 &  NESSI  &  Howell  & 2019/01/20 & 2.574 & 5.778 & 2.037 & 5.423 \\
445070560 &  `Alopeke  &  Howell  & 2024/01/23 & 5.072 & 7.072 & 4.325 & 6.962 \\
''  &  ''  &  Hartman  & 2025/02/17 & 5.072 & 7.072 & 4.325 & 6.962 \\
445258206 &  `Alopeke  &  Burt  & 2020/12/03 & 4.610 & 5.135 & 4.960 & 6.514 \\
''  &  ''  &  Howell  & 2024/09/26 & 5.309 & 6.065 & 4.769 & 6.459 \\
458445966 &  `Alopeke  &  Hartman  & 2025/02/17 & 4.948 & 6.761 & 4.257 & 6.474 \\
1101124558 &  `Alopeke  &  Hartman  & 2025/02/18 & 5.591 & 7.151 & 5.172 & 7.039 \\
1101124559 &  `Alopeke  &  Hartman  & 2025/02/18 & 4.931 & 6.704 & 4.423 & 6.522 
\enddata
\end{deluxetable}

\begin{deluxetable}{cccc}
\tablewidth{700pt}
\tablehead{
\colhead{TIC ID} & \colhead{$\rho$} & \colhead{$\theta$} & \colhead{$\Delta m$} \\
\colhead{-} & \colhead{[\arcsec]} & \colhead{[\degree]} & \colhead{[mag]} }
\tablecaption{\label{tab:specklebinresults}Speckle Detected Binary Results}
\startdata
259291108 & 0.883 & 63.300 & 7.500 \\
367631379 & 0.022 & 2.374 & 0.480
\enddata
\tablecomments{Only 832\,nm filter results shown. No companions were found in the 562\,nm observations.}

\end{deluxetable}

\begin{longrotatetable}
\begin{deluxetable}{ccccccccccccccccc}
\tablewidth{700pt}
\tabletypesize{\tiny}
\tablehead{
\colhead{TIC ID} & \colhead{Gaia DR3 ID} & \colhead{R.V. Error} & \colhead{RUWE} & \colhead{IPD\_fmp} & \colhead{NSS} & \colhead{IPD\_gof} & \colhead{RV\_gof} & \colhead{RV\_chi} & \colhead{RV\_nb} & \colhead{Source} & \colhead{Gaia $\mathrm{ID}_2$}& \colhead{Gaia $\mathrm{ID}_3$}& \colhead{$\rho_1$}& \colhead{$\Delta m_1$} & \colhead{$\rho_2$}& \colhead{$\Delta m_2$} \\
\colhead{-} & \colhead{-} & \colhead{[km/s]} & \colhead{-} & \colhead{-} & \colhead{-} & \colhead{-} & \colhead{-} & \colhead{-} & \colhead{-} & \colhead{-} & \colhead{-} & \colhead{-} & \colhead{[\arcsec]} & \colhead{[mag.]}& \colhead{[\arcsec]}& \colhead{[mag]}}
\tablecaption{\label{tab:gaiaresults}Relevant Gaia Information}
\startdata
3962869 & 2473608009504466688 & 0.12 & 1.03 & -- & -- & 0.07 & -- & 0.33 & 12 &  & -- & -- & -- & -- & -- & -- \\
8915802 & 798068905726303232 & 0.119 & 1.01 & -- & 1 & 0.05 & -- & 0.25 & 27 & 1. & 798068940086479104 & -- & 7.134 & -5.982 & -- & -- \\
11310083 & 5663558453071724288 & 0.129 & 0.85 & -- & -- & 0.01 & -- & 0.0 & 32 &  & -- & -- & -- & -- & -- & -- \\
21535479 & 777254360337133312 & 0.123 & 0.85 & -- & -- & 0.01 & -- & 0.1 & 17 &  & -- & -- & -- & -- & -- & -- \\
23969522 & 821455861647126656 & 0.115 & 0.8 & -- & -- & 0.01 & -- & 0.71 & 9 &  & -- & -- & -- & -- & -- & -- \\
27136704 & 3407611182744289152 & 0.125 & 0.88 & -- & -- & 0.06 & -- & -- & -- &  & -- & -- & -- & -- & -- & -- \\
38511251 & 5164120762333028736 & -- & 2.99 & -- & -- & 0.03 & -- & 0.24 & 19 &  & -- & -- & -- & -- & -- & -- \\
46125330 & 3374052988354372992 & 0.139 & 1.18 & -- & -- & 0.04 & -- & 0.99 & 15 &  & -- & -- & -- & -- & -- & -- \\
47346402 & 3400292798990117888 & 0.138 & 0.82 & -- & -- & 0.15 & -- & 0.38 & 12 & 2. & 3394298532176344960 & -- & 707.219 & 2.693 & -- & -- \\
60716322 & 2719475542667772416 & 0.124 & 3.27 & -- & -- & 0.01 & -- & 0.01 & 35 & 1.& 2719475542667351552 & -- & 11.042 & -6.68 & -- & -- \\
67772871 & 3195919528989223040 & 0.12 & 1.94 & -- & -- & 0.01 & -- & 0.01 & 18 & 2. & 3195919254111314816 & 3195919254111315712 & 78.097 & 5.595 & 83.337 & 5.362 \\
72748794 & 5117974602912370432 & 0.126 & 1.07 & 2 & -- & 0.07 & -- & 0.04 & 12 &  & -- & -- & -- & -- & -- & -- \\
80226651 & 3381727613874753536 & 0.135 & 0.95 & -- & -- & 0.05 & -- & 0.69 & 16 &  & -- & -- & -- & -- & -- & -- \\
95431211 & 624684408880100992 & 0.136 & 1.1 & -- & -- & 0.09 & -- & 0.14 & 21 &  & -- & -- & -- & -- & -- & -- \\
101641846 & 4025850731201819392 & 0.139 & 1.14 & -- & -- & 0.04 & -- & 0.61 & 29 &  & -- & -- & -- & -- & -- & -- \\
113710966 & 96331172942614528 & 0.126 & 1.26 & -- & -- & 0.01 & -- & 0.14 & 24 &  & -- & -- & -- & -- & -- & -- \\
116988032 & 436648129327098496 & 0.117 & 2.67 & -- & -- & 0.01 & -- & 0.49 & 11 &  & -- & -- & -- & -- & -- & -- \\
117979951 & 2980492387558410112 & 0.117 & 1.07 & 1 & -- & 0.04 & -- & 0.83 & 12 &  & -- & -- & -- & -- & -- & -- \\
118572803 & 5164707970261890560 & -- & 2.72 & -- & -- & 0.04 & -- & 0.83 & 16 &  & -- & -- & -- & -- & -- & -- \\
135656809 & 4345775217221821312 & 0.121 & 0.98 & -- & -- & 0.02 & -- & 0.95 & 55 &  & -- & -- & -- & -- & -- & -- \\
141810080 & 5264749303462961280 & 0.12 & 1.36 & -- & -- & 0.03 & -- & 0.25 & 31 & 1.& 5264749303457104384 & -- & 3.29 & -7.465 & -- & -- \\
147407292 & 6564091190988411520 & 0.116 & 0.88 & -- & -- & 0.02 & -- & 0.58 & 16 &  & -- & -- & -- & -- & -- & -- \\
155315739 & 6553614253923452800 & 0.133 & 0.93 & -- & -- & 0.06 & -- & 0.56 & 14 &  & -- & -- & -- & -- & -- & -- \\
156890613 & 974887555341533440 & 0.122 & 1.11 & -- & -- & 0.09 & -- & 0.23 & 23 &  & -- & -- & -- & -- & -- & -- \\
157364190 & 1384195086792268800 & 0.121 & 0.95 & -- & -- & 0.04 & -- & 0.58 & 12 &  & -- & -- & -- & -- & -- & -- \\
157966796 & 1284257623084916480 & 0.123 & 1.18 & -- & -- & 0.03 & -- & 0.92 & 30 &  & -- & -- & -- & -- & -- & -- \\
165602000 & 1872046609345556480 & 0.117 & 1.2 & -- & -- & 0.07 & -- & 0.66 & 44 & 2. & 1872046574983497216 & -- & 31.593 & 0.684 & -- & -- \\
166646191 & 762815470562110464 & 0.131 & 0.96 & -- & -- & 0.03 & -- & 0.78 & 27 &  & -- & -- & -- & -- & -- & -- \\
165602023 & 1872046574983497216 & 0.118 & 0.96 & 1 & -- & 0.01 & -- & 0.14 & 17 & 2. & 1872046609345556480 & -- & 31.593 & 0.684 & -- & -- \\
172954294 & 746545172372256384 & 0.122 & 0.94 & -- & -- & 0.04 & -- & 0.0 & 17 & 2. & 746545859566712832 & -- & 134.386 & 8.919 & -- & -- \\
176521059 & 3023711269067191296 & 0.121 & 1.02 & -- & -- & 0.02 & -- & 0.65 & 10 &  & -- & -- & -- & -- & -- & -- \\
179348425 & 6009538585839374336 & 0.122 & 0.94 & -- & -- & 0.02 & -- & 0.53 & 13 & 2. & 6009537829925128064 & -- & 14.456 & 7.167 & -- & -- \\
189576919 & 348020482735930112 & 0.126 & 7.25 & -- & -- & 0.01 & -- & 0.58 & 12 & 2. & 348020242217448576 & -- & 55.618 & 8.531 & -- & -- \\
202380743 & 1661568850770761600 & 0.122 & 0.99 & -- & -- & 0.09 & -- & 0.0 & 34 &  & -- & -- & -- & -- & -- & -- \\
206686962 & 6604147121141267712 & 0.124 & 0.92 & -- & -- & 0.06 & -- & 0.8 & 35 &  & -- & -- & -- & -- & -- & -- \\
213041474 & 3211461469444773376 & 0.121 & 0.97 & -- & -- & 0.01 & -- & 0.42 & 24 &  & -- & -- & -- & -- & -- & -- \\
219709102 & 1068982214258617216 & 0.117 & 0.86 & -- & -- & 0.02 & -- & 0.34 & 18 & 1.& 1068982214258617856 & -- & 4.466 & -3.623 & -- & -- \\
229092427 & 4934923028038871296 & 0.117 & 1.38 & -- & -- & 0.02 & -- & 0.85 & 14 &  & -- & -- & -- & -- & -- & -- \\
229902025 & 1264630412816366720 & 0.13 & 1.08 & -- & -- & 0.02 & -- & 0.02 & 19 &  & -- & -- & -- & -- & -- & -- \\
231698181 & 6412595290592307840 & 0.129 & 1.15 & -- & -- & 0.03 & -- & 0.67 & 30 &  & -- & -- & -- & -- & -- & -- \\
238432056 & 2860924621205256704 & 0.117 & 0.86 & -- & -- & 0.01 & -- & 0.98 & 22 &  & -- & -- & -- & -- & -- & -- \\
257393898 & 2552925644460225152 & 0.118 & 1.08 & -- & -- & 0.02 & -- & 0.0 & 21 &  & -- & -- & -- & -- & -- & -- \\
259237827 & 2261614264931057664 & 0.128 & 1.68 & 1 & -- & 0.01 & -- & 0.21 & 9 &  & -- & -- & -- & -- & -- & -- \\
259291108 & 6408551802222368384 & 0.143 & 2.42 & 2 & -- & 0.04 & -- & 0.7 & 12 & 1.& 6408551802221816576 & -- & 23.597 & -7.357 & -- & -- \\
261136679 & 4623036865373793408 & 0.118 & 0.81 & -- & -- & 0.03 & -- & 0.98 & 15 &  & -- & -- & -- & -- & -- & -- \\
279649049 & 4722135642226902656 & 0.128 & 1.2 & -- & -- & 0.03 & -- & 0.48 & 28 & 2. & 4722111590409480064 & -- & 309.115 & 0.265 & -- & -- \\
279649057 & 4722111590409480064 & 0.122 & 0.98 & -- & -- & 0.02 & -- & 0.88 & 23 & 2. & 4722135642226902656 & -- & 309.115 & 0.265 & -- & -- \\
283722336 & 2009481748875806976 & 0.119 & 1.0 & -- & -- & 0.06 & -- & 0.37 & 20 &  & -- & -- & -- & -- & -- & -- \\
285544488 & 427322415301550208 & 0.123 & 1.08 & -- & -- & 0.0 & -- & 0.7 & 20 &  & -- & -- & -- & -- & -- & -- \\
289673491 & 2416611663182821120 & 0.128 & 1.69 & -- & -- & 0.02 & -- & 0.94 & 28 &  & -- & -- & -- & -- & -- & -- \\
301051051 & 4847957293278177024 & 0.127 & 1.98 & -- & -- & 0.03 & -- & 0.98 & 22 &  & -- & -- & -- & -- & -- & -- \\
302158903 & 438829629114680704 & 0.127 & 3.17 & -- & -- & 0.04 & -- & 0.01 & 15 & 1.& 438829835272390784 & -- & 20.932 & -5.157 & -- & -- \\
309599261 & 4034171629042489088 & 0.122 & 0.88 & -- & -- & 0.03 & -- & 0.09 & 14 &  & -- & -- & -- & -- & -- & -- \\
311063391 & 263916708025623680 & 0.126 & 0.96 & -- & -- & 0.02 & -- & 0.19 & 13 & 2. & 263916742385357056 & -- & 98.035 & 2.928 & -- & -- \\
311092847 & 3263836568394170880 & 0.116 & 3.06 & 1 & -- & 0.02 & -- & 0.0 & 16 &  & -- & -- & -- & -- & -- & -- \\
328324648 & 976893923544104576 & 0.128 & 1.49 & -- & -- & 0.01 & -- & 0.08 & 21 &  & -- & -- & -- & -- & -- & -- \\
332064670 & 704967037090946688 & 0.122 & 0.86 & -- & -- & 0.03 & -- & -- & -- & 2. & 704966762213039488 & -- & 84.826 & 5.95 & -- & -- \\
343813545 & 3269362645115584640 & 0.148 & 1.34 & -- & -- & 0.03 & -- & 0.24 & 30 &  & -- & -- & -- & -- & -- & -- \\
355127594 & 714116137767540096 & 0.119 & 0.87 & -- & -- & 0.0 & -- & 0.63 & 13 &  & -- & -- & -- & -- & -- & -- \\
366661076 & 3796442680948579328 & 0.296 & 2.81 & -- & -- & 0.06 & -- & 0.09 & 14 &  & -- & -- & -- & -- & -- & -- \\
367631379 & 1146337324038514176 & 0.636 & 9.92 & 1 & 1 & 0.03 & -- & 0.0 & 27 & 1.& 1146337324038513536 & -- & 13.495 & -7.604 & -- & -- \\
371520835 & 823773494718931968 & 0.119 & 0.94 & -- & -- & 0.02 & -- & 0.98 & 26 &  & -- & -- & -- & -- & -- & -- \\
373694425 & 512167948043650816 & 0.12 & 0.98 & -- & -- & 0.02 & -- & 0.41 & 5 &  & -- & -- & -- & -- & -- & -- \\
373765355 & 3936909723803146368 & 0.123 & 0.95 & -- & -- & 0.02 & -- & 0.21 & 34 & 2. & 3936909723803146496 & -- & 7.671 & 2.572 & -- & -- \\
377415363 & 1193030490492925824 & 0.127 & 1.54 & -- & -- & 0.01 & -- & 0.94 & 24 &  & -- & -- & -- & -- & -- & -- \\
389853353 & 1521686782461765888 & 0.124 & 0.83 & -- & -- & 0.02 & -- & 0.37 & 21 &  & -- & -- & -- & -- & -- & -- \\
399665349 & 3288921720025503360 & -- & 3.92 & -- & -- & 0.07 & -- & 0.23 & 9 &  & -- & -- & -- & -- & -- & -- \\
409104974 & 188771135583573504 & 0.12 & 0.89 & -- & -- & 0.01 & -- & 0.7 & 17 &  & -- & -- & -- & -- & -- & -- \\
416519065 & 853819947756949120 & 0.123 & 0.89 & -- & -- & 0.03 & -- & 0.03 & 29 & 2. & 853820948481913472 & -- & 122.865 & 3.411 & -- & -- \\
417762326 & 1092545710514654464 & 0.117 & 1.17 & -- & -- & 0.04 & -- & 0.0 & 14 &  & -- & -- & -- & -- & -- & -- \\
419015728 & 2452378776434477184 & -- & 2.63 & -- & -- & 0.05 & -- & 1.0 & 11 &  & -- & -- & -- & -- & -- & -- \\
425935521 & 4900108950849461248 & 0.129 & 3.21 & -- & -- & 0.12 & -- & 0.26 & 6 &  & -- & -- & -- & -- & -- & -- \\
434210589 & 2802397960855105920 & 0.126 & 0.88 & -- & -- & 0.04 & -- & -- & -- &  & -- & -- & -- & -- & -- & -- \\
437886584 & 3331979901042416512 & 0.131 & 0.98 & -- & -- & 0.06 & -- & 0.45 & 13 &  & -- & -- & -- & -- & -- & -- \\
441709021 & 1604859511344724864 & 0.135 & 3.38 & -- & -- & 0.19 & -- & 0.31 & 8 & 2. & 1604859408265509632 & -- & 69.674 & 6.581 & -- & -- \\
445070560 & 1460229442589223424 & 0.127 & 2.46 & -- & -- & 0.13 & -- & 0.2 & 10 &  & -- & -- & -- & -- & -- & -- \\
445258206 & 425040000962559616 & -- & 3.12 & -- & -- & 0.04 & -- & 0.0 & 23 & 1.& 425040000962497792 & -- & 13.327 & -3.442 & -- & -- \\
458445966 & 1534011998572555776 & 0.117 & 2.11 & -- & -- & 0.08 & -- & 0.58 & 14 &  & -- & -- & -- & -- & -- & -- \\
1101124558 & 1237090738916392704 & 0.124 & 1.1 & -- & -- & 0.05 & -- & 0.47 & 24 & 1.& 1237090738916392832 & -- & 5.656 & -1.95 & -- & -- \\
1101124559 & 1237090738916392832 & 0.125 & 1.51 & -- & -- & 0.09 & -- & -- & -- & 1. & 1237090738916392704 & -- & 5.656 & -1.95 & -- & -- 
\enddata
\tablecomments{1. - \citet{2021elbadry}, 2. - \citet{2021gcns}}
\end{deluxetable}
\end{longrotatetable}

\begin{longrotatetable}
\begin{deluxetable}{cccccccc}
\tablewidth{700pt}
\tabletypesize{\tiny}
\tablehead{\colhead{TIC ID} & \colhead{WDS ID} & \colhead{Gaia-resolved Multiple} & \colhead{Elevated Gaia Metrics} & \colhead{Speckle Detection} & \colhead{Lit. Source(s)}& \colhead{Multiplicity Status}& \colhead{Notes on Multiplicity}\\
\colhead{-} & \colhead{-} & \colhead{-}& \colhead{-}& \colhead{-}& \colhead{-}& \colhead{-} & \colhead{-}}
\tablecaption{\label{tab:overall}Multiplicity results for 80 observed targets using speckle observations, Gaia, and literature sources}
\startdata
3962869 & -- & -- & -- & -- & 8. & S &  \\
8915802 & 09357+3549 & X & X & -- & 1., 8. & M & \makecell{Triple system. Wide companion in WDS. \\ TIC 8915802 is marked in the Gaia NSS \\ Period = 27.1 days.} \\
11310083 & -- & -- & -- & -- & 8. & S &  \\
21535479 & -- & -- & -- & -- & 8. & S &  \\
23969522 & -- & -- & -- & -- & 8. & S &  \\
27136704 & 05074+1839 & -- & -- & -- & 4. & S & \makecell{Speckle finds no companion. \\ WDS entries marked with X and U.} \\
38511251 & -- & -- & X & -- & 6. & S &  \\
46125330 & 06148+1909 & -- & -- & -- & 4. & S & All components are background stars \\
47346402 & 05244+1723 & X & -- & -- & 2. & M & \makecell{AB is marked as U. \\ Wide companion confirmed by Gaia.}\\
60716322 & 22467+1210 & X & X & -- & 1. & M & No third companion with similar parallax within 500”. \\
67772871 & 04153-0739 & X & -- & -- & 2. & M & Triple \\
72748794 & -- & -- & X & -- & 8. & S &  \\
80226651 & -- & -- & -- & -- & 8. & S &  \\
95431211 & 10197+1928 & -- & -- & -- & 4. & S & \makecell{Not binary according to WDS notes \\ 4. marks as single.} \\
101641846 & 11411+3412 & -- & -- & -- & 8. & S & \makecell{Marked with L \\ no component seen in Gaia.} \\
113710966 & 01425+2016 & -- & -- & -- & 6. & S & \makecell{Gaia does not show either WDS companion. \\ AC is marked as U.}  \\
116988032 & 03091+4937 & -- & X & -- & 8. & S & WDS entry marked as S. \\
117979951 & -- & -- & -- & -- & 8. & S &  \\
118572803 & 03329-0927 & -- & X & -- & 6. & S & Marked as X in WDS. \\
135656809 & 16156-0822 & -- & -- & -- & 8. & S & WDS entry marked as U. \\
141810080 & 6102-7445 & X & -- & -- & 1. & M &  \\
147407292 & 21483-4718 & -- & -- & -- & 8. & S & WDS entry marked as U. \\
155315739 & 23059-3551 & -- & -- & -- & 5., 9. & S & Background source detection in 9. \\
156890613 & -- & -- & -- & -- & 8. & S &  \\
157364190 & -- & -- & -- & -- & 8. & S &  \\
157966796 & 14347+2945 & -- & -- & -- & 8. & S & \makecell{Marked with L and U. \\ Not seen in Gaia.} \\
165602000 & 21069+3845 & X & -- & -- & 6. & M &  \\
165602023 & 21069+3845 & X & -- & -- & 6. & M &  \\
166646191 & -- & -- & -- & -- & 6. & S &  \\
172954294 & 10010+3155 & X & -- & -- & 2., 8. & M & \makecell{Possible Triple due to over-luminosity. \\ However, no confirmed detection.}\\
176521059 & -- & -- & -- & -- & 5. & S &  \\
179348425 & 15475-3755 & X & -- & -- & 2., 5. & M & AC is marked as S in WDS \\
189576919 & 01368+4124 & X & X & -- & 2., 5. & M &  \\
202380743 & -- & -- & -- & -- & 8. & S &  \\
206686962 & 22577-2937 & -- & -- & -- & 6. & M & \makecell{Fomalhaut system \\ $\alpha$ PsA is not in Gaia \\ B-C not recovered by wide binary searches.} \\
213041474 & -- & -- & -- & -- & 6. & S &  \\
219709102 & 09104+6708 & X & -- & -- & 1. & M & Second WDS entry not confirmed in Gaia \\
229092427 & -- & -- & -- & -- & 8. & S &  \\
229902025 & 15073+2452 & -- & -- & -- & 8. & S & All WDS entires marked with U. \\
231698181 & 22034-5647 & -- & -- & -- & 6. & M & Part of triple with brown dwarf binary \\
238432056 & 00066+2901 & -- & -- & -- & 8. & S & All other WDS entries are marked with U. \\
257393898 & 00484+0517 & -- & -- & -- & 6. & S & \makecell{No companions seen in Gaia. \\ Second WDS entry marked with U.} \\
259237827 & 19324+6940 & -- & -- & -- & 6. & S & No companions seen in Gaia. \\
259291108 & 22249-5748 & X & X & X & 1., 7. & M & Triple system with close-in companion. \\
261136679 & -- & -- & -- & -- & 5. & S &  \\
279649049 & 03182-6230 & X & -- & -- & 2. & M & No close binary detected \\
279649057 & 03182-6230 & X & -- & -- & 2. & M & No close binary detected \\
283722336 & 23133+5710 & -- & -- & -- & 6. & S &  \\
285544488 & 00531+6107 & -- & -- & -- & 4. & S & All WDS components marked as U. \\
289673491 & -- & -- & -- & -- & 8. & S &  \\
301051051 & -- & -- & -- & -- & 6. & S &  \\
302158903 & 02442+4914 & X & X & -- & 1. & M & Potential Third companion BD although only seen once. \\
309599261 & -- & -- & -- & -- & 6. & S &  \\
311063391 & 05413+5329 & X & -- & -- & 8. & M & All other WDS entries marked with U.; \\
311092847 & -- & -- & X & -- & 8. & S &  \\
328324648 & 07299+4940 & -- & -- & -- & 4. & S & 4. has single. \\
332064670 & 08526+2820 & X & -- & -- & 2., 5. & M &  \\
343813545 & 03194+0322 & -- & -- & -- & 6. & S & WDS entries marked with U. \\
355127594 & -- & -- & -- & -- & 8. & S &  \\
366661076 & 11507+0146 & -- & X & -- & 8. & S & Both WDS entries listed as L or U. \\
367631379 & 10311+8234 & X & X & X & 1., 3., 4. & M & \makecell{Triple system \\ Close companion detected in our speckle observations. \\ Marked as NSS.} \\
371520835 & 10114+4927 & -- & -- & -- & 6. & S & No Gaia confirmation of companions. \\
373694425 & 01477+6351 & -- & -- & -- & 8. & S & Both WDS entries marked with L or U. \\
373765355 & 13169+1701 & X & -- & -- & 5. & M &  \\
377415363 & 15565+1540 & -- & -- & -- & 8. & S & WDS entries marked with L or S. \\
389853353 & 05191+4006 & -- & -- & -- & 8. & S & All WDS entries are chance alignments \\
399665349 & 04498+0658 & -- & X & -- & 6. & S & \makecell{Potential BD companions; \citet{2007lafreniere}. \\ Other WDS entries are marked with U.} \\
409104974 & -- & -- & -- & -- & 8. & S &  \\
416519065 & 10306+5559 & X & -- & -- & 2. & M & Second WDS entry is U \\
417762326 & -- & -- & -- & -- & 8. & S &  \\
419015728 & 01441-1556 & -- & X & -- & 6. & S & WDS entry marked with L. \\
425935521 & -- & -- & X & -- & 6. & S &  \\
434210589 & 00394+2115 & -- & -- & -- & 5. & M & \makecell{AB entry marked with U. \\ C is brown dwarf companion.} \\
437886584 & 06164+1216 & -- & -- & -- & 4. & S & \makecell{WDS reports two companions. \\ Gaia does not detect these \\ 4. lists the target as single}\\
441709021 & 14252+5151 & X & X & -- & 2. & M & Wide Companion \\
445070560 & 13119+2753 & -- & X & -- & 6. & S & WDS entry marked with U. \\
445258206 & 00491+5749 & X & X & -- & 6. & M & \makecell{Wide companion \\ Other WDS entries marked with U; L; and S} \\
458445966 & 12337+4121 & -- & X & -- & 8. & S & WDS entries marked with X or U. \\
1101124558 & 14514+1906 & X & -- & -- & 6. & M & Other WDS entries marked with U \\
1101124559 & 14514+1906 & X & -- & -- & 6. & M & Other WDS entries marked with U 
\enddata
\tablecomments{1. - \citet{2021elbadry}, 2. - \citet{2021gcns}, 3. - \citet{2023gaiamultiples}, 4. - \citet{2023golovin}, 5. - \citet{2024gonzalez}, 6. - \citet{2025gonzalezpayo}, 7. - \citet{2019kane}, 8. - \citet{2024kirkpatrick}, 9. - \citet{2015wardduong}}
\end{deluxetable}
\end{longrotatetable}

\end{document}